\documentclass[letterpaper, 12pt]{article}
\pdfoutput = 1

\usepackage{shortcuts}

\usepackage[margin = 2.5cm]{geometry}
\titleformat{\subsubsection}
  {\normalfont\normalsize \it}{\thesubsubsection}{1em}{}

\numberwithin{equation}{section}

\usepackage[nottoc]{tocbibind}

\begin{document}

\thispagestyle{empty}

\begin{flushright}
\texttt{BRX-TH-6682}
\end{flushright}

\begin{center}

\vspace*{0.02\textheight}

{\Large \textbf{The Ultraviolet Structure of Quantum Field Theories} \\ \medskip
  Part 3: Gauge Theories}

\vspace{1cm}

{\large \DJ or\dj e Radi\v cevi\'c}
\vspace{1em}

{\it Martin Fisher School of Physics\\ Brandeis University, Waltham, MA 02453, USA}\\ \medskip
\texttt{djordje@brandeis.edu}\\

\vspace{0.04\textheight}

\begin{abstract}
{This paper develops a detailed lattice-continuum correspondence for all common examples of Abelian gauge theories, with and without matter. These rules for extracting a continuum theory out of a lattice one represent an elementary way to rigorously define continuum gauge theories. The focus is on $(2+1)$D but the techniques developed here work in all dimensions.

The first half of this paper is devoted to pure Maxwell theory. It is precisely shown how continuum Maxwell theory emerges at low energies in an appropriate parameter regime of the $\Z_K$ lattice gauge theory at large $K$. The familiar features of this theory --- its OPE structure, the distinction between compact and noncompact degrees of freedom, infrared ``particle-vortex'' dualities, and the Coulomb law behavior of its Wilson loop --- are all derived directly from the lattice.

The rest of the paper studies gauge fields coupled to either bosonic or fermionic matter. Scalar QED is analyzed at the same level of precision as pure Maxwell theory, with new comments on its phase structure and its connections to the topological BF theory. Ordinary QED (Dirac fermions coupled to Maxwell theory) is constructed with an eye toward properly defining spinors and avoiding global anomalies. The conventional continuum QED is shown to arise from a specific restriction of the starting theory to smoothly varying fields. Such smoothing also turns out to be a necessary condition for Chern-Simons theory to arise from integrating out massive fermions in a path integral framework. In an effort to find a canonical origin of Chern-Simons theory that does not rely on path integral smoothing, a simple flux-attached lattice gauge theory is shown to give rise to a Chern-Simons-like action in the confining regime.}
\end{abstract}

\vfill
\textit{Dedicated to all foes of misinformation.}
\end{center}

\newpage


\tableofcontents

\newpage

\section{Introduction}

If you take only one lesson away from this paper, let it be this:
\begin{center}
\begin{minipage}{0.8\textwidth}
  \it \centering
  Defining the continuum limit of a lattice gauge theory has nothing to do with introducing a lattice spacing and letting it be infinitesimally small.
\end{minipage}
\end{center}
This claim bucks decades of tradition. Since the very beginning \cite{Wilson:1974sk, Wilson:2004de}, lattice gauge theories have been defined as quantum systems with $N \gg 1$ degrees of freedom along each direction of space (or spacetime), with continuum limits obtained by judiciously rescaling operators and couplings by powers of the lattice spacing $a \propto 1/N$ so that the formal limit $a \rar 0$
yields Hamiltonians (or actions) with no explicit factors of $a$. While this is often a perfectly cromulent heuristic, it remains silent about what happens when field eigenvalues are of size $O(1/a)$, or when the difference between two neighboring fields does not scale as $O(a)$, or when a continuum coupling is dialed away from $O(a^0)$ values. In other words, the standard discussion of a continuum limit implicitly (and sometimes explicitly \cite{Luscher:1998du}) assumes that continuum fields are well behaved and that continuum couplings are $O(a^0)$. It provides neither a quantifiable justification for this assumption, nor a playbook for what to do when the assumption fails. This paper will fix this shortcoming, and along the way it will show that an infinitesimal lattice spacing is not even needed to define the continuum limit of a lattice gauge theory.

The issue outlined above is not a purely academic question of tidying up the definition of the continuum limit. When couplings in a continuum quantum field theory (cQFT) are not $O(a^0)$, the theory will generically be in a regime where naturalness is violated \cite{tHooft:1979rat}. With evidence piling up that our world is described by a somewhat unnatural cQFT \cite{Weinberg:1988cp, Peccei:2006as, deGouvea:2014xba}, we would be remiss not to examine such theories in a well defined lattice setup. Furthermore, theories in which fields are not necessarily smooth have recently been marketed as a new frontier in the QFT study of condensed matter \cite{Gorantla:2020xap}. This is therefore a timely reminder that questions of smoothness play a nontrivial r$\hat{\trm o}$le even in conventional lattice field theories.

That said, there are also excellent formal reasons to be interested in the  lattice-continuum correspondence. By being very precise about how a lattice gauge theory reaches its continuum limit, one is led to a nonperturbative \emph{definition} of a continuum gauge theory. Rigorously defined cQFTs are generally few and far between, and a lattice-based rigorous definition is especially valuable because it involves only familiar concepts of finite-dimensional quantum mechanics --- no advanced functional analysis necessary! This finitary approach will be particularly illuminating when it comes to phase transitions. In a lattice theory in which the limit $a \rar 0$ is not haphazardly taken, all transitions are crossovers, and it becomes possible to describe the confinement/deconfinement transition in ways that were unavailable in the literature.

\newpage

This novel approach to continuum limits of lattice gauge theories is just one facet of the broader push to understand the lattice-continuum correspondence in general QFTs \cite{Radicevic:1D, Radicevic:2D, Radicevic:4D}. The main idea is the same for all lattice theories.\footnote{As in the rest of the series, a ``lattice theory'' is a quantum theory with a finite-dimensional Hilbert space. In particular, a lattice theory always has an explicitly regulated target space. This means that there is no such thing as a U(1) lattice gauge theory in this paper. All Abelian theories discussed here will have $\Z_K$ target spaces, possibly for extremely large values of $K$.} In short, a lattice theory with an operator algebra $\A$ has a continuum description if two criteria are satisfied:
\begin{enumerate}
  \item There exists a \emph{precontinuum basis} that spans the algebra $\A$. This basis generalizes the usual notion of a Fock basis. It is generated by a large number of ladder operators $c_k$ and $c_k\+$ whose labels $k$ constitute the \emph{momentum space} $\Pbb$. The ladder operators are chosen so that particle number operators $n_k \equiv c\+_k c_k$ commute with each other and with the Hamiltonian, and have integer eigenvalues $\{0, 1, \ldots, J_k - 1\}$ for some $J_k \geq 2$.
  \item The Hamiltonian equips the momentum space with a notion of distance (or topology), so that nearby momenta have similar energy costs of changing the particle numbers. This in turn makes it possible to define a ``piecewise continuous'' space $\Pbb\_S \subset \Pbb$ of momenta at which particle excitations cost a small amount of energy.
\end{enumerate}
If both conditions are met, a \emph{continuum basis} can be defined by removing all high-momentum ($k \notin \Pbb\_S$) basis operators, except for particle number operators $n_k$ and their powers, from the precontinuum basis. This new basis spans the \emph{continuum algebra} $\A\_S$. All pure density matrices in this algebra have a constant amount of entanglement at high momenta. The only ladder operators remaining in $\A\_S$ are associated to low momenta $k \in \Pbb\_S$. When expressed on the Fourier dual of the momentum space $\Pbb$, these ladder operators $c_k$ precisely become the dynamical quantum fields $c(x)$ of a cQFT: they are defined on a lattice, but their lack of high-momentum modes makes them vary smoothly from site to site.

The remarkable fact about these lattice-based continuum fields is that they truly encode all the familiar properties of cQFTs. For example, it is possible to use them to define operator product expansions \cite{Radicevic:2019jfe, Radicevic:2019mle}; and when rephrased in the path integral language, it is possible to precisely see the invariance of their actions under infinitesimal symmetries. The parameter that characterizes the continuum nature of all these phenomena is the ratio $k\_S/N \ll 1$, where $N \gg 1$ is the linear size of the lattice, and $k\_S$ is the linear size of the low-momentum subspace $\Pbb\_S$. Crucially, the limit $a \propto 1/N \rar 0$ does \emph{not} need to be taken before all other limits in order to obtain a continuum theory. This fact allows for a precise analysis of how various extremal couplings or field fluctuations manifest themselves in a cQFT setting.

A general discussion of this construction can be found in the previous part of this series \cite{Radicevic:2D}. \textbf{The goal of this paper} is to systematically develop these ideas in the context of Abelian gauge theories in $d = 2$ spatial dimensions.

Before outlining the structure of this paper, it might be instructive to stress how its approach to the continuum differs from the customary ones in lattice gauge theory.
\begin{itemize}
  \item Monte Carlo lattice calculations typically focus on obtaining a limited set of few-point correlation functions in the continuum. The construction given here instead captures the entire operator algebra of the continuum theory, up to corrections specified by the quantity $k\_S/N$ mentioned above.
  \item The projection $\A \mapsto \A\_S$, here called \emph{smoothing}, effectively smears out position-space gauge fields over $N/2k\_S$ lattice sites in each spatial direction. Almost since their inception, lattice simulations have employed various smearings to more reliably extract continuum correlation functions \cite{Parisi:1983hm, Teper:1985rb, Albanese:1987ds}. Many modern studies employ a particular kind of smearing called the stout link formalism \cite{Morningstar:2003gk}. Smoothing differs from such familiar smearing constructions because it is defined in the Hamiltonian formalism, with temporal smoothing playing an important but logically very distinct r$\hat{\trm{o}}$le. Another difference is that smoothing has a completely analogous generalization to fermions, scalars, and other kinds of fields not necessarily associated to links.
  \item Conceptually, perhaps the biggest novelty of this series is that \emph{universality is not needed}. Conventional intuition says that a continuum description becomes valid when a large lattice system is brought near a second-order phase transition. In this parameter regime correlation lengths diverge and the large-distance correlation functions of certain operators become insensitive to the underlying lattice details. Such operators are then understood to be the fields of the appropriate cQFT. But in order to get a practical handle on this cQFT, one typically postulates an effective theory based on symmetries of the lattice theory, arguing that any deviation from the actual cQFT will be invisible at the large distance scales of interest. In other words, to keep things practical, field theorists restrict themselves to discussing only universal parts of a cQFT.

      The soundness of this  paradigm is strongly supported both by experience and by formal renormalization group (RG) arguments. Still, such analyses sometimes feel more like art than like science: one must find evidence that the phase transition is second order, postulate the right effective cQFT based on trial and error, and then try to maintain an often-slippery distinction between universal and nonuniversal data while exploring the cQFT parameter space.

      In contrast, the present approach constructs a cQFT directly out of lattice fields, using the same procedure for each lattice theory. There is no need to distinguish between universal and nonuniversal data because all data is considered physical. Common  but subtle notions of usual QFT --- counterterms, singular operator products, anomalous dimensions --- can all be given explicit definitions without invoking RG ideas.
  \item The previous comment does not mean that universality has no meaningful manifestation in this finitary paradigm. In fact, it appears in at least three different guises:
  \begin{enumerate}
    \item The lattice gauge theory algebra $\A$ that has a precontinuum basis is \emph{not} the maximal algebra generated by clock and shift operators associated to links. In any bosonic lattice theory, the precontinuum algebra is obtained from the maximal algebra by a procedure called \emph{taming} \cite{Radicevic:1D}. Taming is performed on each link separately. It essentially amounts to replacing a clock algebra with a subalgebra generated by (approximately) canonically conjugate pairs of position and momentum operators. Loosely speaking, tame operators act only on the space of small fluctuations in the gauge theory. A kind of universality, studied in detail in \cite{Radicevic:1D}, ensures that common lattice Hamiltonians have tame low-energy eigenstates.
    \item An interacting theory with a precontinuum basis, such as lattice quantum electrodynamics (QED), may induce a topology on the momentum space $\Pbb$ that significantly differs from the topology induced when the interactions are turned off. Consequently, the space $\Pbb\_S$ can nontrivially change as the matter charge is varied.  This was reviewed in some detail in \cite{Radicevic:2D}. This paper will assume that the change in $\Pbb\_S$ due to interactions is negligible. However, a proper treatment would integrate out high-momentum modes and identify universal properties of fixed points in the space of all possible low-momentum spaces $\Pbb\_S$.
    \item The transfer matrix formalism can be used to define path integrals for cQFTs constructed using smoothing. The variables of such path integrals are smooth along spatial directions only. Smoothing them along the temporal direction gives the familiar path integrals with nice properties like Euclidean rotation invariance. This temporal smoothing has no canonical analogue, and it is not a justified approximation. However, it can be understood to preserve a certain universal part of the original partition function \cite{Radicevic:1D}.
  \end{enumerate}
      While this will not be the focus of this paper, the above list shows that the smoothing paradigm splits the general notion of universality into several logically distinct concepts. Each of them can be studied separately.

  \item Each of the above three points featured a pair of ``before'' and ``after'' theories related by a type of universality. Even without universality, however, the ``after'' theories remain well defined. For example, a temporally smooth path integral is a well defined lattice object with desirable continuum properties, even if it cannot be shown to encode the same universal data as some quantum theory defined in the Hamiltonian formalism. This point of view makes it straightforward to rigorously define many interesting cQFT path integrals  in ways that were not attempted before.
\end{itemize}

\subsection*{Summary of the paper}

\textbf{Section \ref{sec Maxwell}} will illustrate the smoothing paradigm on one of the simplest lattice gauge theories around: the $\Z_K$  Maxwell gauge theory on a square toric lattice. The ultimate goal is to define the $(2+1)$D Maxwell cQFT and to present a rather precise picture of its dynamics.

This Section will cover some basic details that are often glossed over in QFT textbooks. It will start by defining the operator algebra of a $\Z_K$ gauge theory on a general $d = 2$ spatial lattice $\Mbb$. Even in this general setup it is possible to obtain a novel understanding of the phase structure of the Maxwell Hamiltonian \eqref{def H}. The standard lore is that this theory exhibits a confinement/deconfinement crossover as the gauge coupling $g \in \R^+$ is varied.\footnote{Since $\Mbb$ is always large but manifestly finite in this work, there are strictly speaking no phase transitions. Phase transitions can be \emph{defined} as those crossovers whose width in parameter space goes to zero as $N$ is increased.} In a slightly more sophisticated language, this crossover corresponds to a spontaneous breaking of a $\Z_K$ one-form symmetry, with the one-form symmetry fully broken at $g \rar 0$ and unbroken at $g \rar \infty$ \cite{Gaiotto:2014kfa}. Here it will be argued that the crossover between these extrema can be viewed in a more refined way, as a cascade of symmetry breakings, so that at each point along the crossover a subgroup $\Z_{K'} \subset \Z_K$ is approximately broken,  with $K'$ changing as $g$ is dialed. This picture will first be described by drawing on an analogy with the $\Z_K$ clock model. (In $(1+1)$D, the corresponding crossover in the clock model takes places over an $O(1)$ length of parameter space \cite{Radicevic:2D}; this is the BKT line of fixed points \cite{Berezinsky:1970fr, Kosterlitz:1973xp}.) Later in the Section, this crossover in the $(2+1)$D $\Z_K$ Maxwell theory will be argued to happen over a $O(1/\sqrt N)$ length in parameter space, on the basis of duality and other dynamical considerations. The position of the crossover will be shown to be at $g \sim 1/\sqrt N$ when $K \gg 1$.

The major part of Section \ref{sec Maxwell} will be dedicated to exploring the interior of this crossover, defined as the regime where a one-form symmetry $\Z_{K'} \subset \Z_K$, $1 \ll K' \ll K$, is spontaneously broken. This is where the U(1) Maxwell cQFT can be plausibly argued to arise as the low-energy description of the lattice theory. A square lattice is used from here on out.

Perhaps surprisingly, it will be shown that both the ``compact'' and the ``noncompact'' versions of the Maxwell cQFT arise from this one parametric regime. The noncompact cQFT comes from a tame subspace of the lattice theory that is defined w.r.t.\ a single \emph{taming background} --- essentially, a subspace of small fluctuations around a single background $\Z_{K'}$ gauge field. The compact cQFT is obtained from the lattice theory that includes all taming backgrounds that minimize the magnetic term in the Hamiltonian.

The noncompact theory can be further divided into two versions. The ``basic noncompact'' theory contains only photons. It is dual to a scalar cQFT. The ``standard noncompact'' theory contains both photons and classical background charges. Roughly, the gap of these background charges describes the position along the confinement/deconfinement crossover.

The basic noncompact theory has a precontinuum basis and can be smoothed following the general prescription given earlier in the Introduction. The full nitty-gritty of this smoothing will be presented. The procedure is straightforward, with the primary complication coming from the fact that the precontinuum ladder operators represent nontrivial (``gauge-invariant'') linear combinations of the original vector potential and electric fields.

Speaking of gauge invariance, it will be treated like any other symmetry. The lattice theory \eqref{def H} is simply viewed as a theory with an extensive number of symmetries. The symmetry generators here are the familiar Gauss operators, and their eigenvalues are precisely the background charges mentioned above.  The Gauss operators generate the ``local'' part of the one-form $\Z_K$ symmetry. The fact that the $\Z_{K'} \subset \Z_K$ symmetry is approximately broken means that background charges can be changed in units of $K/K'$ with negligible energy cost. This makes it consistent to work in a single sector of this $\Z_{K'}$ symmetry, restricting the allowed background charges to $K/K'$ possible values. The operators that measure these remaining background charges generate ``infinitesimal'' gauge transformations that form the coset $\Z_K/\Z_{K'}$. They are unable to take a given gauge field configuration outside of the tame subspace. This makes the entire picture of partial one-form symmetry breaking pleasantly consistent with the notion that gauge field configurations are tame, and it justifies small field expansions like $\e^{\i A_{\b x}^i} \approx \1 + \i A_{\b x}^i$ that most authors (unjustifiably) take as a given.

Readers who have found this paragraph too cryptic should rest assured that the same story will be repeated in full detail in Subsection \ref{subsec tame gauge transf}. It may also be helpful to first study the symmetry breaking analysis of the clock model \cite{Radicevic:2D}. While not advised, it is largely possible to skip this set of ideas altogether, take it for granted that gauge field fluctuations are small, and jump straight to Subsection \ref{subsec dynamics on torus}.

After presenting the smoothing construction, it will be shown how operator product expansions (OPEs) of various operators can be computed using the smoothed lattice fields. All the familiar results are reproduced using just elementary numerical methods to perform Fourier sums. In particular, it will be shown that single-plaquette Wilson loops in the noncompact theory display precisely the same product structure as vertex operators in the $(1+1)$D scalar CFT.  This makes it easy to calculate the expectation value of a large Wilson loop and verify that it shows the expected Coulomb law behavior.

Finally, this Section will end with an extensive derivation of the familiar ``particle/vortex'' (or photon/scalar) duality, starting from the Kramers-Wannier duality of the $\Z_K$ gauge theory and the $\Z_K$ clock model \cite{Kramers:1941kn}. This issue is nontrivial because Kramers-Wannier duality is ``singlet-singlet,'' which means that it does not naturally map tame operators to tame operators. This paper will offer a new and detailed derivation of an alternative Kramers-Wannier duality that only maps tame theories. The resulting duality between cQFTs ultimately serves as a useful consistency check for the symmetry breaking scenario presented here.

Path integrals for the various versions of Maxwell cQFT will be constructed in \textbf{Section \ref{sec Maxwell path int}}. For the most part, this will be a straightforward exercise in the transfer matrix formalism. The derivation will keep track of the various approximations invoked along the way, and in particular the allowed values of various path integral variables will be explicitly stated. The action for the basic noncompact theory will be shown to be precisely the action of a noncompact scalar. The action for the standard noncompact theory will be shown to be precisely the familiar Maxwell action. This derivation will stress that configurations with a nontrivial value of the Polyakov loop do not arise from the canonical formalism. Adding them to the path integral in principle changes the theory, and it becomes necessary to invoke universality to argue that these extra configurations do not affect the various universal parts of the path integral value.

The final part of this Section will also stress the connection between the canonically obtained path integrals and the temporally smooth ones that many analytic lattice studies implicitly work with. In particular, it will be shown how, after temporal smoothing, the path integral variables can be rescaled to give precisely the familiar continuum actions.

The $\Z_K$ Maxwell theory coupled to a $\Z_K$ clock model will be studied in \textbf{Section \ref{sec Higgs}}. The resulting theory \eqref{def H Higgs} can be understood as scalar quantum electrodynamics (QED). Through a slight abuse of nomenclature, it will be simply called the \emph{Higgs model}. This theory has a rich phase diagram with multiple kinds of crossovers (Fig.\ \ref{fig phases}). Its phase structure will be analyzed in some detail, and then the continuum limit will be constructed by assuming that both gauge and scalar fields are tame in an appropriate part of parameter space. (This assumption is supported by the self-duality of the theory.) This yields the cQFT that will be called \emph{scalar cQED}. This is an interesting example of a continuum theory because it is intrinsically gapped due to the Higgs mechanism. Despite the fact that this is clearly not a scale-invariant theory, there are no obstacles to defining it as a cQFT.

A slight detour at the end of this Section will discuss the Abelian BF theory. This is a topological QFT that cannot, at face value, be fit into the present cQFT paradigm. (In the absence of boundaries, topological QFTs have a flat spectrum of gapped excitations that prevents one from placing any natural structure on the momentum space and identifying the low-momentum subspace $\Pbb\_S$.) This paper will confront the fact that many authors, starting with \cite{Banks:2010zn}, treat BF theory as a particular limit of scalar cQED, which \emph{is} a cQFT. Based on the known properties of the phase diagram of the Higgs model, it will be argued that it is plausible that this limit of the cQFT does have the same universal behavior as the topological BF theory, but that it is nevertheless unnatural to try to rigorously think of BF theory as a cQFT. Instead, here it will be proposed that BF theory can be defined by a  path integral constructed directly from the $g \rar 0$ limit of pure Maxwell theory on the lattice.

\newpage

\textbf{Section \ref{sec QED}} is devoted to studying what could simply be called \emph{lattice QED} in $(2 + 1)$D. This is a theory that couples Dirac fermions to Maxwell theory. The Section starts slow, reviewing the construction of a pure Dirac cQFT from the lattice, using the staggered fermion formalism \cite{Kogut:1974ag, Susskind:1976jm} and showing how the usual operator products are defined in this context. Then things get significantly less mainstream as the focus shifts to the question of how, exactly, is one supposed to couple fermions (with a $\Z_2$ target space) to Maxwell theory (with a $\Z_K$ target space). This question is nontrivial because for any $K > 2$ there exists an underappreciated global anomaly that can prevent gauging a $\Z_K$ symmetry in a system with a $\Z_2$ target space \cite{Radicevic:2018zsg}. Clarifying this points to a way towards defining a consistent lattice QED, which is given by the simple but nontrivial Hamiltonian \eqref{def H QED}.

The generators of gauge symmetries in QED consist of a $\Z_2$ fermionic density and a $\Z_K$ Gauss operator. This mismatch in group orders can be shown to imply a novel local constraint on gauge fields \eqref{QED consistency conditions}. Tame gauge fields form one set of solutions to this constraint, and this tameness is assumed for the rest of the paper. The resulting theory of fermions and tame gauge fields is interacting, and it is not a priori obvious what its precontinuum basis is. This paper will obviate this question by restricting to an interacting theory of low-momentum fermions and gauge fields, the \emph{conventional QED} of eq.\ \eqref{def H QED conv}. This is not a cQFT in the strict sense, as it does not possess high-momentum particle number operators in its algebra. Still, this is a reasonable (if na\"ive) lattice construction of continuum QED. Basic observations on its properties and phase structure will be given at the close of the Section.

\textbf{Section \ref{sec CS}} aims to find a finite Hamiltonian theory that yields a pure Chern-Simons (CS) theory upon suitable restrictions. The objective is not achieved in full generality, but there is still significant progress to report on.

Two promising directions are explored here. First, it is known that integrating out massive fermions gives rise to CS actions \cite{Niemi:1983rq, Redlich:1983dv, Redlich:1983kn}. Studying the massive Dirac cQFT shows that the effects associated to CS theory, such as a ``parity-odd'' structure in current correlators, are all present in the canonical formalism without any reference to CS theory. It is only by switching to temporally smoothed path integrals for fermions that it becomes necessary to include ad hoc CS terms to get the right answers. This is illustrated completely explicitly in $(0+1)$D, while the $(2+1)$D case is presented without fixing an overall normalization. This approach ultimately leads to a definition of a noncompact CS action in a path integral context only.

The second, more satisfactory path to CS comes from studying recently discovered flux-attached lattice gauge theories \cite{Chen:2017fvr}. The confined regime of one such $\Z_K$ gauge theory, given by Hamiltonian \eqref{def H CS}, will be shown to \emph{almost} give rise to a U(1)$_{2K}$ CS action, cf.\ \eqref{def S CS}. The appearance of such CS-like actions is a generic property of the confined regime of flux-attached theories. The example given here serves to merely initiate their study.

\newpage

\section{Pure Maxwell theory} \label{sec Maxwell}

\subsection{Lattice preliminaries}

Gauge theories feature local symmetries. This paper will focus on $\Z_K$ lattice gauge theories. To define them, start with a lattice $\Mbb$. Place a $K$-state clock degree of freedom (a ``qudit'') on each link $\ell \in \Mbb$. The algebra of operators on each link is generated by a clock operator $Z_\ell$ and a shift operator $X_\ell$ \cite{Radicevic:1D}. An operator is \emph{gauge-invariant} if it commutes with \emph{Gauss operators}
\bel{\label{def G}
  G_v \equiv \prod_{\ell \in \del_{-1} v} X_\ell, \quad v \in \Mbb.
}
A gauge-invariant Hamiltonian will thus have a $\Z_K$ symmetry generator at each site $v$.

A notational interlude is in order before proceeding. Labels like $v$ or $\ell$ do not contain information about any further structure, such as orientations or spin structures, that can be put on $\Mbb$. Instead, such structure will be embedded into the definitions of various special chains. For example, the object
\bel{
  \del_{-1} v \equiv \sum_{\ell \supset v} \sigma_v^\ell \, \ell
}
in eq.\ \eqref{def G} is a one-chain formed by all links $\ell$ that contain the vertex $v$. The weights $\sigma_v^\ell \in \{\pm 1\}$ then encode the choice of link orientation. They must be chosen so that for each link $\ell$ one has
\bel{
  \sum_{v \subset \ell} \sigma_v^\ell = 0.
}
Note that $\del_{-1}$ is the Poincar\'e dual of the boundary operator $\del_1 \equiv \del$. On $\Mbb$, this boundary operator acts on one-chains as
\bel{\label{def del}
  \del \ell = \sum_{v \subset \ell} \sigma^v_\ell \, v
}
for weights $\sigma^v_\ell \in \{\pm 1\}$ that satisfy $\sum_{v \subset \ell} \sigma_\ell^v = 0$.

Products of commuting operators over one-chains $c = \sum_{\ell \in \Mbb} \sigma^\ell \, \ell$  for any $\sigma^\ell \in \Z$ will be defined as
\bel{
  \prod_{\ell \in c} \O_\ell \equiv \prod_{\ell \in \Mbb} \O_\ell^{\sigma^\ell}.
}
Sums over one-chains, on the other hand, will be defined as
\bel{
  \sum_{\ell \in c} \O_\ell \equiv \sum_{\ell \in \Mbb} \sigma^\ell\, \O_\ell.
}
Analogous definitions hold for chains of other ranks and for operators $\del_n$ with $n \in \Z$ \cite{Radicevic:2019vyb}.

Back to physics. The algebra of gauge-invariant operators is generated by shift operators $X_\ell$ and by \emph{Wilson loops} defined on closed one-chains (i.e.\ one-cycles) $c$ via
\bel{
  W_c \equiv \prod_{\ell \in c} Z_\ell.
}
Recall that a closed chain satisfies $\del c = 0$. The minimal generating set of Wilson loops consists of two kinds of operators, local and nonlocal. The local ones come from cycles around each face (plaquette) $f \in \Mbb$, and they will be denoted by $W_f \equiv W_{\del f} = \prod_{\ell \in \del f} Z_\ell$. The nonlocal ones come from homologically distinct noncontractible one-cycles on $\Mbb$.

All Gauss operators \eqref{def G} commute with each other. They generate the center of the gauge-invariant operator algebra. The Hilbert space associated to this algebra has superselection sectors labeled by eigenvalues $\e^{\frac{2\pi\i}K \varrho_v}$ of $G_v$. The integers $\varrho_v \in \{0, \ldots, K - 1\}$, subject to $\sum_{v \in \Mbb} \varrho_v \in K \Z$, will be called \emph{background charges}. It is customary to fix the background charges and only ever refer to one sector. The default choice is $\varrho_v = 0$. The resulting theory then has a \emph{gauge constraint} $G_v = \1$ for each $v$.

The archetypical $\Z_K$ gauge theory at large $K$ is the \emph{Maxwell theory}. Its Hamiltonian is \cite{Kogut:1974ag}
\bel{\label{def H}
  H = \frac{g^2}{2 (\d A)^2} \sum_{\ell \in \Mbb} \left(2 - X_\ell - X_\ell\+ \right) + \frac1{2g^2} \sum_{f \in \Mbb} \left(2 - W_f - W_f\+ \right),
}
where
\bel{
  \d A \equiv \frac{2\pi}K \ll 1.
}
Like the clock model \cite{Radicevic:2D}, the Maxwell theory has a well studied phase structure \cite{Banks:1977cc, Ukawa:1979yv}. The following description is restricted to the $\varrho_v = 0$ sector. At $g \rar \infty$, the theory is in the \emph{confined} phase with a unique ground state. At $g \rar 0$, the theory is in the \emph{topological} phase with $K^{b_1}$ ground states, where $b_1$ is the first Betti number of $\Mbb$. (If $\Mbb$ is a $d$-torus, the first Betti number is $b_1 = d$.) If $K$ is large enough, these two phases are separated by a region in $g$-space where the low-energy states can be described by a cQFT with a linear dispersion.

In a clock model, the ground state degeneracy in the ordered (ferromagnetic) phase is often viewed as the result of the spontaneous breaking of its $\Z_K$ shift symmetry. The gapless degrees of freedom found at larger couplings can then be understood as Nambu-Goldstone bosons. In a gauge theory, the ordered (topological) phase features a spontaneously broken $\Z_K$ one-form symmetry. This symmetry is generated by products of shift operators along links that pierce a homologically nontrivial cycle on the dual lattice. (These are sometimes called 't Hooft operators.) The corresponding Nambu-Goldstone bosons (photons) have multiple components (polarizations) as a consequence of the nonzero rank of the broken symmetry from which they originate \cite{Gaiotto:2014kfa}.

\subsection{How to tame your gauge theory} \label{subsec tame gauge transf}

The goal of this Section is to understand the cQFTs that emerge from the Maxwell theory \eqref{def H}. In the absence of an exact solution, a self-consistent treatment must suffice. The basic strategy was already deployed to analyze the clock model \cite{Radicevic:2D}: assume that every low-energy state is tame with respect to some taming background, construct a precontinuum algebra out of tame operators, find that it can be reduced to a continuum algebra, and notice that along the way no inconsistencies arise. (A famous example of the kind of inconsistency that may arise is expressed by the CHMW theorem in $(1+1)$D scalar theories \cite{Mermin:1966fe, Hohenberg:1967zz, Coleman:1973ci}.)

The presence of gauge constraints adds an important conceptual wrinkle to this strategy. Taming does not preserve the Gauss operators \eqref{def G}. Said another way, taming a clock variable on a given link is not a gauge-invariant operation. The reason is that, by definition, taming restricts the angle of any clock variable to a small subinterval --- but clock angles are not gauge-invariant and can always be taken out of any subinterval by applying a high enough power of Gauss operators.

How is taming to be defined in a gauge theory, then? There are two different options:
\begin{enumerate}
  \item Fix the gauge. Most standard gauge choices replace the starting system by one in which certain links do not host dynamical degrees of freedom. The remaining clock variables are unconstrained and can be tamed with impunity.
  \item Tame the original (gauge-variant) degrees of freedom, but limit the kind of gauge transformations that are considered when defining tameness.
\end{enumerate}
The first option remains true to the gauge principle. It may appear ``obviously'' right. Its downside is that it depends on a choice of gauge. This causes significant complications downstream, not least of which is the need to prove that a particular phenomenon is independent of the gauge choice.\footnote{To list just one concrete example, the need to sum over magnetic fluxes when computing partition functions of Chern-Simons-matter theories \cite{Aharony:2012ns, Jain:2013py} is not a sign of a deep nonperturbative phenomenon, but merely a consequence of a particular choice of gauge \cite{Radicevic:2015yla}.} For this reason, this paper will pick the second option. This route will ultimately lead to a fully gauge-invariant notion of a tame state.

Some precise notation is needed now. On each link, define the clock eigenstates $\qvec{\e^{\i A_\ell}}$ via
\bel{
  Z_\ell \qvec{\e^{\i A_\ell}} = \e^{\i A_\ell} \qvec{\e^{\i A_\ell}}, \quad A_\ell = n_\ell \, \d A, \quad 0 \leq n_\ell < K.
}
A global field theory state will be denoted by
\bel{
  \qvec{\e^{\i A}} \equiv \bigotimes_{\ell \in \Mbb} \qvec{\e^{\i A_\ell}}.
}

Taming is done on each link separately, in complete analogy with the clock model \cite{Radicevic:1D, Radicevic:2D}. (More details can be found in these earlier papers.) A wavefunctional $\greek y[A]$ is \emph{tame} if
\bel{\label{target smoothness}
  \Big| \, \greek y[A + \delta^{(\ell)}\d A] -  \greek y[A] \, \Big| = O\left(\frac{E\_S}K\right) \quad \trm{for all }\ell \in \Mbb
}
and
\bel{\label{target compactness}
  \greek y[A]  = 0 \quad \trm{whenever}\quad |A_\ell - A_\ell\^{cl}| > A\_T \quad \trm{for all }\ell \in \Mbb,
}
where $A\_T \equiv \frac{2\pi}{2E\_S} n\_T$, $1 \ll n\_T \ll E\_S \ll K$, and $\delta^{(\ell)}$ is a Kronecker delta supported at $\ell$. The field $A_\ell\^{cl}$ will be called a \emph{taming background}.

Operators that preserve the tame subspace, defined relative to some taming background, will also be called tame. The projection from the full operator algebra to the tame one will be called \emph{taming}. The taming of an operator $\O$ will be denoted $\O\_T$. A tamed product of operators is generally \emph{not} equal to the product of corresponding tamed operators. In what follows, it will always be assumed that all operators are multiplied first and tamed second.

Two important tame operators are the \emph{vector potential} and \emph{electric field},
\bel{\label{def A E}
  \widehat A_\ell \equiv \frac{1}{2\i}\left(\e^{-\i A_\ell\^{cl}} Z_\ell - \e^{\i A_\ell\^{cl}} Z_\ell\+\right)\_T, \quad \widehat E_\ell \equiv \frac{1}{2\i\, \d A } \left(X_\ell - X_\ell\+ \right)\_T.
}
The carets will be dropped whenever possible. These gauge-theoretic analogs of position and momentum operators obey the canonical commutation relation
\bel{\label{comm rels A E pos}
  [A_\ell, E_{\ell'}] \approx \i \, \delta_{\ell, \, \ell'}
}
when acting on tame states. To get this result it is crucial to multiply first and tame second.

The tame subspace on each link $\ell$ is spanned by $2n\_T$ vector potential eigenstates, denoted $\qvec{A_\ell}$, with approximate eigenvalues $A_\ell = n_\ell \frac{2\pi}{2E\_S}$ for $-n\_T \leq n_\ell < n\_T$. In states with $|A_\ell| \ll A\_T$, the electric field acts as a formal derivative w.r.t.\ $A_\ell$ up to $1/E\_S$ corrections \cite{Radicevic:1D},
\bel{\label{E action}
  E_\ell \approx -\i\, \hat \del_{A_\ell}.
}

Now it is time to turn to gauge constraints. As already stated, Gauss operators are not tame. Each $G_v$ shifts the clock positions on links $\ell \supset v$ by $\pm \d A$, and so powers of $G_v$ obey
\bel{\label{G tamed}
  (G^n_v)\_T = 0 \quad \trm{for} \quad 2n\_T < n < K - 2n\_T.
}
In a gauge theory without background charges, every gauge-invariant state $\qvec\psi$ must satisfy $G^n_v \qvec \psi = \qvec \psi$ for all $n$. Comparing this to \eqref{G tamed} shows that, as anticipated, no gauge-invariant state can be tame. The most urgent task now is to make sense of taming despite this finding.

Delving deeper into the structure of the theory \eqref{def H} will show a natural way to proceed. First, note that any gauge-invariant state (i.e.\ a state in the $\varrho_v = 0$ sector) can be written as
\bel{\label{general ginv state}
  \qvec\psi =  \trm P\_{Gauss} \sum_{\{A\} \in \Gamma} \greek y[A]\, \qvec{\e^{\i A}}, \quad \trm P\_{Gauss} \equiv \prod_{v \in \Mbb} \left(\frac1K \sum_{n = 1}^K G_v^n \right),
}
where the sum runs over configurations $\{A_\ell\}_{\ell \in \Mbb} \equiv \{ A\}$ in any \emph{gauge slice} $\Gamma$, i.e.\ over any set of clock eigenstates that cannot be related to each other by gauge transformations. Any rule for choosing $\Gamma$ constitutes a gauge choice, but the above expression for $\qvec \psi$ is manifestly independent of it. The coefficients $\greek y[A]$ depend only on sums of $A_\ell$ over closed chains.

Next, define the global Gauss operator
\bel{\label{def G n}
  G[n] \equiv \prod_{v \in \Mbb} G_v^{n_v}, \quad 1 \leq n_v \leq K.
}
It acts on a clock eigenstate in a familiar fashion,
\bel{
  G[n]\, \qvec{\e^{\i A}} = \qvec{\e^{\i (A - \delta n\, \d A)}},
}
where the coboundary operator $\delta$ acts on the zero-cochain $n_v$ to give the one-cochain
\bel{
  (\delta n)_\ell \equiv \sum_{v \in \del \ell} n_v = \sum_{v \subset \ell} \sigma^v_\ell \, n_v.
}
Then the state \eqref{general ginv state} can be written as
\bel{\label{general ginv state 2}
  \qvec \psi = \frac1{K^{N\_V}} \sum_{\{n\}}  \sum_{\{A\} \in \Gamma} \greek y[A] \, \qvec{\e^{\i (A - \delta n \, \d A)}}.
}
This is the explicit expression of any gauge-invariant state as a superposition of states related by (local) symmetry transformations $G[n]$. Note that constant gauge parameters $n_v$ do not lead to different states $\qvec{\e^{\i (A - \delta n \, \d A)}}$. If desired, these gauge parameters can be removed from the sum at the expense of rescaling the normalization factor by $K^{b_0}$, where the zeroth Betti number $b_0$ counts disconnected components of $\Mbb$.

No state of the form
\bel{\label{degenerate gvar states}
   \sum_{\{A\} \in \Gamma} \greek y[A] \, \qvec{\e^{\i (A - \delta n \, \d A)}}
}
is gauge-invariant. Nevertheless, there is nothing ill defined about such gauge-variant states. They are simply superpositions of states from sectors with different background charges. In fact, these states will play a fundamental r$\hat{\trm o}$le in the formulation of tame gauge theories.

Considering states \eqref{degenerate gvar states} and thinking of Gauss operators as generators of ordinary symmetries is useful because it leads to a direct parallel with the clock model \cite{Radicevic:2D}. There the only relevant symmetry is the global $\Z_K$ shift. In the ordered phase with the spontaneously broken $\Z_K$, found at $g \rar 0$, the $K$ ground states can be chosen to have the form analogous to \eqref{degenerate gvar states},
\bel{\label{degenerate clock states}
  \qvec{\e^{\i \phi}} = \bigotimes_{v \in \Mbb} \qvec{\e^{\i\phi_v}} \quad \trm{for} \ \phi_v = \frac{2\pi}K n \ \trm{at each } v.
}
The situation changes as the coupling increases. In some part of coupling space, only a subgroup $\Z_{K'} \subset \Z_K$ should be regarded as spontaneously broken \cite{Radicevic:2D}. Each of the $K' \ll K$ degenerate ground states can be understood to provide a \emph{taming background}, with the natural taming parameter being $\varphi\_T = \pi/K'$. In this regime, every low-energy state is tame w.r.t.\ one of the $K'$ backgrounds. As $g$ increases further, the appropriate $K'$ decreases until the system enters a disordered phase in which there is a unique ground state, given by the equal superposition of all states $\qvec{\e^{\i\phi}}$. In this phase, no low-energy states are tame w.r.t.\ any taming background. In particular, none of the individual states \eqref{degenerate clock states} are energy eigenstates.

An analogous story holds in the Maxwell theory once all the background charge sectors are taken into account. Start by considering the topological phase at $g \rar 0$. Each state of the form \eqref{degenerate gvar states}, with
\bel{\label{wf topo}
  \greek y[A] \propto \prod_{f \in \Mbb} \delta_{(\delta A)_f, \, 0},
}
is a ground state of the Hamiltonian \eqref{def H} at vanishing coupling.\footnote{To avoid confusion, note that the $\delta$ in $(\delta A)_f$ is a coboundary operator that acts on cochains, while $\delta_{a, b}$ is a Kronecker delta that is unity when $a = b$ and zero otherwise.} Beside the $K^{b_1}$-fold degeneracy due to noncontractible one-cycles, in this phase there are $K^{N\_V - b_0}$ degenerate states labeled by $\{n_v\}_{v \in \Mbb}$. (The number of sites in $\Mbb$ is $N\_V$.) This additional degeneracy comes from the fact that the ground states in all background charge sectors have the same energy.

At the other extreme, $g \rar \infty$, is the confining phase. Here there is a globally unique ground state, and it lies in the sector with background charge $\varrho_v = 0$. (Any nonzero $\varrho_v$ forces electric flux lines to exist in the ground state for that sector, and these cost energy.) The overall ground state has
\bel{\label{wf confined}
  \greek y[A] = \trm{const}.
}
None of the states \eqref{degenerate gvar states} are energy eigenstates, but their superpositions are --- just like in the disordered phase of the clock model.

A natural conjecture now is that, like in the clock model, intermediate couplings show a crossover between the topological and confined phases. It is simplest to posit that the total ground state degeneracy is $(K')^{N\_V - b_0 + b_1}$ at a generic point along this crossover. This means that the degenerate ground states are mapped to each other via operators $G'_v \equiv G_v^{K/K'}$.

This plausible conjecture implies that states of the form
\bel{\label{degenerate partial states}
  \sum_{\{n\}/\Z_{K'}}  \sum_{\{A\} \in \Gamma} \greek y[A] \, \qvec{\e^{\i (A - \delta n \, \d A)}}
}
are all ground states at this crossover point. The notation $\{n\}/\Z_{K'}$ means that each gauge parameter $n_v$ is restricted to
\bel{\label{def tame gauge param}
  -\frac{K}{2K'} \leq n_v < \frac K{2K'}.
}
There is some freedom in the exact choice of this interval. The important property is that it only involves $K/K'$ sequential values of $n_v$. If $n_v$ is thus restricted, a generic gauge parameter can be written as
\bel{
  m_v \frac{K}{K'} + n_v, \quad 1\leq m_v \leq K'.
}

There are, in effect, two kinds of gauge transformations that can be discussed separately:
\begin{enumerate}
  \item The first kind are the ``tame'' transformations \eqref{def tame gauge param}. They do not form a group. Precisely speaking, they take values in the coset $\Z_K/\Z_{K'}$. Loosely speaking, they live in the tangent space of the group $\Z_K \approx \trm U(1)$, and they can be informally viewed as noncompact gauge transformations. In the coupling range of interest, this part of the gauge symmetry behaves as if it were confined. This means that there is a nonzero energy cost to insert background charges for this gauge symmetry, i.e.\ to change $\varrho_v$ by an amount smaller than $K/K'$.
  \item Gauge transformations of the second kind are generated by $G_v'$. They form the gauge group $\Z_{K'}$. At intermediate couplings this gauge symmetry appears topologically ordered, as all states of form \eqref{degenerate partial states} are degenerate. Inserting a background charge for this symmetry --- that is, changing $\varrho_v$ by a integer multiple of $K/K'$ --- does not cost energy, as it simply transforms one superposition of these degenerate states to another.
\end{enumerate}

The distinction between zero and nonzero gaps between states \eqref{degenerate gvar states}, alluded to in the above two paragraphs, is rather imprecise at this point. The precise definition is that a set of states \eqref{degenerate partial states} is degenerate if the gaps between them are much smaller than the gaps associated to the other excitations in the theory. These other gaps, which are all inversely proportional to the linear size of the system, will be calculated in Subsections \ref{subsec basic nc Maxwell} and \ref{subsec standard nc Maxwell}.

Unlike gauge-invariant states, the states \eqref{degenerate partial states} can be tame. When $n_v$ is in the interval \eqref{def tame gauge param},  $|(\delta n)_\ell|$ is bounded by $K/K'$. It is thus reasonable to pick $\pi/A\_T = E\_S/n\_T = K/K'$ as the ratio of taming parameters, just as it was done in the clock model. Any values of $E\_S$ and $n\_T$ satisfying $n\_T \ll E\_S \ll K$ should ensure that all states of sufficiently low energy are tame. Results in $(0 + 1)$D suggest that the number of tame states will be maximized for $E\_S = \sqrt{K n\_T}$ \cite{Radicevic:1D}.

It is now straightforward to construct fully gauge-invariant combinations of tame states. They are equal superpositions of states \eqref{degenerate partial states} at all values of the $\Z_{K'}$ gauge parameters $m_v$,
\bel{\label{general ginv tame state}
  \trm P'\_{Gauss} \sum_{\{n\}/\Z_{K'}}  \sum_{\{A\} \in \Gamma} \greek y[A] \, \qvec{\e^{\i (A - \delta n \, \d A)}}, \quad \trm P'\_{Gauss} \equiv \prod_{v \in \Mbb} \left( \frac1{K'} \sum_{m = 1}^{K'} (G'_v)^m \right).
}
Up to normalization, these are precisely the gauge-invariant states \eqref{general ginv state}. This way the definition of tameness developed over the last few pages can be extended to bona fide gauge-invariant states. Of course, at a generic coupling, $\trm P'\_{Gauss}$ superposes tame states with wildly different energies. Thus, defining tame states this way may not be stable under perturbations. A gauge-invariant state only has a \emph{natural} notion of tameness if the background charges associated to tameness-violating gauge transformations cost no energy to insert.

This establishes a precise definition of tame states in the Maxwell theory. Now one can ask whether a given gauge-invariant wavefunctional $\greek y[A]$ is such that an eigenstate of the form \eqref{degenerate partial states} is tame. For ground states at the extremes of the coupling space, the answer is an unsurprising ``no.'' In the topological phase, the state \eqref{degenerate partial states} with wavefunctional \eqref{wf topo} is compact but not smooth in the target space. In the confined phase, the state with wavefunctional \eqref{wf confined} is smooth but not compact --- the sum over configurations $\{A\} \in \Gamma$ necessarily takes $A_\ell - (\delta n)_\ell \, \d A$ out of any interval of width $2A\_T$ despite the boundedness of $(\delta n)_\ell$. It is thus only at the intermediate couplings where $\greek y[A]$ stands a chance of satisfying the tameness conditions \eqref{target smoothness} and \eqref{target compactness}. This is, of course, consistent with the fact that the states of the form \eqref{degenerate partial states} are assumed to be ground states only at such couplings.

Finally, consider again the tame operators defined in eq.\ \eqref{def A E}. The vector potential and electric field operators defined here are unable to alter the tameness of any fully gauge-invariant state of form \eqref{general ginv tame state}. Even more importantly, these operators cannot take any individual state of form \eqref{degenerate partial states} out of the tame subspace. They can thus be understood as operators in a separate gauge theory in which gauge parameters take values in the coset $\Z_K/\Z_{K'}$. At $K \gg K' \gg 1$, this theory will be called the \emph{noncompact Maxwell theory}.

The tame Gauss operator $(G_v)\_T$ that generates gauge transformations in a noncompact Maxwell theory becomes zero when applied more than $2n\_T$ times. Sometimes it is said that this theory has gauge group $\R$, but this is an abuse of nomenclature; the gauge transformations in this theory simply \emph{do not form a group}. The tame Gauss operator acts as
\bel{\label{tame gauge transf}
  (G_v)\_T \approx \1 + \i\, \d A \sum_{\ell \in \del_{-1}v} E_\ell \equiv \1 + \i \, \d A\, (\nabla E)_v.
}
in all tame states with $|A_\ell| \ll A\_T$. Thus the familiar gauge constraint $(\nabla E)_v = 0$ is applicable only in this limited set of tame states.

\subsection{Maxwell dynamics on a toric lattice}  \label{subsec dynamics on torus}

The analysis so far was mostly based on an analogy with the $\Z_K$ clock model. To put this story on firmer footing, it is necessary to explicitly solve the tamed Maxwell theory and check for consistency. To do this in a concrete setting, let $\Mbb$ be a cubic spatial lattice in $d$ dimensions, with periodic boundary conditions and $N$ sites in each direction. Each link $\ell$ can then be labeled by a site coordinate $\b x$ and a direction label $i = 1, \ldots, d$. Define the spatial derivative to be
\bel{
  \del_i f(\b x) \equiv f(\b x + \b e_i) - f(\b x),
}
where $\b e_i$ is the unit vector in the $i$'th direction.

The microscopic Hamiltonian \eqref{def H} can be rewritten as
\bel{\label{def H Maxwell}
  H = \frac{g^2}{2 (\d A)^2} \sum_{\b x, \, i} \left(2 - X_{\b x}^i - (X_{\b x}^i)\+\right) + \frac1{2g^2} \sum_{\b x, \, i > j} \left(2 - W_{\b x}^{ij} - (W_{\b x}^{ij})\+ \right).
}
Now consider taming this theory relative to a background $(A\^{cl})_{\b x}^i$. To leading order in the taming parameters, the Hamiltonian acts on tame states as
\bel{
  H\_T \big[B\^{cl}\big] \approx \frac{g^2}{2} \sum_{\b x, \, i} (E^i_{\b x})^2 + \frac1{g^2} \sum_{\b x, \, i, \, j} \sin^2 \frac{(B\^{cl})_{\b x}^{ij} \, \1 + B_{\b x}^{ij}}2.
}
The \emph{tame magnetic field} is
\bel{
  B_{\b x}^{ij} \equiv \del_i A_{\b x}^j - \del_j A_{\b x}^i.
}
Note that $B^{ij} = -B^{ji}$. The \emph{background magnetic field} $(B\^{cl})^{ij}_{\b x}$ is defined analogously in terms of $(A\^{cl})_{\b x}^i$. Note that taming backgrounds are c-numbers, unlike the vector potentials.

A priori, the taming background can be any configuration of clock positions on links. However, all backgrounds related to each other by a gauge transformation will be included when constructing gauge-invariant states of form \eqref{degenerate partial states} or \eqref{general ginv tame state}. It is thus only necessary to consider gauge-inequivalent taming backgrounds.

As in the clock model, a plausible assumption is that the only taming backgrounds needed for capturing the low-energy spectrum minimize the potential term. Such configurations obey
\bel{\label{taming bkgd eq}
  \del_i \sin \left( (B\^{cl})^{ij}_{\b x}\right) = 0 \quad \trm{(summation implied)}.
}
If $(A\^{cl})^i_{\b x}$ varies slowly as a function of $\b x$, the background magnetic field is small and the familiar Maxwell equation arises,
\bel{
  \del_i (B\^{cl})^{ij}_{\b x} \approx 0.
}
This approximation hides the angular nature of $(A\^{cl})^i_{\b x}$, and so it should be treated with care.

With these assumptions for taming backgrounds, the Hamiltonian becomes
\bel{
  H\_T \big[B\^{cl} \big] \approx \frac{g^2}{2} \sum_{\b x, \, i} (E^i_{\b x})^2 + \frac1{4g^2} \sum_{\b x, \, i, \, j} \left((B\^{cl})_{\b x}^{ij} + B_{\b x}^{ij} \right)^2.
}
This Subsection will focus on the case $(B\^{cl})_{\b x}^{ij} = 0$. Just like in the scalar case, the way to proceed is by Fourier transforming,
\bel{
  A_{\b x}^i \equiv \frac1{N^{d/2}} \sum_{\b k \in \Pbb} A_{\b k}^i \, \e^{\frac{2\pi\i}{N} \b k \b x}, \quad E_{\b x}^i \equiv \frac1{N^{d/2}} \sum_{\b k \in \Pbb} E_{\b k}^i \, \e^{\frac{2\pi\i}{N} \b k \b x}.
}
The momentum space is
\bel{
  \Pbb \equiv \{\b k = (k^1, \ldots, k^d)\}, \quad -\frac N2 \leq k^i < \frac N2,
}
and $\b k \b x$ denotes the usual scalar product of $d$-vectors. The commutation relations \eqref{comm rels A E pos} imply
\bel{\label{comm rels A E mom}
  [A_{\b k}^i, E_{\b l}^j] \approx \i\, \delta_{ij}\, \delta_{\b k + \b l, \, 0}
}
when these operators act on tame states.

Recall that gauge transformations act on tame operators as
\bel{
  \O \mapsto \left(G[n]^{-1}\, \O\, G[n]\right)\_T,
}
for gauge parameters $n_v$ given by \eqref{def tame gauge param}. If $n_v$ is also small compared to other taming parameters, then for the natural choice of orientations, $(\delta n)_{\b x}^i = -(\del_i n)_{\b x}$, the gauge transformation is approximately
\bel{\label{tame gauge transf O}
  \O \mapsto \O - \i \sum_{\b x,\, i} (\del_i n)_{\b x} \, [\O, E_{\b x}^i]\, \d A.
}
To simplify the notation, let
\bel{\label{def lambda}
  \lambda_{\b x} \equiv n_{\b x} \, \d A.
}
Then, using \eqref{comm rels A E pos} and \eqref{comm rels A E mom}, the vector potential operators transform as
\bel{\label{tame gauge transf A}
  \widehat A_{\b x}^i \mapsto \widehat A_{\b x}^i + (\del_i \lambda)_{\b x} \, \1, \quad
  \widehat A_{\b k}^i \mapsto \widehat A_{\b k}^i + \big(\e^{\frac{2\pi\i}{N} k^i} - 1\big) \lambda_{\b k}\, \1,
}
to leading order in the taming parameters. Keep in mind that transformations with $|(\del_i \lambda)_{\b x}| > 2A\_T $ are guaranteed to violate tameness and so cannot be included in this discussion.

The operator map should be distinguished from the gauge transformation of a state $\qvec{\e^{\i A}}$. A global Gauss operator $G[\lambda]$ always maps the state label via $A_{\b x}^i \mapsto A_{\b x}^i + (\del_i \lambda)_{\b x}$. This holds for all values of $\lambda_{\b x}$ subject to the definition \eqref{def lambda}, including $O(1)$ ones.

The momentum space Hamiltonian is thus
\bel{\label{def HT A E}
  H\_T \approx \frac{g^2}{2} \sum_{\b k, \, i} (E^i_{\b k})\+ E^i_{\b k}  + \frac{1}{2g^2} \sum_{\b k, \, i,\, j} \Omega^{ij}_{\b k} (A^i_{\b k})\+ A^j_{\b k},
}
with
\gathl{
  \Omega^{ii}_{\b k} \equiv \sum_{j \neq i} 4\sin^2 \frac{\pi k^j}{N}, \quad
  \Omega^{ij}_{\b k} \equiv - \left(\e^{\frac{2\pi \i}N k^i} - 1\right) \left(\e^{-\frac{2\pi \i}N k^j} - 1\right).
}
More compactly, let
\bel{\label{def kappa}
  \greek k_{\b k}^i \equiv \e^{\frac{2\pi \i}N k^i} - 1.
}
Then the kernel in the harmonic potential is
\bel{
  \Omega_{\b k}^{ij} = |\greek{\textbf k}_{\b k}^2| \delta_{ij} - \greek k^i_{\b k} \greek k^j_{-\b k},
}
or, at low momenta,
\bel{
  \Omega^{ij}_{\b k} \approx \frac{4\pi^2}{N^2} \left(\b k^2 \delta_{ij} - k^i k^j \right).
}

The matrix $\Omega^{ij}_{\b k}$ is singular. At $\b k = 0$, it simply equals zero. At $\b k \neq 0$, it has one null vector, $\greek{\textbf k}_{\b k}$, and $d - 1$ degenerate eigenvectors, e.g. $\greek k^1_{-\b k} \b e_i - \greek k^i_{-\b k} \b e_1$ for $i = 2, \ldots, d$, with eigenvalues
\bel{\label{def omega k}
  \omega_{\b k}^2 \equiv |\greek{\bf k}^2\_{\b k}| = \sum_{i = 1}^d 4\sin^2 \frac{\pi k^i}{N}.
}

The situation is analogous to the clock model \cite{Radicevic:2D}. There, the zero-momentum position fields do not contribute to the harmonic potential. The corresponding momentum fields label a set of approximate superselection sectors. In string theory, these are called ``momentum modes,'' while in condensed matter they are known as the ``Anderson tower of states.''

In the present case, the $\b k = 0$ sector gives rise to $d$ electric zero-modes $E_0^i$. They are straightforward generalizations of the scalar momentum mode. In addition to these, however, there also exists a single zero-potential mode at each nonzero momentum. This state space is spanned by eigenstates of
\bel{
  \sum_{i = 1}^d \greek k^i_{-\b k} E_{\b k}^i.
}
In position space, this operator is the divergence of the electric field,
\bel{\label{def nabla E}
  \sum_{i = 1}^d \del_i E^i_{\b x - \b \e_i} \equiv (\nabla E)_{\b x}.
}
This is precisely the generator of tame gauge transformations identified in \eqref{tame gauge transf}.

The tame operators that enter the Hamiltonian without a potential term thus come in two classes. The local ones, $(\nabla E)_{\b x}$ from \eqref{def nabla E}, generate gauge transformations at each point. The nonlocal ones, $E_0^i$, generate global rotations of all clocks on links pointing in the $i$'th direction. Just like zero modes in the clock model, all of these operators commute with the Hamiltonian to leading order in the taming parameters. They will thus be called \emph{tame symmetries}, and their eigenvalues can be understood to label superselection sectors.

The eigenstates of local tame symmetries correspond to different background charges in the noncompact gauge theory. The eigenstates of nolocal tame symmetries correspond to different amounts of electric flux along noncontractible cycles of $\Mbb$. Eq.\ \eqref{def HT A E} shows that energy gaps between different sectors of both kinds of tame symmetries are set by the same quantity, $g^2$. This is the first quantitative hint at the symmetry breaking scenario outlined in Subsection \ref{subsec tame gauge transf}. There it was proposed that the ground state degeneracies associated to background charges and to noncontractible one-cycles change in lockstep as $g$ is dialed, so that the approximate ground state degeneracy is $(K')^{N\_V - b_0 + b_1}$ for a $g$-dependent $K'$.

An elegant approach to both classes of tame symmetry operators views them as joint generators of the noncompact $\Z_K/\Z_{K'}$ one-form symmetry for $K' = \pi/A\_T$. In other words, the regime in which a tame description is valid is precisely the regime in which the original $\Z_K$ one-form symmetry has a spontaneously broken subgroup $\Z_{K'}$. As $g$ is reduced, the gaps between tame symmetry sectors decrease. A reasonable scenario is that, when the smallest gaps between these sectors become comparable to the tiny gaps between the $(K')^{N\_V - b_0 + b_1}$ ground states, the taming assumption breaks down and the spontaneously broken subgroup becomes $\Z_{K''}$ for $K'' > K'$. Such crossovers persist until $g$ becomes so small that the whole $\Z_K$ one-form symmetry is spontaneously broken. In other words, crossovers persist until there are $K^{N\_V - b_0 + b_1}$ approximately degenerate ground states, with the gaps between them being much smaller than the $O(1/N)$ gaps of the basic excitations with dispersions $\omega_{\b k}$.

At this point, it may be wise to take a step back and make a general observation. The entire structure explained above comes solely from the fact that the microscopic model \eqref{def H} couples four clocks at a time using the gauge-invariant operators $W_f$. The same underlying degrees of freedom, with a Hamiltonian that coupled pairs of clocks on different links, could have given an ordinary clock model whose tame sector contained just a scalar cQFT. Gauge invariance of the Hamiltonian ensures a much greater ground state degeneracy in certain parts of the parameter space, but it does not change the fact that \emph{all} clock configurations are equally physical --- as physical and meaningful as they would have been in the clock model.

Sometimes the literature gives the impression that working with gauge-variant states leads to nonunitarity or other paradoxes. One way to reach such conclusions is to try to understand tame gauge-theoretic states without ever leaving the $\varrho_v = 0$ sector. As shown here, though, gauge-variant states are perfectly well defined when handled with care.

\subsection{Photons: the basic noncompact Maxwell theory} \label{subsec basic nc Maxwell}

Every tame sector of the noncompact Maxwell theory   contains $d - 1$ decoupled harmonic oscillators at each $\b k \in \Pbb \backslash \{0\}$. These degrees of freedom are \emph{photons}, and their components are \emph{polarizations}. This Subsection will study photons in $d = 2$. Generalizing to $d > 2$ is easy.

The combination of vector potentials that enters the tame Hamiltonian \eqref{def HT A E} is
\bel{\label{def A tilde}
  A^\square_{\b k} \equiv \hat{\greek k}^1_{\b k} A^2_{\b k} - \hat{\greek k}^2_{\b k} A^1_{\b k}, \quad \hat{\greek k}^i_{\b k} \equiv \frac{{\greek k}^i_{\b k}}{\omega_{\b k}}.
}
Note that $A^\square_{\b k}$ is invariant under tame gauge transformations \eqref{tame gauge transf A}, which act as $A^i_{\b k} \mapsto A^i_{\b k} + \greek k_{\b k}^i \lambda_{\b k} \1$. Indeed, $\omega_{\b k} A^\square_{\b k}$ is simply the momentum space version of the gauge-invariant magnetic field $B^{12}_{\b x}$. When in $d = 2$, the magnetic field will be denoted $B_{\b x} \equiv B^{12}_{\b x}$.

The electric field conjugate to $A^\square_{\b k}$ is
\bel{\label{def E tilde}
  E^\square_{\b k} \equiv \hat{\greek k}^1_{\b k} E^2_{\b k} - \hat{\greek k}^2_{\b k} E^1_{\b k}.
}
The other linear combination of electric fields $E^i_{\b k}$ that enters the Hamiltonian is
\bel{\label{def rho k}
  \frac1{\omega_{\b k}} \rho_{\b k} \equiv - \hat {\greek k}^1_{-\b k} E^1_{\b k} - \hat {\greek k}^2_{-\b k} E^2_{\b k},
}
where $\rho_{\b k}$ is the \emph{tamed} Fourier transform of the background charge density $\varrho_{\b x}$. (In position space, the tame background charge can be defined as $\rho_{\b x} \equiv (\nabla E)_{\b x}$, with the divergence of the electric field given by \eqref{def nabla E}.) The Hamiltonian is then
\bel{\label{def HT A tilde}
  H\_T
   \approx
  \frac{g^2}{2} \big(E_0^i)\+ E_0^i
   +
  \frac{g^2}{2}
  \sum_{\b k \in \Pbb \backslash \{0\}}
    \frac{\rho_{\b k}\+ \rho_{\b k}}{\omega^2_{\b k}}
   +
  \frac{g^2}{2}
  \sum_{\b k \in \Pbb \backslash \{0\}}
    (E^\square_{\b k})\+ E^\square_{\b k}
   +
  \frac{1}{2 g^2}
  \sum_{\b k \in \Pbb \backslash \{0\}}
    \omega_{\b k}^2 (A^\square_{\b k})\+ A^\square_{\b k}.
}
From now on, the summation convention over repeated indices $i$ will be implied.

The operators in the first two sums in $H\_T$ are the tame symmetries that were discussed a few paragraphs above. Just like the taming backgrounds $(B\^{cl})^{ij}_{\b x}$, their analysis will be postponed to the following Subsections.

The last two sums in \eqref{def HT A tilde} together form the photon sector. These are the only ``quantum'' degrees of freedom in the tame regime: all other modes only contribute mutually commuting operators to the Hamiltonian. These degrees of freedom, with associated operators $A^\square_{\b k}$ and $E^\square_{\b k}$, constitute the \emph{basic noncompact Maxwell theory}. Its dynamics is governed by
\bel{\label{def H0}
  H_0 = \frac{g^2}{2} \sum_{\b k \in \Pbb \backslash \{0\}} (E^\square_{\b k})\+ E^\square_{\b k}  + \frac{1}{2 g^2} \sum_{\b k \in \Pbb \backslash \{0\}} \omega_{\b k}^2 (A^\square_{\b k})\+ A^\square_{\b k}.
}

The basic noncompact Maxwell theory has a precontinuum basis that is defined just like in the clock model \cite{Radicevic:2D}. Define the ladder operators as
\bel{\label{def bk}
  A^\square_{\b k} \equiv \frac g{\sqrt{2 \omega_{\b k}}} \left(b_{\b k} + b_{-\b k}\+ \right), \quad
  E^\square_{\b k} \equiv \frac{\sqrt{2 \omega_{\b k}}}{2\i g} \left(b_{\b k} - b_{-\b k}\+ \right),
}
for $\omega_{\b k}$ given by \eqref{def omega k}. The photon Hamiltonian then becomes
\bel{
  H_0 \approx \sum_{\b k \in \Pbb \backslash \{0\}} \omega_{\b k}\, \left( b_{\b k}\+ b_{\b k} + \frac12 \right).
}
To get a properly defined precontinuum basis, the ladder operators must be modified so that they cannot create more than $J_{\b k} \gg 1$ excitations at each momentum. As in the clock model, since $H_0$ is approximately free, this modification is not going to be detectable in states with few excitations, at least to first order in the taming parameters.

To define the continuum basis, start from the precontinuum basis generated by the set $\{b_{\b k}, b_{\b k}\+\}_{\b k \in \Pbb \backslash \{0\}}$ and remove individual ladder operators at $|k^i| > k\_S$, but keep the particle number operators $n_{\b k} \equiv b_{\b k}\+ b_{\b k}$ at all momenta. This results in the generating set
\bel{
  \{b_{\b k}, b_{\b k}\+ \}_{\b k \in \Pbb\_S \backslash \{0\}} \cup \{n_{\b k} \}_{\b k \notin \Pbb\_S}, \quad \Pbb\_S \equiv \{\b k\}_{-k\_S \leq k^i \leq k\_S}.
}
The high-momentum number operators generate the center of the continuum algebra. The ground state belongs to the superselection sector labeled by $n_{\b k} = 0$ for all $\b k \notin \Pbb\_S$. It is also possible to smoothe in a more rotation-invariant way by removing ladder operators at $|\b k| > k\_S$. The choice is immaterial to leading order in $k\_S/N$.

The operators $A^\square_{\b k}$ are related to Fourier transforms of the magnetic field by eq.~\eqref{def A tilde}. Indeed, smoothing can be understood as the projection
\bel{\label{def B(x)}
  B_{\b x} \mapsto B(\b x) \equiv \frac1N \sum_{\b k \in \Pbb\_S \backslash \{0\}} \omega_{\b k} A^\square_{\b k} \, \e^{\frac{2\pi\i}N \b k \b x}.
}
As usual when smoothing, this ensures that the field $B(\b x)$ varies slowly,
\bel{
  B(\b x + \b e_i) = B(\b x) + \hat\del_i B(\b x) + O \left( k\_S^2/N^2 \right),
}
where $\hat\del_i$ is the formal derivative satisfying $\hat \del_i \e^{\i \boldsymbol \alpha \b x} \equiv \i \alpha^i \e^{\i \boldsymbol \alpha \b x}$. It is tempting to further write
\bel{\label{B(x) via A(x)}
  B(\b x) = \del_1 A^2(\b x) - \del_2 A^1(\b x).
}
There is, however, no immediate way to choose smooth fields $A^i(\b x)$. In particular, the space of smooth states does \emph{not} contain two independent degrees of freedom per $\b k \in \Pbb\_S \backslash \{0\}$.

The situation with electric fields is different. It \emph{is} possible to naturally define both components of the electric field as smooth fields $E^i(\b x)$. All that is needed is to assume that the theory is in the sector with tame background charges $\rho_{\b x}$. Then \eqref{def E tilde} and \eqref{def rho k} imply
\bel{
  E^1_{\b k} = - \hat{\greek k}^1_{\b k} \frac{\rho_{\b k}}{\omega_{\b k}} - \hat{\greek k}^2_{-\b k} E^\square_{\b k}, \quad
  E^2_{\b k} = - \hat{\greek k}^2_{\b k} \frac{\rho_{\b k}}{\omega_{\b k}}  + \hat{\greek k}^1_{-\b k} E^\square_{\b k}.
}
The basic noncompact theory is usually defined to be in the sector with no background charge. In this case, both electric field components are given as reasonably simple smearings of the same microscopic field $E^\square_{\b y}$ (the Fourier transform of $E^\square_{\b k}$),
\bel{\label{def E(x)}
  E^1(\b x) = \sum_{\b y \in \Mbb} f^2_{\b x - \b y} \, E^\square_{\b y}, \quad E^2(\b x) = -\sum_{\b y \in \Mbb} f^1_{\b x - \b y} \, E^\square_{\b y}, \quad f^i_{\b r} \equiv \frac1{N^2} \sum_{\b k \in \Pbb\backslash\{0\}} \frac{\i k^i}{|\b k|} \e^{\frac{2\pi\i}N \b k \b r}.
}
This shows that both $E^1(\b x)$ and $E^2(\b x)$ act on the \emph{same} space of spatially smooth states, with one $\sim 2n\_T$-dimensional Hilbert space at each of the $(2k\_S)^2$ momentum modes in $\Pbb\_S \backslash \{0\}$.

The different statuses of continuum fields $A^i(\b x)$ and $E^i(\b x)$ --- one undefined, the other defined for a fixed $\rho_{\b x}$ --- should not be surprising. Indeed, information beyond the gauge-invariant sector is accessed by considering all possible background charge configurations. These are precisely the eigenstates of operators $\rho_{\b x}$ that were discarded when passing from $H\_T$ to the photon Hamiltonian $H_0$. It is in principle possible to define the smooth background charge density $\rho(\b x)$ as a smearing of the microscopic quantity $\rho_{\b x}$. However, the hitch is that the Hamiltonian \eqref{def HT A E}, viewed as a function of a ``particle number operator'' $\rho_{\b x}$, does not induce a continuum basis \cite{Radicevic:2D}. In other words, changing $\rho_{\b x}$ by a small amount at two points generally leads to a large change in the energy. (This is consistent with the picture presented in Subsection \ref{subsec tame gauge transf}, where the $\Z_K/\Z_{K'}$ gauge transformations generated by $\rho_{\b x}$ were associated to a confining gauge theory.) This is ultimately why it is impossible to naturally define continuum fields $A^i(\b x)$ by smoothing the basic noncompact Maxwell theory.

One way to proceed is to generalize the smoothing procedure. The lattice fields $A_{\b k}^i$ can be written as
\bel{\label{Ak via A tilde}
  A_{\b k}^1 \equiv - \hat{\greek k}_{-\b k}^2 A^\square_{\b k} + \hat{\greek k}_{\b k}^1 A^\times_{\b k}, \quad
  A_{\b k}^2 \equiv  \hat{\greek k}_{-\b k}^1 A^\square_{\b k} + \hat{\greek k}_{\b k}^2 A^\times_{\b k}.
}
The $A^\times_{\b k}$'s are operators conjugate to background charges $\rho_{\b k}/\omega_{\b k}$. They do not act on the Hilbert space of the basic noncompact Maxwell theory. Smoothing can thus be \emph{defined} to project them away together with high-momentum modes, giving expressions analogous to \eqref{def E(x)}:
\bel{\label{def A(x)}
  A_{\b x}^1 \mapsto A^1(\b x) \equiv - \sum_{\b y \in \Mbb} f^2_{\b x - \b y} A^\square_{\b y}, \quad
  A_{\b x}^2 \mapsto A^2(\b x) \equiv \sum_{\b y \in \Mbb} f^1_{\b x - \b y} A^\square_{\b y}.
}
With this prescription, the continuum fields $A^i(\b x)$ become familiar textbook objects.

The gauge transformations of the smooth vector potentials \eqref{def A(x)} are subtle. On the one hand, their expression explicitly depends only on gauge-invariant operators $A^\square_{\b x}$. On the other hand, decades of intuition say that these operators \emph{should} be gauge-variant.

The resolution of this puzzle lies in the fact that smoothing and operator multiplication do not commute. The tame gauge transformation \eqref{tame gauge transf O}, performed on the operator $A^i_{\b x}$ for gauge parameter $n_{\b y}$, involves the commutator $n_{\b y} [A^i_{\b x}, \rho_{\b y}]$. By \eqref{Ak via A tilde}, $A^i_{\b x}$ contains in it the operator $A^\times_{\b x}$, whose commutator with $\rho_{\b y}$ is proportional to the identity operator when acting on tame states. This identity operator survives the generalized smoothing \eqref{def A(x)} and shows up as the gauge transformation of the smooth vector potential,
\bel{\label{tame gauge transf A(x)}
  A^i(\b x) \mapsto A^i(\b x) + \del_i \lambda (\b x)\, \1,
}
where the c-number $\lambda(\b x)$ is the smoothing of the gauge parameter $\lambda_{\b x} = n_{\b x} \, \d A$. However, if one were to smoothe first and then perform the gauge transformation \eqref{tame gauge transf O}, $A^i(\b x)$ would turn out gauge-invariant.

Two points deserve emphasis now:
\begin{itemize}
  \item The operators $A^\times_{\b k}$ belong to the tame microscopic algebra, but not to the algebra of the basic noncompact Maxwell theory. The transformation \eqref{tame gauge transf A(x)} is defined in terms of smooth operators in the basic noncompact gauge theory, but \emph{it cannot be generated by any operator in this theory}. It is an echo of the tame symmetry transformation \eqref{tame gauge transf A}.
  \item When considering products of vector potentials in the basic noncompact Maxwell cQFT, it is \emph{not} necessary to specify whether these products are taken before or after projecting out the $A^\times$'s. The reason is that no product of $A^\times$'s can give an identity operator. Worrying about the order of operations is only important when there are also $\rho$'s in play --- which is never the case in the basic gauge theory.
\end{itemize}

While it may not be necessary to keep track of when the $A^\times$'s are dropped during an operator multiplication, it \emph{is} important to keep track of the order of smoothing and multiplication of other operators within the basic noncompact Maxwell theory. The ``commutator'' of these two operations is the operator product expansion (OPE) \cite{Radicevic:2D, Radicevic:2019jfe}.

Recall that, for two lattice operators $\O_{\b x}$ and $\~\O_{\b y}$, the OPE is
\bel{
  \O_{\b x} \times \~\O_{\b y} \equiv \O \~\O(\b x, \b y) - \O(\b x) \~\O(\b y).
}
This is a smooth operator that captures all the nontrivial short-range correlations of ladder operators that are contained in $\O_{\b x}$ and $\~\O_{\b y}$. Its ``singular'' behavior at decreasing $|\b x - \b y|$  furnishes the cQFT operators with a notion of a scaling dimension.

\newpage

The basic object to compute is the OPE of two gauge-invariant parts of vector field operators. Using \eqref{def bk}, this OPE can be expressed as
\bel{
  A^\square_{\b x} \times A^\square_{\b y}
  = \frac{g^2}{2N^2} \sum_{\b k \notin \Pbb\_S}  \frac1{\omega_{\b k}} \e^{\frac{2\pi \i}N \b k(\b x - \b y)} \avg{b\+_{-\b k} b_{-\b k} + b_{\b k} b_{\b k}\+} = \frac{g^2}{2N^2} \sum_{\b k \notin \Pbb\_S}  \frac1{\omega_{\b k}} \e^{\frac{2\pi \i}N \b k(\b x - \b y)}.
}
This derivation uses the fact that the vacuum expectation value of $n_{\b k} = b\+_{\b k} b_{\b k}$ for all $\b k \in \Pbb \backslash \{0\}$ is zero. A straightforward numerical exercise shows that
\bel{\label{OPE A tilde}
  A^\square_{\b x} \times A^\square_{\b y} = \frac{g^2}{4\pi} \frac1{|\b x - \b y|} + O\left(\frac{|\b x - \b y|}{\ell\_S} \right) + O\left(\frac1N\right),
}
where the ``string length''  is
\bel{
  \ell\_S \equiv \frac N{2k\_S}.
}
This result says that the gauge-invariant (``physical'') polarization of the $d = 2$ photon is given by the continuum field
\bel{
  \frac{\sqrt{4\pi}}g A^\square(\b x)
}
which has a well defined scaling dimension $\Delta_{\sqrt{4\pi} A^\square/g} = 1/2$. (For more on writing continuum fields using such rescalings, see \cite{Radicevic:2019mle} where this was discussed in detail for the Ising CFT.)

An unsavory feature of this continuum limit is that the natural gauge-theoretic variables --- vector potentials $A^i(\b x)$ from \eqref{def A(x)} and magnetic fields $B(\b x)$ from \eqref{def B(x)} --- are not local functions of the gauge-invariant field $A^\square(\b x)$ that exhibits the nice scaling property, \eqref{OPE A tilde}. Of course, this does not mean these operators are ill defined. They are simply not easy to relate to the field $A^\square(\b x)$ and its derivatives. The momentum space relation $\omega_{\b k} A^\square_{\b k} = B_{\b k}$ implies that, schematically,
\bel{\label{B(x) via A tilde}
  B(\b x) \sim \sqrt{\del_1^2 + \del_2^2} A^\square(\b x).
}
These square roots are, ultimately, the reason why the Maxwell theory is a scale- but not conformal-invariant cQFT \cite{ElShowk:2011gz}.

A numerical calculation reveals the OPE of two magnetic fields to be
\bel{\label{OPE B}
  B_{\b x} \times B_{\b y} = \frac{g^2}{2N^2} \sum_{\b k \notin \Pbb\_S}  \omega_{\b k} \e^{\frac{2\pi \i}N \b k(\b x - \b y)} \approx -\frac{g^2}{4\pi} \frac1{|\b x - \b y|^{2\Delta_B}}, \quad \Delta_B \approx 1.5.
}
Reassuringly, the same answer follows from acting on \eqref{OPE A tilde} by the product of square roots of Laplacians w.r.t.\ $\b x$ and $\b y$.
To get this agreement it is crucial to take $|\b x - \b y| \gg 1$ (for practical purposes, $|\b x - \b y| \gtrsim 10$ is sufficient). It is only at such ``large'' distances that the discrete derivatives agree with continuum ones.


Smooth Wilson loops of charge $q$ that surround individual lattice plaquettes,
\bel{\label{def W(x)}
  W^q(\b x) \equiv \e^{\i q B}(\b x),
}
are complete analogues of vertex operators $\e^{\i p \varphi}(x)$ from the noncompact scalar cQFT \cite{Radicevic:2D}. Here is a brief recap of their properties. They can be defined for arbitrary $q \in \R$, but only $O(n\_T)$ of them are linearly independent. They are ``eigenstates'' of the smoothing operation, i.e.\ they satisfy
\bel{
  W^q(\b x) = \N \e^{\i q B(\b x)}, \quad \N \equiv \e^{-\frac{q^2}2 B_{\b x} \times B_{\b x}}.
}
Finally, they obey the product structure
\algns{
  W^{q_1} \cdots W^{q_n}(\b x_1, \ldots, \b x_n)
  &= \e^{-\frac12 \sum_{i \neq j} q_i q_j B_{\b x_i} \times B_{\b x_j}} \, W^{q_1}(\b x_1) \cdots W^{q_n} (\b x_n) \\
  &= \e^{-\frac12 \sum_{i, j = 1}^n q_i q_j B_{\b x_i} \times B_{\b x_j}} \, \e^{\i \sum_{i = 1}^n q_i B(\b x_i)}.
}

This product structure implies that the expectation of a smooth Wilson loop around a region $\Vbb$ is
\bel{\label{Wq expression}
 \avg{W^q(\del \Vbb)} =  \e^{-\frac{q^2}2 \sum_{\b x, \, \b y \in \Vbb} B_{\b x} \times B_{\b y}} \, \avg{\e^{\i q \sum_{\b x \in \Vbb} B(\b x)}}.
}
To calculate this, assume the region --- or the fluctuation size --- is small enough, so that
\bel{
  q^2 \sum_{\b x,\, \b y \in \Vbb} \avg{B(\b x) B(\b y)}  \ll 1.
}
Schematically, if $|\Vbb|$ is the number of sites in the region $\Vbb$, this smallness is assured if
\bel{\label{Wilson loop area constraint}
  q |\Vbb| \ll \frac{\ell\_S}{A\_T}.
}
Under this assumption, the expectation value on the r.h.s.\ of \eqref{Wq expression} is unity to leading order in taming/smoothing parameters. Then, using \eqref{OPE B}, the Wilson loop can be expressed as
\bel{
  \avg{W^q(\del \Vbb)} \approx \exp\left\{-\frac{q^2g^2}{4N^2} \sum_{\b k \notin \Pbb\_S} \omega_{\b k} \left|\sum_{\b x \in \Vbb} \e^{\frac{2\pi\i}N \b k \b x} \right|^2 \right\}.
}
For a square region of linear size $r$, subject to $r \ll \sqrt{\ell\_S/q A\_T}$ as per \eqref{Wilson loop area constraint}, this can be numerically evaluated to be
\bel{\label{Wq result}
  \avg{W^q(\del \Vbb)} \approx \e^{-c (qg)^2/r}
}
for positive $c = O(1)$, and up to some nonessential technicalities regarding corners. This is the ubiquitous Coulomb behavior of the Wilson loop in the noncompact Maxwell theory.

\subsection{Photons and electric backgrounds: the standard noncompact Maxwell theory} \label{subsec standard nc Maxwell}

The previous Subsection solved the photon sector of the Maxwell theory and showed how its lattice operators can be smoothed to give elements of a cQFT algebra. The main lesson was that the photon theory is free, with gaps between energies at each momentum $\b k \in \Pbb\_S$ given by
\bel{
  \omega_{\b k} \approx \frac{2\pi}N |\b k|.
}
The gap size $\E\_{photon} = 2\pi/N$ sets the benchmark for all other energy gaps.

This idea is based on the assumed tameness of low-energy states. The reasoning is roughly as follows. If the low-energy eigenstates of the lattice theory \eqref{def H} are indeed tame at some $g$, then this theory must have an approximate ground state degeneracy $(K')^{N\_V - b_0 + b_1}$. This number counts the superselection sectors of the theory whose energy differences are much smaller than all other scales in the theory. In particular, their energy differences must be much smaller than the photon gap $\E\_{photon}$. If other degrees of freedom, e.g.\ eigenstates of tame symmetries, have gaps much smaller than $\E\_{photon}$, then it may be the case that the taming assumption is wrong, and that the ground state degeneracy is actually $(K'')^{N\_V - b_0 + b_1}$ for $K'' > K'$. Indeed, it is natural to expect that as $g$ is reduced, the gaps of these other degrees of freedom will indeed decrease until the taming assumption is violated and the theory enters a new regime in which the spontaneously broken group is $\Z_{K''}$ and not $\Z_{K'}$. On the other hand, as long as $g$ is such that all degrees of freedom have gaps comparable to $\E\_{photon}$, the tameness assumption cannot be faulted (from the ground state degeneracy perspective).

With this in mind, consider again the tame symmetries that enter the tame Hamiltonian \eqref{def HT A tilde}, which can be written as
\bel{
  H\_T \approx \frac{g^2}{2} \big(E_0^i)\+ E_0^i + \frac{g^2}{2} \sum_{\b k \in \Pbb \backslash \{0\}} \frac{\rho_{\b k}\+ \rho_{\b k}}{\omega^2_{\b k}} + \sum_{\b k \in \Pbb \backslash \{0\}} \omega_{\b k} \left(b_{\b k}\+ b_{\b k} + \frac12 \right).
}
Different eigenvalues of $E_0^i$ and $\rho_{\b k}$ (or $\rho_{\b x}$) correspond to different superselection sectors of the tame theory. These sectors can be interpreted as different \emph{classical} excitations in the theory governed by $H\_T$. They can be promoted to fully quantum degrees of freedom by  allowing perturbations of $H\_T$ to include operators that map one tame symmetry sector to another. These excitations will be called \emph{electric backgrounds}. The theory that includes both photons and electric backgrounds will be called the \emph{standard noncompact Maxwell theory}. As the name suggests, this is what is typically meant by the phrase ``noncompact Maxwell/U(1) theory'' in the literature. Note that there is also a hybrid model, in which only the nonlocal electric backgrounds are dynamical. All electric backgrounds will be treated equally here.

What are the energy gaps between sectors with different electric backgrounds? In other words, what are the eigenvalues of the (linearly independent) operators $E_0^i$ and $\rho_{\b k}$?

To start, focus on the ``local'' term
\bel{\label{local el bkgd energy}
  \frac{g^2}2 \sum_{\b k \in \Pbb \backslash \{0\}} \frac{\rho_{\b k}\+ \rho_{\b k}}{\omega^2_{\b k}} \equiv \frac{g^2}2 \sum_{\b x, \, \b y \in \Mbb} \rho_{\b x} \rho_{\b y} D(\b x - \b y),
}
where the interaction between background charges is mediated by
\gathl{\label{def D(x - y)}
  D(\b x - \b y) = \frac1{N^2} \sum_{\b k \in \Pbb \backslash \{0\}} \frac1{\omega^2_{\b k}} \e^{\frac{2\pi\i}N \b k (\b x - \b y)} \stackrel {|\b x - \b y| \ll N}\approx -\frac1{2\pi} \log\frac{|\b x - \b y|}N + D, \\
  D(0) = \frac1{N^2} \sum_{\b k \in \Pbb \backslash \{0\}} \frac1{\omega^2_{\b k}} \approx \frac1{2\pi} \log N + D_0.
}
Eigenvalues of each $\rho_{\b x}$ are integers, bounded by $n\_T$ in absolute value. They are independent at each $\b x$, except they must satisfy the net neutrality constraint
\bel{\label{neutrality}
  \sum_{\b x \in \Mbb} \rho_{\b x} \in K \Z
}
in order for the operator $\left(\prod_{\b x \in \Mbb} G_{\b x}\right)\_T$ to always be the identity. (On a general lattice, there is a separate constraint of this kind for each disconnected component of $\Mbb$.)

The result \eqref{def D(x - y)} indicates that the local contribution to the energy, \eqref{local el bkgd energy}, for a pair of opposite background charges ($\rho_{\b x} = \delta_{\b x}^{(\b y_1)} - \delta_{\b x}^{(\b y_2)}$) grows as $\frac{g^2}{4\pi} \log|\b y_1 - \b y_2|$. This familiar result is one way to quantify how ``confined'' the noncompact Maxwell theory is; the fact that background charges cost a nonzero energy to insert is already reassuring, as it was an assumption made when defining taming in the first place. In order for the gaps of the background charges to be comparable to the photon gaps, it is further necessary to have
\bel{\label{scaling of g}
  g^2 \sim \frac1N.
}

This is a very important result. It means that the cQFT description should not be expected to hold at $g = O(N^0)$ couplings, in stark contrast to the $d = 1$ clock model. Said another way, $d = 2$ Maxwell theory will only exhibit continuum behavior along an interval of length $O(1/\sqrt N)$ in coupling space. In a sense, this is generic: most lattice theories do not display a line of points with continuum descriptions, and instead they only have isolated critical points whose vicinities are described by perturbations of a single CFT. The relatively unusual property of this lattice theory is that the coupling $g^2$ must be taken to be proportional to the lattice spacing in order to get a cQFT. More on this will be said in Subsection \ref{subsec temp smoothing}.

The nonlocal electric backgrounds are captured by the zero-momentum modes of the electric field, $E^i_0$. They encode the electric flux that passes through noncontractible one-cycles of the torus. (In $d > 2$, they encode the flux through noncontractible $(d - 1)$-cycles.) Let
\bel{\label{def Phi}
  \Phi^i_{y} \equiv \sum_{\b x \in \Mbb, \  x^i = y} E^i_{\b x}
}
be the total electric flux that passes through the $x^i = \trm{const}$ cycle of $\Mbb$. The zero-mode is then expressible as
\bel{
  E_0^i = \frac1N \sum_{y = 1}^N \Phi^i_y,
}
and represents the ``average'' flux passing through each of the $N$ cycles that are perpendicular to the $i$'th direction. In $d = 3$, the $\Phi^i$'s are sometimes called \emph{'t Hooft operators}.

The $\Phi^i_y$ operators are useful because their eigenvalues are integer multiples of $K$. This follows from demanding that the noncompact Maxwell theory be dual to a noncompact scalar theory; the demand is reasonable because a $\Z_K$ lattice gauge theory is exactly dual to a $\Z_K$ clock model via Kramers-Wannier duality. This will be reviewed in Subsection \ref{subsec KW}. For now this fact will just be taken for granted. An immediate consequence is that the eigenvalues of each $E_0^i$ are integer multiples of $K/N$. This means that different sectors of the nonlocal tame symmetry have energy gaps proportional to $g^2 (K/N)^2$. If the coupling is chosen according to \eqref{scaling of g}, the gaps due to electric fluxes are
\bel{\label{scaling elect lines}
  \E\_{electric} \sim \frac{K^2}{N^3}.
}
The nonlocal electric backgrounds are thus much heavier than the local ones if $K \gg N$.

A dramatic scenario now presents itself. If $K \gg N$, the nonlocal $\Z_K$ symmetry may be \emph{unbroken} even as the local $\Z_K$ symmetries --- the ones generated by $G_v$ --- break to $\Z_{K'}$ at small values of $g$. This does not seem to invalidate the analysis of photons or tame local symmetries, but it does suggest that the topological ground state degeneracy may be absent for some parameter choices. Even more exotically, at some couplings there may exist $(K'_0)^{N\_V - b_0} (K'_1)^{b_1}$ ground states for $K'_1 < K'_0 < K$. While these symmetry breaking scenarios cannot be ruled out by the present analysis, they are not mandatory. This paper will assume, for simplicity, that both local and nonlocal tame symmetry break the same way as $g$ is dialed across the part of parameter space where \eqref{scaling of g} holds.

The most important aspect of the result \eqref{scaling elect lines} is the $(K/N)^2$ enhancement of the gap $\E\_{electric}$, compared to $\E\_{photon}$. In the $d = 1$ clock model, it was shown that $2\pi / K$ plays the r$\hat{\trm o}$le of $\hbar$ in the compact boson cQFT. By this token, the nonlocal electric backgrounds represent quantum, $O(1/\hbar^2)$, effects in the Maxwell cQFT.


\subsection{Photons, electric backgrounds, magnetic fluxes: the compact Max\-well theory}

The remaining superselection sectors that must be discussed are associated to different taming backgrounds $(A\^{cl})_{\b x}^i$. As discussed in Subsection \ref{subsec dynamics on torus}, the backgrounds of interest are assumed to minimize the magnetic term in \eqref{def H}. The equation they obey, \eqref{taming bkgd eq}, in $d = 2$ becomes
\bel{\label{taming bkgd eq 2d}
  \del_i B\^{cl}_{\b x} \ \trm{mod} \ 2\pi = 0.
}
Since, by definition, $\sum_{\b x \in \Mbb} B\^{cl}_{\b x} = 0$, the only way for this equation to have nontrivial solutions is for $B_{\b x}\^{cl}$ to have jumps by integer multiples of $2\pi$ across some links. Such backgrounds are
\bel{\label{flux backgrounds}
  (A\^{cl})^1_{\b x} = - \frac{2\pi}N m_1 x^2, \quad (A\^{cl})^2_{\b x} = \frac{2\pi}N m_2 x^1, \quad m_i \in \Z.
}
This is the ``symmetric gauge'' choice of $A\^{cl}$ that leads to the background magnetic field
\bel{\label{def magn flux}
  B\^{cl}_{\b x} = \frac{2\pi}N (m_1 + m_2) - 2\pi m_1 \, \delta_{x^2,\, N} - 2\pi m_2 \, \delta_{x^1, \, N},
}
which clearly satisfies \eqref{taming bkgd eq 2d}. As in the clock model \cite{Radicevic:2D}, it will be assumed that the taming backgrounds \eqref{flux backgrounds} vary slowly, so that over a ``string length'' $\ell\_S$ they change by at most $2A\_T$.  This leads to the bound
\bel{\label{bound w}
  |m_i| \lesssim k\_S A\_T.
}

The theory that includes all taming backgrounds subject to \eqref{bound w} will be called the \emph{compact Maxwell theory} because it crucially exploits the periodicity of $A\^{cl}_\ell$. The extra, classical degrees of freedom will be called \emph{magnetic fluxes}. The operators that insert magnetic flux are untame and nonlocal. They are given by
\bel{
  \prod_{\b x \in \Mbb} (X_{\b x}^1)^{-[\frac KN  m_1 x^2]} (X_{\b x}^2)^{[\frac KN  m_2 x^1]}.
}

In the presence of a generic magnetic flux \eqref{def magn flux}, the Hamiltonian contains the term
\bel{
  H\_T[B\^{cl}] = H\_T[0] + \frac{2\pi^2}{g^2} (m_1 + m_2)^2
}
This assumes that $(A\^{cl})^i_{\b x}$ varies slow enough so that $m_i\ll N$ in all sectors. Thus the energy gaps of magnetic flux excitations are
\bel{\label{magn flux bkgd energy}
  \E\_{magnetic} = \frac{2\pi^2}{g^2} (m_1 + m_2)^2.
}

There are two remarkable properties of this result. First, all sectors with $m_1 = -m_2$ have the same energy as the ground state of the corresponding noncompact theory. Indeed, the local magnetic flux $B\^{cl}_{\b x}$ vanishes in this case. (Note, however, that taming backgrounds \eqref{flux backgrounds} with $m_1 = -m_2$ are \emph{not} gauge-equivalent to each other.) This degeneracy does not have immediate physical consequences; taking these sectors into account simply shifts the free energy by a constant. The same degeneracy is actually present at all nonzero $m_i$. Distinct magnetic flux sectors are thus effectively labeled by just one integer $m = m_1 + m_2$.

The second interesting property of \eqref{magn flux bkgd energy} is that the associated gaps are
\bel{
  \E\_{magnetic} \sim N m^2
}
when the scaling \eqref{scaling of g} is assumed. This means that the $m \neq 0$ sectors are always infinitely heavy in the compact Maxwell cQFT. They do not affect the correlation functions in significant ways. For example, the smooth Wilson loop \eqref{Wq expression} in the presence of a nontrivial $B\^{cl}_{\b x}$ simply acquires a phase, so that \eqref{Wq result} changes to
\bel{\label{Wilson loop}
  \avg{W^q(\del \Vbb)} \approx \e^{\frac{2\pi\i}N m q r^2} \e^{-c (qg)^2/r}.
}
Taking the magnetic fluxes into account therefore does \emph{not} alter the Coulomb behavior of the Wilson loop in this theory.

At this point it is necessary to remark on a longstanding piece of lore about Maxwell theory. Using path integral methods, Polyakov has famously shown that this theory confines at every coupling \cite{Polyakov:1975rs}. (See also \cite{Drell:1978hr}.) Precisely, this result claims that a photon has mass
\bel{
  m \propto \e^{-\epsilon/2g^2}, \quad \epsilon = O(1),
}
and that the Wilson loop decays with the area of the region $\Vbb$ as
\bel{
  \avg{W^q(\del \Vbb)} \sim \e^{-m (q g r)^2}.
}
To explain the contradiction with \eqref{Wilson loop}, note that Polyakov's result does not say anything useful about the parametric regime $g^2 \sim 1/N$ of interest here. In this situation, the mass is $m \propto \e^{- \epsilon' N}$ for $\epsilon' = O(1)$, and hence much smaller than the gap $\E\_{photon} \sim 1/N$ of \emph{massless} photons. Said another way, the mass induces a correlation length $1/m \sim \e^{\epsilon' N}$ that is much larger than the number of lattice sites in the system. Thus the photon is effectively massless in this regime. Where Polyakov's mechanism truly shines is in providing a concrete illustration of how increasing $g$ to the point where $\e^{-\epsilon/2g^2} \sim 1/N$, or $g \sim 1/\sqrt{\log N} \ll 1$, causes the theory to confine. The result \eqref{Wilson loop}, or indeed any analysis of Maxwell cQFT in this paper, has no strong claim to correctness in this regime.

\newpage

\subsection{Maxwell-scalar duality} \label{subsec KW}

Subsection \ref{subsec standard nc Maxwell} claimed, without proof, that the tame electric fluxes \eqref{def Phi} had eigenvalues given by integer multiples of $K$. The Subsection will show how this follows from demanding that the Kramers-Wannier dual of Maxwell theory also have a tame low-energy spectrum. Readers not interested in the long but precise treatment of this fundamental subject should \strike{read a different paper} jump to Table \ref{table KW} that summarizes how tame operators dualize.

\begin{figure}[b]
\begin{center}
\begin{tikzpicture}[scale = 1]
  \contourlength{1.5pt}

  \draw[step = 1, gray] (-2.2, -1.2) grid (3.2, 1.2);
  \draw[step = 1, dotted, gray, shift = {(0.5, 0.5)}] (-2.7, -1.7) grid (2.7, 0.7);

  \filldraw[fill=green!60!black, draw=green!60!black, very thick, fill opacity = 0.2]  (-2, -1) rectangle +(1, 1);
  \draw (-2, -0.7) node[left, green!60!black] {\contour{white}{$W^{12}_{\b x}$}};
  \draw[green!60!black, thick] (-2, -1) node {$\bullet$};

  \draw[green!60!black, very thick] (0, 0) -- (0, 1);
  \draw (0, 0) node[below, green!60!black] {\contour{white}{$X_{\b y}^2$}};
  \draw[green!60!black, thick] (0, 0) node {$\bullet$};

  \draw[green!60!black, very thick] (2, 0) -- (3, 0);
  \draw (2, 0) node[left, green!60!black] {\contour{white}{$X_{\b z}^1$}};
  \draw[green!60!black, thick] (2, 0) node {$\bullet$};

  \draw[blue, thick] (-1.5, -0.5) node {$\bullet$};
  \draw (-1.5, -0.5) node[above, blue] {\contour{white}{$X^\vee_{\b x}$}};

  \draw[blue, thick, dotted] (-0.5, 0.5) -- (0.5, 0.5);
  \draw[blue, thick] (-0.5, 0.5) node {$\bullet$};
  \draw[blue, thick] (0.5, 0.5) node {$\bullet$};
  \draw (0.001, 0.5) node[above, blue] {\contour{white}{$Z^\vee_{\b y - \b e_1}\, (Z^\vee_{\b y})\+$}};

  \draw[blue, thick, dotted] (2.5, -0.5) -- (2.5, 0.5);
  \draw[blue, thick] (2.5, -0.5) node {$\bullet$};
  \draw[blue, thick] (2.5, 0.5) node {$\bullet$};
  \draw (2.5, 0.5) node[right, blue] {\contour{white}{$Z^\vee_{\b z} $}};
  \draw (2.5, -0.5) node[right, blue] {\contour{white}{$(Z^\vee_{\b z - \b e_2})\+ $}};

\end{tikzpicture}
\end{center}
\caption{\small Gauge-invariant operators in the Maxwell theory on $\Mbb$ (green), and their dual operators in the clock model on $\Mbb^\vee$ (blue).  For each operator shown, the point in its argument is indicated by a filled circle. Each label like $\b x$ is used to denote both a vertex on the original lattice and a vertex on the dual lattice that is just northeast of the original one.}
\label{fig duality}
\end{figure}
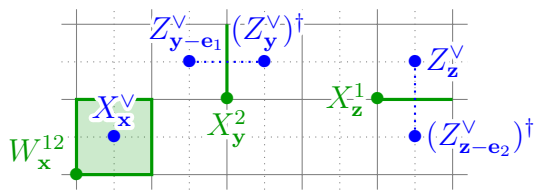

Kramers-Wannier (KW) is an exact duality between the lattice gauge theory \eqref{def H} and another theory with a $\Z_K$ target whose degrees of freedom live on Poincar\'e duals of original plaquettes $f \in \Mbb$. In $d = 2$, the dual theory is simply a clock model on the sites of the dual lattice $\Mbb^\vee$. The basic duality relation between the algebra generators is
\bel{\label{KW basic}
  X_\ell = \prod_{f \in \del_{-1} \ell} Z^\vee_f, \qquad
  W_f = X_f^\vee.
}
Components of $\Mbb^\vee$ will be denoted by the labels of their dual elements from $\Mbb$, so that e.g.\ a site of the dual lattice shares a  label with the site's dual plaquette in the original lattice. On the toric lattice with a particular choice of orientations (Fig.\ \ref{fig duality}),
\bel{
  X_{\b x}^1 = \left( Z^\vee_{\b x - \b e_2}\right)\+ Z^\vee_{\b x}, \qquad
  X_{\b x}^2 = \left( Z^\vee_{\b x} \right)\+ Z^\vee_{\b x - \b e_1}, \qquad
  W^{12}_{\b x} = X^\vee_{\b x}.
}
In this Subsection, however, $\Mbb$ will not be assumed to be a torus.

The KW duality \eqref{KW basic} only maps gauge-invariant operators in the Maxwell theory. Indeed, the dual of any Gauss operator is inferred from the first relation in \eqref{KW basic} to be
\bel{\label{singlet constraint Gv}
  G_v = \prod_{\ell \in \del_{-1} v} \prod_{f \in \del_{-1} \ell} Z^\vee_f  = \prod_{f \in \del_{-2} v} \left(Z^\vee_f \right)\+ Z^\vee_f = \1.
}
This means that only the $\varrho_v = 0$ sector of the original gauge theory maps under this duality.

In fact, not even all gauge-invariant operators map under this version of KW. Consider the $d = 2$ 't Hooft operators
\bel{
  T_{c^\vee} \equiv \prod_{\ell \in c^\vee} X_\ell.
}
Here $c^\vee$ is a noncontractible one-cycle on $\Mbb^\vee$, and $T_{c^\vee}$ is a product of clock operators on links of $\Mbb$ that pierce this cycle. This operator must also dualize to the identity. In particular, no operators that fail to commute with it --- i.e.\ no Wilson loops along noncontractible cycles on $\Mbb$ --- map under \eqref{KW basic}.

In short, this duality is ``singlet-singlet.'' It only maps the singlet sector of the one-form $\Z_K$ symmetry. The dual theory, similarly, must be in the singlet sector of the shift symmetry generated by $\prod_f X_f^\vee$ for any product over faces that belong to a connected component of $\Mbb$.

If the symmetries appearing in a singlet-singlet duality are not anomalous, they can be \emph{twisted} by coupling one or both sides to gauge degrees of freedom \cite{Radicevic:2018okd}. For example, consider mapping the $\Z_K$ gauge theory to a $\Z_K$ clock model on sites coupled to \emph{another} $\Z_K$ gauge theory on links of the dual lattice, with operators $\zeta_\ell^\vee$ and $\xi_\ell^\vee$. This modifies \eqref{KW basic} to
\bel{\label{KW fully twisted}
  X_\ell = \zeta^\vee_\ell \prod_{f \in \del_{-1} \ell} Z^\vee_f, \qquad
  W_f = X_f^\vee.
}
In fact, individual (gauge-variant) clock operators can be mapped to shift operators via
\bel{\label{KW fully twisted g-var}
  \e^{\i \vartheta_\ell} Z_\ell\+ =  \xi_\ell^\vee, \quad \vartheta_\ell \in \d A \, \Z.
}
This is consistent with $W_f = X_f^\vee$ from \eqref{KW fully twisted} only if the dual theory obeys the gauge constraint
\bel{\label{KW gauss 0}
  G^\vee_f \equiv X_f^\vee \prod_{\ell \in \del f} \xi^\vee_\ell = \e^{\i (\delta\vartheta)_f} \1.
}
From the map of $X_\ell$ in \eqref{KW fully twisted}, the singlet constraints $G_v = \1$ and $T_{c^\vee} = \1$ in the original theory are now modified to nontrivial maps
\bel{
  G_v = \omega_v^\vee \equiv \prod_{\ell \in \del_{-1}v} \zeta_\ell^\vee, \qquad T_{c^\vee} = \omega^\vee_{c^\vee} \equiv \prod_{\ell \in c^\vee} \zeta_\ell^\vee.
}
In other words, all eigenstates of the one-form $\Z_K$ symmetry generators are now dualizable. Background charges map to magnetic fluxes, and 't Hooft lines map to Wilson lines.

An astute reader will have already noticed that a requirement like $G_v = \1$ means that a tame theory cannot map under the original duality \eqref{KW basic}. Conversely, assuming the duality is valid in a tame theory would lead to a contradiction because theories without background charges cannot be tamed, as explained in Subsection \ref{subsec tame gauge transf}.

This is where twisting comes in handy. The duality \eqref{KW fully twisted} in which the matter side is fully twisted --- i.e.\ where the full $\Z_K$ symmetry is gauged in the clock model --- \emph{does} map tame gauge theory states of the form \eqref{degenerate partial states}. In fact, a \emph{partial} twist of the matter side of \eqref{KW basic} suffices to make these states dualizable. Since the tame states \eqref{degenerate partial states} are obtained by only superposing sectors with nontrivial background charges of the $\Z_{K'} \subset \Z_K$ part of the gauge symmetry, it is enough to gauge the $\Z_{K'}$ shift symmetry in the dual clock model in order to get a duality that maps tame states. In this partially twisted duality, only Wilson loops and operators that shift original link variables by $2A\_T = \frac{2\pi}{K'} = \frac{K}{K'} \d A$ map,
\bel{\label{KW part twisted}
  X_\ell^{K/K'} = \zeta_\ell^\vee \prod_{f \in \del_{-1} \ell} (Z^\vee_f)^{K/K'},
  \qquad W_f = X_f^\vee.
}
Here $\zeta_\ell^\vee$ are clock operators in a $\Z_{K'}$ gauge theory on $\Mbb^\vee$. The first of these maps implies that powers of Gauss operators map to Wilson loops on the dual plaquettes in the $\Z_{K'}$  gauge theory,
\bel{
  G_v^{K/K'} =  \omega_v^\vee,
}
and so all of their eigenstates have meaningful duals. The same holds for 't Hooft lines $T^{K/K'}_{c^\vee}$.

Instead of \eqref{KW fully twisted g-var}, a consistent duality of $Z_\ell$ now has the general form
\bel{\label{KW part twisted g-var}
  \e^{\i \vartheta_\ell} Z_\ell\+ =  \xi_\ell^\vee, \quad \vartheta_\ell \in \d A\, \Z,
}
where $\xi_\ell^\vee$ is the shift operator in the $\Z_{K'}$ gauge theory on $\Mbb^\vee$ that satisfies $\xi_\ell^\vee \zeta_\ell^\vee = \e^{\frac{2\pi\i}{K'}} \zeta_\ell^\vee \xi_\ell^\vee$. The only states that map under \eqref{KW part twisted} and \eqref{KW part twisted g-var} must obey the operator equation
\bel{\label{KW compactness 1}
  Z_\ell^{K'} = \e^{\i K' \vartheta_\ell} \1.
}
In other words, only $K'$ states on each link map for a given $\vartheta_\ell$. They are of the form
\bel{
  \qvec{\e^{\i A_\ell}} = \qvec{\e^{\frac{2\pi \i}{K'}n\^{cl}_\ell + \i \alpha_\ell}}, \quad  n_\ell\^{cl} \in \{0, 1, \ldots, K' - 1\}.
}
There is a \emph{different} duality for each value $\alpha_\ell = \vartheta_\ell\, \trm{mod}\, \frac{2\pi}{K'} \in \{ -\frac{\pi}{K'}, \ldots, \frac{\pi}{K'} - \d A \}$. The ``integer part'' of $\vartheta_\ell/(\frac{2\pi}{K'})$ plays no important r$\hat{\trm o}$le in the duality and can be set to zero, so that $\alpha_\ell = \vartheta_\ell$.

Like in \eqref{KW gauss 0}, multiplying \eqref{KW part twisted g-var} over $\ell \in \del f$ gives the Gauss law in the dual theory,
\bel{\label{KW gauss 1}
  \e^{\i (\delta \alpha)_f} W_f\+ = \prod_{\ell \in \del f} \xi_\ell^\vee, \quad\trm{or} \quad \e^{\i(\delta\alpha)_f} =  X_f^\vee \prod_{\ell \in \del f} \xi_\ell^\vee.
}
Let $\qvec{p_f^\vee}$, $p_f^\vee \in \{0, 1, \ldots, K - 1\}$, be the eigenstates of $X_f^\vee$. One part of the Gauss law then says that $p_f^\vee - \frac{(\delta \alpha)_f}{\d A} = 0 \, \trm{mod}\, K'$. The other part says that, in most states, $(p\^{cl})^\vee_f \equiv \left[p_f^\vee/K' \right]$ is the electric charge of the $\Z_{K'}$ gauge theory with background charges $(\delta n\^{cl})_f$.



\newpage

The states labeled by $\alpha_\ell \equiv A_\ell \, \trm{mod}\, \frac{2\pi}{K'}$ for a fixed $A\^{cl}_\ell \equiv \frac{2\pi}{K'} n\^{cl}_\ell = 2A\_T \, n_\ell\^{cl}$ can be used to construct the tame subspace relative to a taming background specified by $A\^{cl}_\ell$. It is convenient to define the operator $U_\ell$ that measures just $\alpha_\ell$,
\bel{\label{def U}
  U_\ell \qvec{\e^{\i (A\^{cl}_\ell + \alpha_\ell)}} \equiv \e^{\i \alpha_\ell} \qvec{\e^{\i(A\^{cl}_\ell + \alpha_\ell)}}.
}
Then the duality \eqref{KW part twisted g-var} can be recorded as
\bel{\label{KW part twisted g-var 2}
  Z_\ell U_\ell\+ = (\xi_\ell^\vee)\+,
}
once --- as in the previous paragraph --- it is assumed that $\vartheta_\ell$ lies in the same range as $\alpha_\ell$, namely
\bel{\label{KW range alpha}
  \alpha_\ell \in \left \{- A\_T, - A\_T + \d A, \ldots, A\_T - \d A \right\}, \quad A\_T = \frac\pi{K'}.
}

By \eqref{KW part twisted g-var 2}, a fixed taming background $A_\ell\^{cl}$ is dual to a confined $\Z_{K'}$ gauge theory in which clock operators obey the constraint $\xi_\ell^\vee = \e^{\i A_\ell\^{cl}}\1$. The Gauss law \eqref{KW gauss 1} now fixes $(p\^{cl})^\vee_f \d A = A_\ell\^{cl}$ in most states. (This hedging will be explained below \eqref{KW part twisted remainder}.) The remaining task is to understand the duality of operators that measure or change $\alpha_\ell$ but not $A\^{cl}_\ell$.

The operator that measures $\alpha_\ell$ is simply $U_\ell$ from \eqref{def U}. One operator that changes $\alpha_\ell$ is the original shift operator $X_\ell$, which acts as
\bel{
  X_\ell \qvec{\e^{\i (A\^{cl}_\ell + \alpha_\ell)}} =
  \left\{
    \begin{array}{ll}
      \qvec{\e^{\i (A\^{cl}_\ell + \alpha_\ell - \d A)}},
      & \alpha_\ell \in \left \{- A\_T + \d A, \ldots,  A\_T - \d A \right\};  \\
      \qvec{\e^{\i \left[(A\^{cl}_\ell - 2A\_T) + (A\_T -  \d A)\right]}},
      & \alpha_\ell = - A\_T.
    \end{array}
  \right.
}
However, this also shows that $X_\ell$ does \emph{not} preserve $A\^{cl}_\ell$. Indeed, when applied $\frac K{K'}$ times, this operator preserves $\alpha_\ell$ while only changing $A_\ell\^{cl}$, as already explained in the discussion below \eqref{KW part twisted}. To obtain an operator that only affects $\alpha_\ell$, consider the projection $(X_\ell)\_C$ of $X_\ell$ to the compact space spanned by the $\alpha_\ell$'s at a fixed $A_\ell\^{cl}$. This operator is defined to act as
\bel{
  (X_\ell)\_C \qvec{\e^{\i (A\^{cl}_\ell + \alpha_\ell)}} =
  \left\{
    \begin{array}{ll}
      \qvec{\e^{\i (A\^{cl}_\ell + \alpha_\ell - \d A)}},
      & \alpha_\ell \in \left \{- A\_T + \d A, \ldots, A\_T - \d A \right\};  \\
      0,
      & \alpha_\ell = - A\_T.
    \end{array}
  \right.
}
Projecting this way will be called \emph{compactifying}. This is precisely the projection used to upgrade target space smoothing to taming \cite{Radicevic:1D}. As usual with these projections, multiplication does not always commute with compactifyng. For example,
\bel{
  \1 = (X\+_\ell X_\ell)\_C \neq (X\+_\ell)\_C (X_\ell)\_C.
}
Unless explicitly denoted like on the r.h.s.\ above, it will always be assumed that multiplication comes before compactifying. Note that, in particular, $\left( X_\ell^{K/K'} \right)\_C = 0$.

In a fixed background $A\^{cl}_\ell$, one consistent proposal for the duals of $U_\ell$ and $(X_\ell)\_C$ is then
\bel{\label{KW part twisted remainder}
  (X_\ell)\_C = \prod_{f \in \del_{-1}\ell} (Z_f^\vee)\_S,
  \qquad \e^{\i (\delta A\^{cl})_f} \prod_{\ell \in f} U_\ell = X_f^\vee.
}
\begin{itemize}
  \item The operator $U_\ell$ itself does not map under this duality. This is analogous to how $Z_f^\vee$ or $Z_\ell$ did not map under the original KW duality \eqref{KW basic}. It is necessary to twist one side to get the individual clock operators on the other side to map, like in \eqref{KW fully twisted g-var}.
  \item Compactifying the $X_\ell$'s dualizes to target space smoothing of the $Z_f^\vee$'s on dual links $\ell$. This smoothing is a restriction to the $X^\vee_f$ eigenspace labeled by target momenta from the set
      \bel{\label{KW dual momenta}
         p_f^\vee \in \left\{(p\^{cl})_f^\vee - \frac K{2K'}, (p\^{cl})_f^\vee - \frac K{2K'} + 1, \ldots,  (p\^{cl})_f^\vee + \frac K{2K'}\right\},
      }
      for a fixed \emph{smoothing background} $(p\^{cl})_f^\vee \in \{0, \frac K{K'}, \ldots, K - \frac K{K'} \}$. This is the momentum-space analogue of the taming background $(\varphi\^{cl})^\vee_x$.\footnote{In the $d = 0$ clock model with a harmonic potential, the low-energy spectrum has four taming backgrounds labeled by the target positions $\varphi\^{cl} \in \{0, \pi\}$ and target momenta $p\^{cl} \in \{0, K/2\}$. The nontrivial value of $p\^{cl}$ can be understood as a second \emph{spin structure} that needs to be accounted for in this cQM \cite{Radicevic:1D}.} A more precise term for $(\varphi\^{cl})^\vee_x$ would be \emph{compactification background}, with ``taming background'' then used as an umbrella term that can refer to either $(p\^{cl})^\vee_x$ or $(\varphi\^{cl})^\vee_x$ --- or to both at once.

      Projections to the smooth subspace, such as $(Z_f^\vee)\_S$ and $\left( (Z_f^\vee)\+ \right)\_S$, annihilate a momentum eigenstate instead of taking it outside of the interval \eqref{KW dual momenta}. The smoothing parameter used here is
      \bel{\label{KW pS dual}
         p\_S^\vee \equiv \frac K{2K'} = \frac{K n\_T}{2E\_S}.
      }
  \item The individual $\alpha_\ell$'s can accumulate so that $| \sum_{\ell \in f} \alpha_\ell | > A\_T$. This means that a Wilson loop $\prod_{\ell \in f} U_\ell$ may still have a phase larger than $A\_T$. The Gauss law \eqref{KW gauss 1} states that taming backgrounds are dualized via $(p\^{cl})^\vee_f \, \d A= (\delta A\^{cl})_f$ --- \emph{except} in states in which the $\alpha_\ell$'s accumulate as above. (This explains the earlier weasel words ``in most states.'') However, such deviations are bounded, $|(\delta A\^{cl})_f - (p\^{cl})_f^\vee \, \d A| \leq 2 \d A$, because each plaquette has four links. These \emph{defect states} are thus relatively scarce, and they involve ``large,'' $O(A\_T)$, values of $\alpha_\ell$ that do not appear at the lowest energies.
  \item The gauge constraint in the theory with compactified shift operators is (as usual) obtainable by taking the product of the first map in \eqref{KW part twisted remainder} over all links containing a given site $v$,
      \bel{
        \prod_{\ell \in \del_{-1} v} (X_\ell)\_C = \prod_{f \in \del_{-2} v} \left((Z_f^\vee)\+ \right)\_S (Z_f^\vee)\_S.
      }
      The Gauss law $\prod_{\ell \in \del_{-1} v} (X_\ell)\_C = \1$ only holds if $\alpha_\ell$ is not at the edge of its range \eqref{KW range alpha}.
\end{itemize}

\newpage

This is not the end of the story. Readers distracted by the onslaught of different dualities are gently reminded that the ultimate goal is to find a duality of cQFTs, i.e.\ a duality that maps tame states in the original theory to tame states in the dual one. The duality \eqref{KW part twisted remainder} makes sure that mappable states in the original theory have one crucial aspect of tameness --- target space compactness. This simultaneously ensured that mappable states in the dual theory satisfy the second aspect of tameness --- target space smoothness. However, this duality manifestly does not map compact states in the dual theory. Taking the product of the second relation in \eqref{KW part twisted remainder} over all faces of (a connected component of) $\Mbb$ gives
\bel{
  \1 = \prod_{f \in \Mbb} X_f^\vee.
}
This means that no dual state with a definite taming background $(\varphi\^{cl})^\vee_f$ can map; only equal superpositions of all possible $(\varphi\^{cl})^\vee_f$'s can.

A further partial twist of \eqref{KW part twisted remainder} is evidently needed. The goal is to ensure a nontrivial duality for the operator that generates global shifts by $2\varphi\_T^\vee \equiv 2n\_T^\vee \d\varphi^\vee \equiv K n\_T^\vee/p\_S^\vee$ in the dual clock model. To achieve this, fix $A\^{cl}_\ell$ and amend the duality \eqref{KW part twisted remainder} to get
\bel{\label{KW part twisted twice}
  (X_\ell)\_C = \prod_{f \in \del_{-1}\ell} (Z_f^\vee)\_S, \qquad
  \zeta_f \, \prod_{\ell \in f}  \big(\e^{\i (\delta A\^{cl})_f } U_\ell \big)^{Kn\_T^\vee/p\_S^\vee} = (X_f^\vee)^{Kn\_T^\vee/p\_S^\vee}.
}
The gauge theory on the original lattice is here coupled to a $\Z_{p\_S^\vee/n\_T^\vee}$ one-form gauge theory.\footnote{Recall that a one-form gauge theory has \emph{two-form} degrees of freedom. In other words, its clock variables live on faces of the lattice $\Mbb$. It is the Gauss operators that are one-form --- they live on links.} The clock operator in this theory is denoted $\zeta_f$. The generator of global $2\varphi\_T^\vee$ shifts in the dual theory maps to the Wilson surface in the $\Z_{p\_S^\vee/n\_T^\vee}$ theory,
\bel{
  (X_f^\vee)^{Kn\_T^\vee/p\_S^\vee} = \prod_{f \in \Mbb} \zeta_f \equiv \omega.
}
It is now possible to say that a tame state in the dual theory will correspond to a confined state of the $\Z_{p\_S^\vee/n\_T^\vee}$ gauge theory, in which $\avg{\omega} = 0$.

Before proceeding, it may be useful to tidy up the notation. It is natural to choose the tame spaces on both dual sides to have the same dimensionality. This is implemented by
\bel{
  n\_T = n\_T^\vee.
}
Using \eqref{KW pS dual}, the one-form gauge group is identified as $\Z_{K/2E\_S}$, and the duality \eqref{KW part twisted twice} becomes simply
\bel{\label{KW part twisted twice simplified}
  (X_\ell)\_C = \prod_{f \in \del_{-1}\ell} (Z_f^\vee)\_S, \qquad
  \zeta_f \,  \prod_{\ell \in f} U_\ell^{2E\_S} = (X_f^\vee)^{2E\_S}.
}

The twist of the compactified degrees of freedom in the original theory now allows the individual smoothed clock operators in the dual theory to be mapped, in analogy with \eqref{KW part twisted g-var 2}, by positing the duality
\bel{
  \xi_f\+ = (U^\vee_f)\+ (Z_f^\vee)\_S.
}
Here $\xi_f$ is the shift operator in the $\Z_{K/2E\_S}$ gauge theory, and $U^\vee_f$ is the operator that measures the ``small fluctuations'' in the dual clock model. To describe its action, let
\bel{
  \qvec{\e^{\i \varphi_f^\vee}}, \quad \varphi_f^\vee \in \left\{0, \d \varphi^\vee, \ldots, 2\pi - \d\varphi^\vee  \right\},
}
be the $2p\_S^\vee$ approximate eigenstates of $(Z_f^\vee)\_S$ for a given $(p\^{cl})^\vee_f$, with associated eigenvalues $\e^{\i \varphi_f^\vee}$. (Also, recall from \eqref{KW pS dual} that $\d\varphi^\vee \equiv \pi/p\_S^\vee = K'\, \d A$.) Then $U_f^\vee$ acts as
\bel{
  U_f^\vee \left \qvec{\e^{\i \left((\varphi\^{cl})^\vee_f + \alpha_f^\vee \right)} \right} = \e^{\i \alpha_f^\vee} \left\qvec{\e^{\i \left((\varphi\^{cl})^\vee_f + \alpha_f^\vee \right)}\right}.
}
The backgrounds here are $(\varphi\^{cl})^\vee_f \in \{0, 2\varphi\_T^\vee, \ldots, 2\pi - 2\varphi\_T^\vee\}$, and the range of $\alpha_f^\vee$ is, in analogy to \eqref{KW range alpha},
\bel{
  \alpha_f^\vee \in \left\{ -\varphi\_T^\vee, -\varphi\_T^\vee + \d\varphi^\vee, \ldots, \varphi\_T^\vee - \d\varphi^\vee \right\}, \quad \varphi\_T^\vee = E\_S \, \d A.
}

Note that $U_f^\vee$ only acts on a single smooth subspace with a fixed background $(p\^{cl})_f^\vee$. When further restricted to act on a tame subspace in which both $(\varphi\^{cl})^\vee_f$ and $(p\^{cl})_f^\vee$ are fixed, this operator can be written as
\bel{
  U_f^\vee = \e^{-\i (\varphi\^{cl})^\vee_f} (Z_f^\vee)\_T.
}

What is the dual of $U_f^\vee$, or of $(Z_f^\vee)\_T$, according to \eqref{KW part twisted twice simplified}? First, note that $(X_\ell)\_C$ has $K/K' = 2p\_S^\vee$ approximate eigenstates; each is  a superposition of original clock eigenstates $\qvec{E_\ell}$ around values $E_\ell \in \{0, K', \ldots, K-K'\}$. At fixed $(\varphi_f\^{cl})^\vee$, by the first relation in \eqref{KW part twisted twice simplified}, the compact states that dualize are those approximate eigenstates of $(X_\ell)\_C$ that are centered around $E_\ell \in \{E_\ell\^{cl} - n\_T K', E_\ell\^{cl} -n\_T K' + K', \ldots, E_\ell\^{cl} + n\_T K' - K'\}$. This set has $2n\_T K' = 2E\_S$ elements. Compactifying in the clock model is thus dual to target space smoothing, i.e.\ to restricting target momenta to $E_\ell \in \{E\^{cl}_\ell - E\_S, \ldots, E\^{cl}_\ell + E\_S - \frac{E\_S}{n\_T})\}$, with $E\^{cl}_\ell \, \d A = (\delta \varphi\^{cl})^\vee_f$ in most states:
\bel{\label{KW part twisted twice remainder 1}
  (X_\ell)\_T = \e^{\i (\delta \varphi\^{cl})^\vee_\ell} \prod_{f \in \del_{-1} \ell} U^\vee_f = \prod_{f \in \del_{-1}\ell} (Z_f^\vee)\_T.
}

As before, for a small number of states, $|\sum_{f\in \del_{-1} \ell} \alpha^\vee_f|$ may exceed $\varphi\_T^\vee$. Duals of such states then belong to sectors with smoothing background $E\^{cl}_\ell \, \d A \neq (\delta \varphi\^{cl})^\vee_f$. However, since each link only touches two faces, these defect states have $|E\^{cl}_\ell \, \d A - (\delta \varphi\^{cl})^\vee_f| \leq 1$, and they feature fluctuations of size $\alpha^\vee_f = O(\varphi\_T)$ on the dual lattice. Such configurations will not be the lowest-energy states and can be ignored to the first approximation.

The remaining operators that need addressing are $\e^{\i (\delta A\^{cl})_f } \prod_{\ell \in \del f} U_\ell$ and $X_f^\vee$. These operators do not map under the duality  \eqref{KW part twisted twice simplified}: only their $2E\_S$'th powers do. But their tamed counterparts \emph{can} map. If $\e^{\i A\^{cl}_\ell} U_\ell$ is projected to an operator $\e^{\i A\^{cl}_\ell}  (U_\ell)\_S$ that never changes $E\^{cl}_\ell$, and if $X_f^\vee$ is projected to an operator $(X_f^\vee)\_C$ that never changes $(\varphi\^{cl})^\vee_f$, these projections can be consistently dualized to each other. Indeed, these operators are tamings of the original clock and shift operators, $(Z_\ell)\_T$ and $(X_f^\vee)\_T$. Their duality can thus be recorded simply as
\bel{\label{KW part twisted twice remainder 2}
  \prod_{\ell \in \del f} (Z_\ell)\_T = (X_f^\vee)\_T.
}

The two dualities, \eqref{KW part twisted twice remainder 1} and \eqref{KW part twisted twice remainder 2}, are the central results of this Subsection. They bear a striking resemblance to the starting duality \eqref{KW basic}. It is therefore imperative to recognize that these tame dualities \emph{do not follow} from the standard KW duality, even if they formally look similar to it. Instead, the tame dualities follow from a twice-twisted version of \eqref{KW basic}, with various gauge fields floating around on both sides of the duality.

The tame dualities also implicitly contain data that was not even present in the standard KW duality. Specifically, the standard duality  knows nothing about  the taming parameters $E\_S$ and $n\_T$, or about the taming backgrounds $E\^{cl}_\ell$ and $A\^{cl}_\ell$. All of them have nontrivial duals, and these dualities depend on the properties of the $\Z_{E\_S/n\_T}$ and $\Z_{K/E\_S}$ ``twisting fields,'' which were never parts of the original map \eqref{KW basic}. (The details of these dualities lie scattered across the above pages, and for easy reference they are collected in Table \ref{table KW}.)

The relations between taming backgrounds, used to derive \eqref{KW part twisted twice remainder 1} and \eqref{KW part twisted twice remainder 2}, are incorrect for a small number of tame states. In these defect states, the accumulation of phases from the tame degrees of freedom is large enough to change the taming background of the dual state. It is actually possible to take the tame dualities as the starting point, and to then \emph{derive} the relations between taming backgrounds that are correct even in defect states. This does not fully remove the subtleties here: tame shift operators in one theory may still effect the change of a taming background in the dual theory.

Yet another subtlety comes from the fact that, schematically, $(Z\+)\_T Z\_T \neq (Z\+ Z)\_T = \1$. (Note that $(Z\_T)^2 = (Z^2)\_T$; only inverses cause taming not to commute with multiplication.) Dualizing products of tame operators does not generically yield simple results like \eqref{singlet constraint Gv}. For example, recall that taking the product of \eqref{KW part twisted twice remainder 1} over $\ell \in \del_{-1} v$ does not yield an identity in the dual clock model, and hence the duality does not imply the familiar Gauss law. This effect becomes more drastic when dualizing higher powers of operators. At its extreme, the $(2n\_T)$'th power of these operators is guaranteed to be identically zero on both sides. Thus many results in what follows can only be trusted when operator powers are low and when all states have support far away from the edges of the tame subspace.

\begin{table}
\begin{center}
\begin{tabular}{ccc}
  \textbf{Maxwell theory} && \textbf{Clock model} \\
  \hline \\
  $(X_\ell)\_T \approx \e^{\i\, \d A (E_\ell\^{cl} + E_\ell)}$
  & $\quad = \quad$
  & $\displaystyle \prod_{f \in \del_{-1}\ell} (Z_f^\vee)\_T  \approx \e^{\i \left[(\delta \varphi\^{cl})^\vee_\ell + (\delta \varphi^\vee)_\ell\right]}$
  \\
  \\
  $\displaystyle \prod_{\ell \in \del f} (Z_\ell)\_T \approx \e^{\i \left[ (\delta A\^{cl})_f + (\delta A)_f \right]} $
  & $\quad = \quad$
  & $(X_f^\vee)\_T \approx \e^{\i\, \d A \left[ (p\^{cl})^\vee_f + \pi_f^\vee \right]}$ \bigskip
  \\
  \hline
  \\
  $E\^{cl}_\ell \, \d A$
  & $\quad \trm{``$=$''} \quad$
  & $(\delta \varphi\^{cl})^\vee_\ell \ \trm{mod} \ 2\pi$
  \\ \\
  $(\delta A\^{cl})_f \ \trm{mod} \ 2\pi$
  & $\quad \trm{``$=$''} \quad$
  & $(p\^{cl})^\vee_f \, \d A$
  \\ \\
  $E_\ell \, \d A$
  & $\quad = \quad$
  & $(\delta \varphi^\vee)_\ell \ \trm{mod} \ 2\varphi\_T^\vee$
  \\ \\
  $(\delta A)_f \ \trm{mod} \ 2A\_T$
  & $\quad = \quad$
  & $\pi^\vee_f \, \d A$ \bigskip
  \\
  \hline
  \\
  $\displaystyle G_v^{K/K'} \equiv \prod_{\ell \in \del_{-1} v} X_\ell^{2p\_S^\vee}$
  & $\quad = \quad$
  & $\displaystyle \omega^\vee_v \equiv \prod_{\ell \in \del_{-1} v} \zeta_\ell^\vee$
  \\ \\
  $\displaystyle T_{c^\vee}^{K/K'} \equiv \prod_{\ell \in c^\vee} X_\ell^{2p\_S^\vee}$
  & $\quad = \quad$
  & $\displaystyle \omega^\vee_{c^\vee} \equiv \prod_{\ell \in c^\vee} \zeta_\ell^\vee$
  \\ \\
  $\e^{-\i A_\ell\^{cl}}$
  & $\quad = \quad$
  & $\xi_\ell^\vee$
  \\ \\
  $\displaystyle \omega \equiv \prod_{f \in \Mbb} \zeta_f$
  & $\quad = \quad$
  & $\displaystyle \prod_{f \in \Mbb} (X_f^\vee)^{2E\_S}$
  \\ \\
  $\xi_f$
  & $\quad = \quad$
  & $\e^{-\i (\varphi\^{cl})^\vee_\ell}$ \bigskip
  \\ \hline
\end{tabular}
\end{center}
\caption{\small Dualities between various operators and taming backgrounds for $\Z_K$ models on arbitrary $d = 2$ lattices. Both microscopic theories feature a $\Z_K$ model coupled to appropriate ($\Z_{E\_S/n\_T}$ or $\Z_{K/2E\_S}$) ``twisting'' gauge fields, which ensure that individual taming sectors are dualizable. \\
\hspace*{1ex} The first two lines show the basic \emph{tame dualities} from \eqref{KW part twisted twice remainder 1} and \eqref{KW part twisted twice remainder 2}. When acting on tame states, $X\_T$'s and $Z\_T$'s can be approximated as exponentials of canonical position and momentum fields, to leading order in taming parameters. Note that the taming parameters satisfy the relations $2E\_S/K = n\_T^\vee/p\_S^\vee$ and $1/K' = n\_T/E\_S = 2p\_S^\vee/K$. \\
\hspace*{1ex} The next four lines follow from the tame dualities. The first two are relations between taming backgrounds that are only correct in tame states with no defects. The other two are relations between canonical fields that are most commonly quoted as the continuum Maxwell-scalar duality. \\
\hspace*{1ex} The final five lines are dualities involving twist fields and taming backgrounds. None of these operators act on the tame degrees of freedom. The clock operators $\xi_f$ and $\xi_\ell^\vee$ measure the taming backgrounds, while the Wilson lines $\omega^\vee_v$, $\omega^\vee_{c^\vee}$ and surfaces $\omega$ change these backgrounds.}
\label{table KW}
\end{table}

\newpage

At long last, with Table \ref{table KW} in hand, it is possible to study global constraints that tame operators must obey. Consider first the product of tame shift operators $(X_\ell)\_T$ along a dual one-cycle $c^\vee$,
\bel{
  \prod_{\ell \in c^\vee} (X_\ell)\_T = \e^{\i \, \d A \sum_{\ell \in c^\vee} E_\ell\^{cl}} \prod_{\ell \in c^\vee} \left[\e^{\i \, \d A\, E_\ell} + O\left(E\_S^2/K^2\right) \right].
}
If the taming corrections are small enough compared to the length $N$ of the cycle, so e.g.\ $N \gg \frac K{E\_S}$, this becomes
\bel{
  \prod_{\ell \in c^\vee} (X_\ell)\_T \approx \e^{\i \, \d A \sum_{\ell \in c^\vee} \left[ E_\ell\^{cl} + E_\ell \right]}.
}
By \eqref{KW part twisted twice remainder 1}, the dual is
\bel{
  \e^{\i \sum_{\ell \in c^\vee} \left[(\delta \varphi\^{cl})^\vee_\ell + (\delta \varphi^\vee)_\ell\right] } \approx \1.
}
This means that the sum of electric fields along $c^\vee$ must be an integer multiple of $K$,
\bel{
  \sum_{\ell \in c^\vee} \left[ E_\ell\^{cl} + E_\ell \right] \approx q K \1, \quad q \in \Z.
}
In the Maxwell theory it will always be assumed that the low-energy states have $E_\ell\^{cl} = 0$. If this is the case, the sum of electric fields --- which can be recognized as the flux operator \eqref{def Phi} generalized to an arbitrary lattice --- satisfies
\bel{\label{quantization Phi}
  \Phi_{c^\vee} \equiv \sum_{\ell \in c^\vee} E_\ell \approx q K \1.
}
This proves the claim made way back in Subsection \ref{subsec standard nc Maxwell}.

In the more general case with $E\^{cl}_\ell \neq 0$, it will still be true that $E\^{cl}_\ell \in \{2E\_S, \ldots, K - 2E\_S\}$. Thus global consistency more generally forces the flux operator to be
\bel{
  \Phi_{c^\vee} \approx 2q E\_S \1, \quad q \in \Z.
}

It is good to keep in mind that the ``quantization'' of electric flux \eqref{quantization Phi} only holds if
\bel{\label{target space rel 1}
  N \gg \frac K{E\_S}.
}
If this is false, there will exist tame states where the flux is so large that \eqref{quantization Phi} gets modified to
\bel{
  \Phi_{c^\vee} +  \sum_{\ell \in c^\vee} \left(\alpha_2 E_\ell^2 \d A + \alpha_3 E_\ell^3 \d A^2 + \ldots \right) \approx q K \1
}
for some $O(1)$ numbers $\alpha_i$. Thus \eqref{target space rel 1} should be understood as an important relation between target  and position spaces that is implicit in familiar cQFTs.

One more comment is important here. Assume that the smoothing background is $E\^{cl}_\ell = 0$, so that \eqref{quantization Phi} holds. This result is true for \emph{any} taming background $(\varphi\^{cl})^\vee_f$. In particular, it is \emph{not} necessary to assume that this background minimizes the potential by satisfying $(\delta \varphi\^{cl})^\vee_\ell = \trm{const}$. However, if this extra assumption is made, then the only taming backgrounds of interest in the dual model are winding configurations with $(\delta \varphi\^{cl})^\vee_\ell \in \frac{2\pi}N \Z$, at least as long as $\Mbb$ is a torus. In this case one can imagine a correspondence between the winding number and the integer $q$ in \eqref{quantization Phi}. Indeed, the usual Hamiltonians will associate the same energy gaps to winding sectors and to electric fluxes in a dual pair of theories. Because of this, it is often said that the winding of the compact scalar is ``dual'' to the electric flux in the Maxwell theory. This paragraph is here to remind you that this flux-winding correspondence does \emph{not}, strictly speaking, follow from the microscopic duality.

The other global constraint follows from the second tame duality, \eqref{KW part twisted twice remainder 2}, when it is multiplied over all faces of $\Mbb$, which is assumed to have only one connected component. As above, this gives
\bel{
  \prod_{f \in \Mbb} X_f^\vee \approx \e^{\i\, \d A \sum_{f \in \Mbb} \left[ (p\^{cl})^\vee_f + \pi^\vee_f \right] } \approx \1.
}
Here it is assumed that
\bel{
  N\_V \gg \frac{E\_S}{n\_T}.
}
On an $N\times N$ torus, this becomes
\bel{\label{target space rel 2}
  N \gg \sqrt{\frac{E\_S}{n\_T}},
}
which is the second important relation between the various large numbers that appear in the definitions of relevant cQFTs. (The first one was \eqref{target space rel 1}.)

In a trivial smoothing background, $(p\^{cl})^\vee_f = 0$, it now follows that the momentum zero-mode must satisfy
\bel{\label{quantization pi vee}
  \pi^\vee_0 \equiv \frac1N \sum_{f \in \Mbb} \pi^\vee_f \approx m^\vee \frac KN \1, \quad m^\vee \in \Z.
}
As before, this a priori has no relation to the magnetic backgrounds $B\^{cl}_f$ in the Maxwell theory. However, once they are assumed to minimize the potential and obey \eqref{taming bkgd eq 2d}, there is an immediate correspondence between the total magnetic flux $m \equiv m_1 + m_2$ from \eqref{def magn flux} and the parameter $m^\vee$ in \eqref{quantization pi vee}. In this situation there are more magnetic flux backgrounds than there are momentum zero-modes, which makes it apparent that there is no intrinsic flux-momentum duality  --- just a natural correspondence between their quantizations.

The labels $q$ and $m^\vee$ in \eqref{quantization Phi} and \eqref{quantization pi vee} do not range over all $\Z$. The crudest bound for e.g.\ $m^\vee$ comes from the fact that the largest eigenvalue of $\pi_0^\vee$ in a tame subsector is $p\_S^\vee = K \frac{n\_T}{E\_S}$, so $|m^\vee| \leq N^2 \frac{n\_T}{E\_S}$, which is large according to \eqref{target space rel 2}. A more realistic bound assumes that $\pi_f^\vee$ varies only across lengths greater than the smearing scale $N/k\_S$, giving $|m^\vee| \lesssim \big(\frac{N}{k\_S} \big)^2 \frac{n\_T}{E\_S}$.

\newpage

\section{Continuum path integrals for Maxwell theory} \label{sec Maxwell path int}

\subsection{A cautionary remark}

Even though they were first introduced using canonical methods \cite{Wegner:1984qt}, lattice gauge theories are traditionally defined \cite{Wilson:1974sk} and numerically studied \cite{Rothe:1992nt} in the path integral formalism. A standard path integral approach to the $\Z_K$ Maxwell theory would be to consider a $(d + 1)$-dimensional lattice with clock variables $A_\ell$ on all links. The action would be, for example,
\bel{
  \frac1\hbar S[A] = \frac1\hbar \sum_f \left[ 1 - \cos (\delta A)_f \right].
}
When the coupling constant $\hbar$ is sufficiently small, it can be argued that only configurations with $(\delta A)_f \, \trm{mod}\, 2\pi  \approx 0$ contribute to the path integral. Then the small fluctuations of variables on temporal links can be integrated out. This serves to impose gauge constraints, and the degrees of freedom that remain are small, slowly varying, gauge-invariant fluctuations of $A_\ell$. This way the path integral accesses the basic noncompact Maxwell cQFT. Including suitable background configurations further allows the path integral to compute computables in the other versions of the Maxwell cQFT that were described in Section \ref{sec Maxwell}.

The analysis of fermion path integrals and their doubling problems earlier in this series \cite{Radicevic:2D} has already indicated that one should be wary when taking path integrals as starting points. It is reasonable to expect the above construction to capture certain universal features of the Maxwell cQFT after suitable processing. On the other hand, it would be quite \emph{un}reasonable to expect this lattice path integral to have the same microscopics as the starting theory \eqref{def H}. It is by no means clear how much the path integral knows about the ``string scale'' $k\_S$, or about the different taming parameters and bounds on allowed flux backgrounds in the cQFT. The underlying Hamiltonian is ultimately needed to specify the temperatures or couplings at which a path integral can capture the thermodynamics of a true quantum theory.

Variables on temporal links provide a particularly lucid example of the dissonance between path integrals and underlying Hamiltonian theories. As explained above, small fluctuations of these variables dynamically impose the Gauss law. This much is expected in the microscopic theory. But why are these variables $\Z_K$-valued to begin with? There is no canonical reason for this, and in fact their angular nature may give rise to spurious degrees of freedom.\footnote{A related observation was made off the cuff in \cite{Banerjee:2013mca}.}

Instead of dealing with universality and trying to make sense of these conventional kinds of path integrals, this paper will derive continuum path integrals for gauge theories starting from the Hamiltonian. The procedure is broadly the same as in clock models \cite{Radicevic:1D, Radicevic:2D}. This will in turn clarify the limits of validity for the path integral.

\subsection{The basic noncompact theory}

A path integral based on a Hamiltonian theory is defined by expressing the partition function as
\bel{
  \Zf \equiv \Tr\, \e^{-\beta H} = \Tr \prod_{\tau = \d\tau}^{\beta} \e^{-\d\tau H}, \quad \d\tau \equiv \frac\beta{N_0},
}
and then inserting a decomposition of unity of the form
\bel{\label{decomposition of unity}
  \1 = \sum_{\greek s,\, \greek f} \qproj{\greek f; \greek s}{\greek f; \greek s}
}
at each time step. These states are chosen so that $H$ can only change the labels $\greek f$. The partition function becomes a sum over paths $\greek f_\tau$ in spacetimes labeled by sectors $\greek s$,
\bel{
  \Zf = \sum_{\greek s} \sum_{\{\greek f_\tau\}} \prod_{\tau = \d\tau}^\beta \qmat{\greek f_{\tau + \d\tau}; \greek s}{\e^{-\d\tau H}}{\greek f_\tau; \greek s} \equiv \sum_{\greek s} \e^{- \beta \E(\greek s)} \sum_{\{\greek f_\tau\}} \e^{-S[\greek f; \, \greek s]}.
}

A \emph{continuum path integral}, as defined in this series, is obtained by using a particular \emph{undercomplete} basis of states $\qvec{\greek f; \greek s}$ in place of the decomposition \eqref{decomposition of unity}, and by \emph{fixing} the value of $\greek s$. Different path integrals are defined by different choices of which symmetry eigenvalues to include in $\greek s$ (and therefore fix), and which to include in $\greek f$ (and therefore sum over). For the basic noncompact Maxwell theory on the torus, the fixed labels are
\bel{
  \greek s = \left( \left\{(E\^{cl})^i_{\b x}\right\}_{\b x \in \Mbb}, \left\{ (A\^{cl})^i_{\b x}\right\}_{\b x \in \Mbb}, E^i_0 , \left\{ \rho_{\b x} \right\}_{\b x \in \Mbb} , \left\{ n_{\b k}\right\}_{\b k \notin \Pbb\_S} \right).
}
Here $(E\^{cl})^i_{\b x}$ and $(A\^{cl})^i_{\b x}$ are smoothing and compactification backgrounds that together form the taming background, as discussed in Subsection \ref{subsec KW}. Assuming that only the backgrounds that minimize the respective potentials need to be taken into account, it is enough to focus on
\bel{
  (E\^{cl})^i_{\b x} = 0, \quad (A\^{cl})^1_{\b x} = - \frac{2\pi}N m_1 x^2, \quad (A\^{cl})^2_{\b x} = \frac{2\pi}N m_2 x^1,
}
for some integers $m_i \in \Z$. The other three kinds of labels in $\greek s$ are eigenvalues of tame symmetries (see Subsection \ref{subsec standard nc Maxwell}) and the occupation numbers of nonsmooth photon modes.

Most of the labels that enter $\greek s$ are only symmetries of the tame Hamiltonian, and so fixing them makes sense only for sufficiently low temperatures $1/\beta$ where untame states do not matter.  In the specific case of the basic noncompact theory, if the goal is to get a result $\Zf\_{bnc}$ approximating the microscopic answer $\Zf$, the sector $\greek s = \greek s_0$ must be the one that includes the ground state, and $\beta$ must be large enough to suppress contributions from all other sectors $\greek s$. The ground state sector $\greek s_0$ is characterized by $m_i = 0$, $E^i_0 = 0$, $\rho_{\b x} = 0$, and $n\_{\b k} = 0$.

The above paragraphs tacitly assumed that there was a single ground state, and hence a single sector $\greek s_0$. This was not true in the clock model, where the microscopic theory had $K'$ degenerate ground states in the regime where it exhibited continuum behavior \cite{Radicevic:2D}. This is even less true in the Maxwell theory, where the backgrounds $(A\^{cl})^i_{\b x}$ can be changed by Gauss and 't Hooft operators $G^{K/K'}_{\b x}$ and $T^{K/K'}_{c^\vee}$ to get new sectors. As discussed in Subsection \ref{subsec tame gauge transf}, there are $(K')^{N\_V - b_0 + b_1}$  degenerate ground states, each with its own taming background, and hence with its own sector $\greek s$. Just as in the clock model, however, this degeneracy does not affect the physics beyond contributing an overall multiplicative factor to the partition function. These states will be disregarded without much comment from now on.

The next step is to choose the states $\qvec{\greek f; \greek s}$ to be the approximate non-null eigenstates of smooth vector potentials $\widehat A^i(\b x)$ at points
\bel{\label{def xi}
  \b x = \ell\_S \boldsymbol \xi, \quad \xi^i \in \{1, 2, \ldots, 2k\_S\}.
}
However, recall that the operators $\widehat A^1(\b x)$ and $\widehat A^2(\b x)$ from \eqref{def A(x)} are linearly dependent at a fixed $\b x$. Only $\widehat A^\square(\b x)$ is defined in terms of the continuum basis generators. The precise proposal is thus to use the approximate non-null eigenstates of $\widehat A^\square(\b x)$ for $\b x$ of form \eqref{def xi}. These are the tame states
\bel{\label{tame fields}
  \qvec A \equiv \bigotimes_{\substack{\b x \in \Mbb \\ i \in \{1, 2\} }} \qvec{A^i_{\b x}}
}
(cf.\ the definitions above \eqref{E action}) in which $A_{\b x}^i$ has no $\b k = 0$ mode and satisfies the spatial smoothness condition
\bel{\label{smooth constr fields}
  A^i_{\b x + \b e_j} = A^i_{\b x} + \hat \del_j A^i_{\b x} + O\left(k\_S^2/N^2 \right).
}

Unlike in a clock model, not all of these states can be included in the path integral. The extra constraint comes from the background charges $\rho_{\b x}$, or rather their low-momentum modes $\rho_{\b k}$ at $\b k \in \Pbb\_S \backslash \{0\}$, cf.\ \eqref{def rho k}. These modes can be assembled into a smooth and tame field
\bel{
  \rho(\b x) = \frac1N \sum_{\b k \in \Pbb\_S \backslash \{0\}} \rho_{\b k} \, \e^{\frac{2\pi\i}N \b k \b x} = \nabla E(\b x).
}
The continuum path integral for the basic noncompact theory is defined in a single eigensector of $\rho(\b x)$. The states $\qvec{\greek f; \greek s}$ should thus be approximate eigenstates of both $\widehat A^\square(\b x)$ and $\rho(\b x)$.

Since the ground state has $\rho_{\b x} = 0$, it is enough to focus on the states with $\rho(\b x) \approx 0$ but with nonzero eigenvalues of $\widehat A^\square(\b x)$. Such states can be expressed as equal superpositions of gauge-equivalent states obeying the smoothness constraint \eqref{smooth constr fields},
\bel{\label{tame ginv state}
  \frac1{\N(\Gamma^\perp\_S)} \sum_{\{ A \} \in \Gamma^\perp\_S} \qvec A.
}

\newpage

The normalization factor $\N(\Gamma^\perp\_S)$ in \eqref{tame ginv state} is equal to the square root of the volume of the gauge orbit $\Gamma\_S^\perp$, which is simply the number of tame and smooth configurations $A_{\b x}^i$ that are gauge-equivalent to each other. It can be estimated as a function of taming parameters in the following sense.  Not all gauge orbits have the same volume, but orbits whose magnetic fields $B(\b x)$ are much smaller than $A\_T$ --- the dominant configurations at low energies --- all have approximately the same volume. At each of the $(2k\_S)^2$ momenta $\b k \in \Pbb\_S$, there are approximately $n\_T K/E\_S$ different gauge transformations that can be applied. This implies that
\bel{
  \N(\Gamma\_S^\perp) \approx \left(\frac{n\_T K}{E\_S} \right)^{2k\_S^2}.
}

These states can also be recorded as sums over gauge transformations of the form
\bel{\label{tame ginv state 2}
  \qvec{A^\square} \equiv \frac1{\N(\Gamma\_S^\perp)} \sum_{\{\lambda\}} \qvec{A + \delta \lambda}.
}
The sum, to leading order in taming parameters, runs over all smooth functions $\lambda(\b x)$ whose derivatives are absolutely bounded by $2A\_T$. Here $A = \{A_{\b x}^i\}_{\b x, i}$ is any configuration on $\Gamma^\perp\_S$ whose vector potentials are very small, $|A^i_{\b x}| \ll A\_T$. Any particular choice of $A$ corresponds to a gauge-fixing. The state \eqref{tame ginv state 2} is clearly independent of this choice. One can think of $A$ here as encoding the gauge-invariant data, i.e.\ the eigenvalues of $\widehat A^\square(\b x)$ or $\widehat B(\b x)$. This is the reason for labeling the state with $A^\square$.

It is straightforward to explicitly verify that $\rho(\b x)$ annihilates each state $\qvec{A^\square}$,
\bel{
  \rho(\b x) \qvec{A^\square} \approx 0.
}
The key insight here is that the tameness of states allows the approximation
\bel{
  \qvec{A + \left[\delta^{(1)} (\delta^{\b x} - \delta^{\b x - \b e_1}) + \delta^{(2)} (\delta^{\b x} - \delta^{\b x - \b e_2}) \right] \d A} \approx \qvec A + \d A \left[\hat \del_{A_{\b x}^1} - \hat \del_{A_{\b x - \b e_1}^1} + \hat \del_{A_{\b x}^2} - \hat \del_{A_{\b x - \b e_2}^2} \right] \qvec A.
}

The continuum path integrals for the basic noncompact theory will be constructed by inserting states $\qvec{A^\square}$ at each time step. This discussion demonstrates that it takes some work to \emph{precisely} define the states used in the path integral. In particular, using the relatively simple states of form \eqref{tame ginv state 2} crucially hinges on the assumption that magnetic fields of order $A\_T$ are suppressed, so that the shape of the gauge orbit is the same (to leading order in taming parameters) for all configurations of interest.

One final subtlety must be noted. As in the clock model, the tame states used to construct the basic noncompact path integral do not include $\b k = 0$ modes. This is consistent with the fact that $\rho_{\b k = 0}$ is kinematically constrained to equal zero.

The basic noncompact partition function is thus given by the path integral
\bel{
  \Zf\_{bnc} \equiv \e^{-\beta \E(\greek s_0)} \sum_{\{A_\tau\} \in \Gamma\_S} \prod_{\tau = \d\tau}^\beta \qmat{A^\square_{\tau + \d \tau}}{\e^{-\d\tau H_0}}{A^\square_\tau}.
}
The Hamiltonian $H_0$ involves only photons and is given by \eqref{def H0}. The ground state sector has energy $\E(\greek s_0) = 0$ in this case. At each $\tau$, the sum runs over all smooth configurations $A_\tau \equiv \{A_{\b x, \tau}^i\}_{\b x, i}$ on a gauge slice $\Gamma\_S$. As discussed above, these correspond to different nonzero eigenvalues of $\widehat A^\square(\b x)$ for $\b x = \ell\_S \boldsymbol \xi$.

Inserting the definition \eqref{tame ginv state 2} into the above expression gives
\bel{
  \Zf\_{bnc} = \sum_{\{A_\tau\} \in \Gamma\_S} \sum_{\{\lambda_\tau, \, \lambda'_\tau\}}
  \prod_{\tau = \d\tau}^\beta \frac1{\N(\Gamma\_S^\perp)^2}
  \qmat{A_{\tau + \d \tau} + \delta\lambda'_{\tau + \d\tau}}{\e^{-\d\tau H_0}}{A_\tau + \delta \lambda_\tau}.
}
Since $H_0$ is gauge-invariant, the matrix element imposes $\lambda'_{\tau + \d\tau} = \lambda_\tau$. The sum over gauge transformations cancels against the normalization factor $\N(\Gamma\_S^\perp)^2$. Thus the partition function takes the simple form
\bel{
  \Zf\_{bnc} = \sum_{\{A_\tau\} \in \Gamma\_S}
  \prod_{\tau = \d\tau}^\beta
  \qmat{A_{\tau + \d \tau}}{\e^{-\d\tau H_0}}{A_\tau}.
}

This result means that all the gauge-variant data just disappear from consideration. No matter what configuration $A_\tau$ one chooses to represent the given gauge-invariant data at a given $\tau$, the transfer matrix elements will be the same. In other words, it is possible to just cut out the gauge-theoretic middleman and think of $\Zf\_{bnc}$ as representing the partition function of a basic noncompact \emph{scalar}, living on vertices of $\Mbb$ with position operators $\widehat A^\square_{\b x}$. The familiar procedure from the clock model analysis \cite{Radicevic:2D} then gives, with $\Sbb \equiv \{\d\tau, \ldots, \beta\}$ and $\alpha^2 \equiv \frac{g^2}{4\pi^2} \d\tau (2E\_S)^2$,
\bel{\label{def S bnc}
  \Zf\_{bnc}
   \approx
  \frac{1}{(2\pi \alpha^2)^{\frac{N_0}2(2k\_S)^2}}
  \sum_{\{A^\square_\tau \}}
    \e^{-S[A^\square\, ]},
  \quad
  S\big[A^\square \, \big]
   \equiv
  \frac1{2g^2}
  \sum_{\b x \in \Mbb} \sum_{\tau \in \Sbb}  \left[
    \big(\del_\tau A^\square\big)^2_{\b x, \tau} + \big(\del_i A^\square\big)^2_{\b x, \tau}
  \right]\d\tau.
}

There is nothing that distinguishes this action from a scalar one. One may thus say that photons and scalars are dual to each other in $d = 2$. (Of course, this already follows from the much more general analysis of duality in Subsection \ref{subsec KW}, where it was shown that other Maxwell degrees of freedom also have meaningful duals.) The result \eqref{def S bnc} further means that the familiar gauge theory action does \emph{not} naturally arise in the basic noncompact case. It only appears in continuum path integrals for the standard noncompact theory, after \emph{all} background charges are taken into account.

\subsection{The standard noncompact theory}

When constructing the continuum path integral for the standard noncompact Maxwell theory, tame electric backgrounds are not fixed but instead must be summed over. This immediately presents the path integrator with a choice: should \emph{all} eigenstates of $\widehat  \rho_{\b k}$ be included in the sum, or should the eigenstates of $\widehat \rho_{\b k}$ for $\b k \notin \Pbb\_S$ still be kept fixed? (It is only at this late stage that it becomes useful to distinguish $\widehat \rho_{\b k}$ from its eigenvalues.) This paper will take the latter approach. This means that the fixed sectors for this path integral are
\bel{
  \greek s = \left( \left\{(E\^{cl})^i_{\b x}\right\}_{\b x \in \Mbb}, \left\{ (A\^{cl})^i_{\b x}\right\}_{\b x \in \Mbb}, \left\{ \rho_{\b k}\right\}_{\b k \notin \Pbb\_S}, \left\{ n_{\b k}\right\}_{\b k \notin \Pbb\_S} \right).
}
As before, fixing a ground state sector $\greek s_0$ will correspond to setting all of these to zero.

With this choice, the states $\qvec{\greek f; \greek s_0}$ can now be arbitrary non-null eigenstates of $\widehat A^\square(\b x)$; they no longer need to be constrained to be specific eigenstates of $\widehat \rho(\b x)$. Recall that eigenstates of $\widehat A^\square(\b x)$ are the tame states $\qvec{A}$ from \eqref{tame fields}, subject to  \eqref{smooth constr fields}. To be explicit, for each $\b x = \ell\_S \boldsymbol \xi$ and $i$, these eigenstates are labeled by integer multiples of $\frac{2\pi}{2E\_S}$ in $-A\_T \leq A_{\b x}^i < A\_T$.

The states $\qvec{\greek f; \greek s_0}$ must also be eigenstates of $\widehat E_0^i$. There are thus roughly $(2n\_T)^{2(2k\_S + 1)^2}$ linearly independent states inserted at each time step. They have the form $\qvec{\greek f; \greek s_0} = \qvec A \qvec{E_0}$.

Another way to label the $\qvec{A}$ states is using eigenvalues of the momentum space operators $\widehat A_{\b k}^i$ for $\b k \in \Pbb\_S \backslash \{0\}$. (The states $\qvec A$ are already eigenstates of these operators; this would just be a relabeling.) In fact, it is even more convenient to use as labels the eigenvalues $(A_{\b k}^\square, \lambda_{\b k})$ of the operators $\widehat  A^\square_{\b k}$ from \eqref{def A tilde} and $\widehat A^\times_{\b k}$ from \eqref{Ak via A tilde},
\bel{\label{path int states decomp}
  \qvec{A}
   \equiv
  \left\qvec{ \left\{A^\square_{\b k}, \lambda_{\b k} \right\}_{\b k \in \Pbb\_S \backslash \{0\}} \right}
   =
  \qvec{ \{A^\square_{\b k}\}} \,
  \qvec{ \{ \lambda_{\b k}\}}.
}
Explicitly, the relations between the two kinds of labels are
\bel{\label{def A lambda}
  A_{\b k}^\square = \hat{\greek k}^1_{\b k} A^2_{\b k} - \hat{\greek k}^2_{\b k} A^1_{\b k}, \quad
  \lambda_{\b k} = \hat{\greek k}^1_{-\b k} A^1_{\b k} + \hat{\greek k}^2_{-\b k} A^2_{\b k}.
}
States with different $\{\lambda_{\b k}\}$ correspond to linearly independent states $\qvec{A + \delta \lambda}$ from  \eqref{tame ginv state 2}.

The tameness of states $\qvec{A}$ means that they can be rewritten as superpositions of target momentum eigenstates $\qvec E \equiv \bigotimes_{\b x, i} \qvec{E_{\b x}^i}$ with each $E_{\b x}^i$ bounded by $E\_S$,
\algns{
  \qvec{A }
  &= \frac1{(2E\_S)^{N^2}} \sum_{E_{\b x}^i = -E\_S}^{E\_S - 1} \e^{\i \sum_{\b x, i} E_{\b x}^i A_{\b x}^i} \qvec{E}
  \equiv \frac1{(2E\_S)^{N^2}} \sum_{\{E,\, \rho\}} \e^{\i \sum_{\b k \in \Pbb\_S \backslash \{0\}} \left[ \frac1{\omega_{\b k}} \rho_{\b k}\+ \lambda_{\b k} + E_{\b k}\+ A_{\b k}^\square \right]} \qvec{E}.
}
Note that the second line follows only if $A_{\b x}^i$ obeys the smoothness condition \eqref{smooth constr fields}. Here $E_{\b k}$ and $\rho_{\b k}$ label eigenvalues of operators $\widehat E^\square_{\b k}$ from \eqref{def E tilde} and $-\widehat \rho_{\b k}$ from \eqref{def rho k}, with $\b k \in \Pbb \backslash \{0\}$.

Now consider the building block of the path integral --- the matrix element
\bel{
  \qmat{A_{\tau + \d \tau}, E_{0, \tau + \d\tau}}{ \e^{-\d\tau H\_T}}{A_\tau, E_{0, \tau}}.
}
As usual, take $\d\tau$ to be much smaller than the inverse largest energy scale accessed by the tame Hamiltonian, so that
\bel{
  \d\tau (2k\_S)^2 \frac{n\_T}N \ll 1.
}
(This assumes that $g^2 \sim 1/N$, so that photons and background charges have comparable energies.) The exponential can then be expanded, evaluated, and then reexponentiated. Using \eqref{path int states decomp}, the matrix element decomposes as
\algns{\label{mat elem three factors}
  \qmat{A_{\tau + \d \tau}, E_{0, \tau + \d\tau}}{\e^{-\d\tau H\_T}}{A_\tau, E_{0, \tau}}
  &=
  \qmat{\{A^\square_{\b k, \tau + \d \tau}\}}{\e^{-\d \tau H_0}}{\{A^\square_{\b k, \tau}\}}  \\
  & \qquad \times \qmat{\{\lambda_{\b k, \tau + \d \tau}\}}{\e^{-\d \tau H_\rho}}{\{\lambda_{\b k, \tau}\}} \\
  &\qquad \times \qmat{E_{0, \tau + \d\tau}}{\e^{-\d\tau H_\Phi}}{E_{0, \tau}},
}
where
\bel{
  H_\rho \equiv \sum_{\b k \in \Pbb\_S \backslash \{0\}} \frac{g^2}{2\omega_{\b k}^2} \widehat\rho_{\b k}\+ \widehat\rho_{\b k}, \qquad
  H_\Phi \equiv \frac{g^2}2 \widehat E_0^i \widehat E_0^i.
}

The photon Hamiltonian $H_0$ gives the free scalar contribution to the Lagrangian, as described in \eqref{def S bnc}. The new nontrivial ingredient in the standard noncompact theory is the contribution from $H_\rho$. The Lagrangian associated with background charges is determined by matrix elements $\qmat{\lambda_{\tau + \d \tau}}{ H_\rho \d\tau} {\lambda_{\tau}}$, where $\qvec{\{\lambda_{\b k, \tau}\}}$ is written as $\qvec{\lambda_\tau}$ for simplicity. Using the target space Fourier transform, these matrix elements are
\bel{
   \frac{g^2 \d\tau}{2} \sum_{\b k \in \Pbb\_S \backslash \{0\} }
   \qmat{\lambda_{\tau + \d\tau} \big}{\frac{\widehat\rho_{\b k}\+ \widehat\rho_{\b k}}{\omega_{\b k}^2} \big} {\lambda_\tau}
   = \frac{g^2\d\tau}{2(2E\_S)^{(2k\_S)^2}}
   \sum_{\{\rho\}}
   \e^{-\i \sum_{\b l \in \Pbb\_S \backslash \{0\}} \frac1{\omega_{\b l}} \rho_{\b l}\+ \, \del_\tau \lambda_{\b l} \d\tau} \!\!\!
   \sum_{\b k \in \Pbb\_S \backslash \{0\} }
   \frac{\rho_{\b k}\+ \rho_{\b k}}{\omega^2_{\b k}}.
}
Note that $\qvec{\lambda_\tau}$ contains states at $\b k \notin \Pbb\_S$ momenta, but they merely change the normalization factor. This now means that the matrix element is a sum over all values of $\rho_{\b k}$ for $\b k \in \Pbb\_S \backslash \{0\}$,
\bel{\label{mat elem from bkd charges}
  \qmat{\lambda_{\tau + \d \tau}}{\e^{-\d\tau H_\rho}}{\lambda_\tau} =
  \frac1{(2E\_S)^{N^2}} \sum_{ \{\rho\} }
  \exp\left\{
    -\d\tau \!\!\! \sum_{\b k \in \Pbb\_S \backslash \{0\}}
    \left[
    \i \frac{\rho_{\b k}\+}{\omega_{\b k}} \del_\tau \lambda_{\b k} + \frac{g^2}2  \frac{\rho_{\b k}\+ \rho_{\b k}}{\omega^2_{\b k}} \right]
  \right\}.
}
The sum over each $\rho_{\b k}$ can be replaced by a Gaussian integral if, roughly, $n\_T \gg k\_S/\sqrt{\d\tau}$. Assuming this, the matrix element \eqref{mat elem from bkd charges} becomes, with $\alpha^2 \equiv \frac{g^2}{4\pi^2} \d\tau (2E\_S)^2$ as in \cite{Radicevic:2D},
\bel{
  \qmat{\lambda_{\tau + \d \tau}}{\e^{-\d\tau H_\rho}}{\lambda_\tau} \approx \frac{1}{(2\pi \alpha^2)^{(2k\_S)^2/2}} \e^{-\frac{\d\tau}{2g^2} \sum_{\b k \in \Pbb \backslash \{0\}} \del_\tau \lambda_{\b k}\+ \del_\tau \lambda_{\b k}}.
}

The third factor in the matrix element \eqref{mat elem three factors} does not contain local degrees of freedom. It is possible to define the states $\qvec{E_0}$ as superpositions of locally defined states analogous to $\qvec{A}$, but the resulting action would necessarily be nonlocal. For the time being, instead of looking for a nice Lagrangian for this term, it will simply be noted that summing over all eigenstates of $H_\Phi$ rescales the partition function by
\bel{
  \sum_{\b q} \e^{-\beta \frac{g^2}2 \frac{K^2}{N^2} \b q^2}.
}
Here the sum goes over all integers $q^i$ that parameterize the ``quantization'' of $E^i_0$ discussed in Subsections \ref{subsec standard nc Maxwell} and \ref{subsec KW}. This sum looks like it can be approximated by a Gaussian integral that gives a factor of $2\pi N^2/\beta g^2 K^2$ in front of the partition function. However, here it will be assumed that $\beta \E\_{electric} \gg 1$, cf.\ \eqref{scaling elect lines}, so that nontrivial flux eigenstates are so energetic that only the $E_0^i = 0$ state contributes to the partition function. In this case the entire Hamiltonian $H_\Phi$ can be ignored when computing the partition function.

Putting the photon and local background charge Lagrangians together --- and ignoring the nonlocal background charge contribution as explained in the previous passage --- gives the partition function
\bel{\label{def Zf snc}
  \Zf\_{snc} \approx
    \frac{1}{(2\pi \alpha^2)^{N_0 (2k\_S)^2}}
    \sum_{\{A^\square_{\b k, \tau},\, \lambda_{\b k, \tau}\}}
    \e^{- S[A^\square,\, \lambda]}
}
with
\algns{\label{def S snc}
  S[A^\square, \lambda]
  &\equiv
    \frac{\d\tau}{2g^2}
    \sum_{\b k \in \Pbb\_S \backslash\{0\}}
    \sum_{\tau \in \Sbb}
    \left[
      \del_\tau \lambda_{\b k, \tau}\+ \del_\tau \lambda_{\b k, \tau} +
      \del_\tau (A^\square_{\b k, \tau})\+ \del_\tau A^\square_{\b k, \tau} +
      \omega_{\b k}^2 (A^\square_{\b k, \tau})\+  A^\square_{\b k, \tau}
    \right] \\
  &=
    \frac{\d\tau}{2g^2}
    \sum_{\b x \in \Mbb}
    \sum_{\tau \in \Sbb}
    \left[
      (\del_\tau \lambda)^2_{\b x, \tau} +
      (\del_\tau A^\square)^2_{\b x, \tau} +
      (\del_i A^\square)^2_{\b x, \tau}
    \right].
}
Both $\lambda_{\b x, \tau}$ and $A^\square_{\b x, \tau}$ satisfy spatial smoothness constraints \eqref{smooth constr fields} and have no $\b k = 0$ modes. To precisely list the values of $\lambda_{\b x, \tau}$ and $A^\square_{\b x, \tau}$ included in the sum \eqref{def Zf snc}, one has to go back to their definition \eqref{def A lambda} in terms of the manifestly tame quantities $A_{\b x, \tau}^i$. In fact, doing so gives a more familiar form of the action $S[A^\square, \lambda]$, which will be denoted as
\bel{\label{def S snc 2}
  S[A^i] =
    \frac{\d\tau}{2g^2}
    \sum_{\b x \in \Mbb}
    \sum_{\tau \in \Sbb}
    \left[
      (\del_\tau A^i)^2_{\b x, \tau} +
      B^2_{\b x, \tau}
    \right],
}
where $B_{\b x, \tau} = \del_1 A^2_{\b x, \tau} - \del_2 A^1_{\b x, \tau}$ as defined below \eqref{def A tilde}.

NB: gauge parameters are decoupled, physical variables $\lambda_{\b x, \tau}$ in \eqref{def Zf snc}. Integrating them out is the same as setting $\del_\tau \rho_{\b k} = 0$ in \eqref{mat elem from bkd charges} and then summing over background charges.

The most ubiquitous form of the action for the Maxwell theory is
\bel{
  S[A^\mu] =
    \frac{\d\tau}{4g^2}
    \sum_{\b x \in \Mbb}
    \sum_{\tau \in \Sbb}
    (F^{\mu\nu})^2_{\b x, \tau},
}
where $0 \leq \mu, \nu \leq d$, the path integral sums over variables $A_{\b x, \tau}^\mu$ on spatial and temporal links alike, and the field strength is
\bel{
  F^{\mu \nu} \equiv \del_\mu A^\nu - \del_\nu A^\mu
}
with $\del_0 \equiv \del_\tau$. The action \eqref{def S snc 2} is related to this standard form by
\bel{
  S[A^i] = S[A^\mu] \Big|_{A^0 = 0}.
}
Setting $A^0_{\b x, \tau} = 0$ is often called ``fixing the temporal gauge.'' However, by the philosophy of this series, this is putting the cart before the horse: the Hamiltonian \eqref{def HT A E} and the associated action \eqref{def S snc 2} know nothing of $A_{\b x, \tau}^0$ or any putative gauge transformations that affect it.

It is nevertheless possible to \emph{derive} the notion of gauge transformations that act on $A^0$ by starting from the Hamiltonian framework (and benefiting from hindsight). The logical steps are as follows. To start, generalize the action \eqref{def S snc 2} to
\bel{\label{def S snc generalized}
  S[A^i, A^0] \equiv
    \frac{\d\tau}{2g^2}
    \sum_{\b x \in \Mbb}
    \sum_{\tau \in \Sbb}
    \left[
      (\del_0 A^i - \del_i A^0)^2_{\b x, \tau} +
      B^2_{\b x, \tau}
    \right],
}
with $A^0_{\b x, \tau}$ being a tame spatially smooth field with no $\b k = 0$ modes, just like the variables $A_{\b x, \tau}^i$. This action has precisely the same form as $S[A^\mu]$, but here $A^0_{\b x, \tau}$ is a fixed field.

Next, note that $S[A^i]$ is invariant under $A^i_{\b x, \tau} \mapsto A^i_{\b x, \tau} + \del_i \lambda_{\b x}$, as long as the new configuration does not exceed the tameness bounds. This is of course just the tame shift symmetry of the $\lambda$ degrees of freedom in \eqref{def S snc}. The generalized action $S[A^i, A^0]$ from \eqref{def S snc generalized} has an even larger set of approximate symmetries: the gauge parameters can be time-dependent as long as $A^0_{\b x, \tau}$ also transforms, so that $A^\mu_{\b x, \tau} \mapsto A^\mu_{\b x, \tau} + \del_\mu \lambda_{\b x, \tau}$. These time-dependent gauge transformations have \emph{no} canonical counterparts, unlike transformations with $\del_0 \lambda_{\b x, \tau} = 0$.

The actions of many configurations $\{A^\mu_{\b x, \tau}\}$ are thus equal to actions of some other configurations $\{\bar A^\mu_{\b x, \tau}\}$ in which $\bar A^0_{\b x, \tau} = 0$. The only configurations that cannot be brought to this form have nonvanishing zero-Matsubara-frequency modes, $\sum_{\tau = \d\tau}^\beta A^0_{\b x, \tau} \neq 0$. The upshot is that one can write
\bel{\label{def Zf snc 2}
  \Zf\_{snc} \approx
    \frac{\mathcal C}{(2\pi \alpha^2)^{N_0 (2k\_S)^2}}
    \sum_{\{A^\mu_{\b x, \tau}\}}
    \e^{- S[A^\mu]},
}
where $1/\mathcal C$ is the number of different time-dependent gauge transformations, and the sum excludes zero-frequency modes of all configurations $A_{\b x, \tau}^0$.

Compared to \eqref{def Zf snc}, the path integral \eqref{def Zf snc 2} treats space and time more equally, leading to a more symmetric action and making life easier for most practical intents and purposes. It also does a worse job approximating the correct result than the path integral \eqref{def Zf snc}, as it disregards the behavior of fields near the edges of the tame region. This may seem like a small price to pay for switching to \eqref{def Zf snc 2}. However, the \emph{actual} flip side of this Faustian bargain is that it obfuscates the original quantum degrees of freedom, mixing them up with noncanonical variables $A^0_{\b x, \tau}$ and leading to an infinite proliferation of different points of view on how to work with gauge theories.\footnote{See, for example, the large body of literature dedicated to defining entanglement entropy in gauge theories \cite{Buividovich:2008yv, Donnelly:2011hn, Casini:2013rba, Radicevic:2014kqa, Donnelly:2014gva, Ghosh:2015iwa, Aoki:2015bsa, Soni:2015yga, Lin:2018bud}.} The approach presented here tries to never lose sight of what the physical degrees of freedom really are.

The following cautionary points should be kept in mind whenever using the continuum path integral \eqref{def Zf snc 2}:
\begin{enumerate}
  \item A quantity of great interest in gauge theories is the Polyakov loop \cite{Polyakov:1978vu}
      \bel{
        P_{\b x} \equiv
        \sum_{\tau = \d\tau}^\beta
          \d\tau \, A^0_{\b x, \tau}.
      }
      The above analysis has shown that path integrals coming from the Hamiltonian must have $P_{\b x} = 0$ in all configurations. The path integral that includes $P_{\b x} \neq 0$ configurations does not necessarily approximate the thermal partition function of the starting theory \eqref{def H}. It remains possible that the universal parts of the two path integrals agree.
  \item The variables $A_{\b x, \tau}^0$ are by definition tame. The path integral \eqref{def Zf snc 2} does \emph{not} sum over taming backgrounds $(A\^{cl})^0_{\b x, \tau}$. Including them in the sum is possible, but as with the $P_{\b x} \neq 0$ configurations, this modification of the path integral no longer computes the thermodynamics of the original theory.
  \item Time-independent (zero-frequency) gauge parameters $\lambda_{\b x}$ are physical degrees of freedom that are summed over in \eqref{def Zf snc 2}, even though the zero-frequency parts of gauge fields $A_{\b x, \tau}^0$ are excluded from the sum. They can be expressed in terms of the variables $A^i_{\b x, \tau}$ using \eqref{def A lambda}. However, these configurations all have zero action, so summing over them merely provides yet another multiplicative prefactor in the partition function.
  \item The path integral \eqref{def Zf snc 2} is \emph{not} constrained to be in the $\rho(\b x) = 0$ sector at all times $\tau$, as it is in the basic noncompact theory. In fact, recall that it was necessary to sum over all background charges in \eqref{mat elem from bkd charges} in order to get the simple second-derivative action for the $\lambda$'s. It is thus important to recognize that the standard noncompact theory knows about states with all tame background charges --- not just gauge-invariant states.
\end{enumerate}

\subsection{Temporal smoothing and continuum fields} \label{subsec temp smoothing}

Introducing temporal gauge fields made the standard noncompact action closer to being (approximately) relativistic.  But there still exists one major discrepancy between spatial and temporal behavior: configurations $A_{\b x, \tau}^\mu$ are only smooth along spatial directions. This Subsection will define path integrals that involve only configurations $A^\mu(\b x, \tau)$ that are smooth in both temporal and spatial directions.

This \emph{temporal smoothing} entails changing the path integral \eqref{def Zf snc 2} in a way that has no canonical counterpart. The resulting path integral computes a quantity $\~\Zf\_{snc}$ that does not approximate the partition function $\Zf$ at any temperature. Nevertheless, experience with scalars and fermions suggests that the universal, $N_0$-independent quantities contained in $\~\Zf\_{snc}$ will agree with the corresponding terms in $\Zf\_{snc}$, provided that the right \emph{counterterms} are included in the action after the fields are temporally smoothed out \cite{Radicevic:1D, Radicevic:2D}. In this paper the focus will be on obtaining and studying temporally smooth actions; the study of counterterms is left to the future. The following formul\ae\ work in any dimension $d \geq 2$.

To precisely define temporal smoothing, start by Fourier-transforming the temporal dependence of all gauge fields in \eqref{def S snc generalized}. Define
\bel{
  A_{\b k, \tau}^\mu
    \equiv
  \frac1{\sqrt{N_0}}
  \sum_{n \in \Fbb}
    A_{\b k, n}^\mu \,
    \e^{\i \omega_n \tau},
  \quad
  \Fbb
    \equiv
  \left\{
    -\frac12N_0, \ldots, \frac12N_0 - 1
  \right\},
}
with Matsubara frequencies
\bel{
  \omega_n
    \equiv
  \frac{2\pi}\beta n.
}
Spatially smooth configurations involve only momenta $\b k \in \Pbb\_S \backslash \{0\}$. Temporally smooth configurations are analogously defined to involve only frequencies in
\bel{
  \Fbb\_S
    \equiv
  \{
    -n\_S, \ldots, n\_S
  \}.
}

Note that the temporal gauge fields $A^0_{\b k, \tau}$ are understood to always obey
\bel{
  A^0_{\b k, n = 0}  = 0.
}
Such a requirement is not imposed on the fields $A^i_{\b k, \tau}$. However, zero-frequency modes of these fields do not contribute to the action \eqref{def S snc generalized}. This means that restricting the path integral to configurations with
\bel{
  A^\mu_{\b k, n = 0} = 0
}
will result in an approximately correct answer for $\Zf$ if the factor $\mathcal C$ in \eqref{def Zf snc 2} is rescaled by the number of excluded $n = 0$ configurations, which is roughly $(2n\_T)^{d (2k\_S)^d}$.

With this convention for excluding zero-frequency modes, it is now reasonable to define spatially and temporally smooth configurations as
\bel{
  A^\mu(\b x, \tau)
    \equiv
  \frac1{\sqrt{N_0 N^d}}
  \sum_{\b k \in \Pbb\_S}
  \sum_{n \in \Fbb\_S}
    A_{\b k, n}^\mu \, \e^{\i \omega_n \tau}.
}
These obey the expected smoothness relations
\gathl{\label{smooth constr fields 2}
  A^\mu(\b x + \b e_i, \tau)
    =
  A^\mu(\b x, \tau) + \hat\del_i A^\mu(\b x, \tau)
  + O\left(k\_S^2/N^2\right),
  \\
  A^\mu(\b x, \tau + \d \tau)
    =
  A^\mu(\b x, \tau) + \d\tau\, \hat\del_0 A^\mu(\b x, \tau)
  + O\left(n\_S^2/N_0^2\right).
}
Note that the configurations $A_{\b x, \tau}^\mu$ appearing in \eqref{def S snc 2} or \eqref{def S snc generalized} should really have been denoted $A^\mu_\tau(\b x)$, as they were spatially but not temporally smooth.

In what follows, the spacetime coordinates $(\b x, \tau)$ will be assembled into a vector $x$ with components $\{x^\mu\}_{\mu = 0}^d$. The unit vectors $e_\nu$, each with components $\{e_\nu^\mu\}_{\mu = 0}^d$, are
\bel{
  e_0 \equiv (\b 0, \d \tau), \quad
  e_i \equiv (\b e_i, 0)
}
The vectors $x$ take values in the spacetime lattice
\bel{
  \Ebb \equiv \Mbb \times \Sbb,
}
and the smoothness relations \eqref{smooth constr fields 2} can be written as
\bel{\label{smooth constr fields 3}
  A^\mu(x + e_\nu) \approx A^\mu(x) + e_\nu^{\nu'} \hat\del_{\nu'} A^\mu(x).
}

The action \eqref{def S snc generalized}, projected to the space of temporally smooth fields, is simply
\bel{\label{def S tilde Maxwell}
  \~S[A^\mu]
    \equiv
  \frac1{4g^2}
  \sum_{x \in \Ebb}
    \d\tau F^{\mu\nu}(x) F^{\mu\nu}(x),
  \quad
  F^{\mu\nu}(x)
    \equiv
  \del_\mu A^\nu(x) - \del_\nu A^\mu(x).
}
Even though it is defined on the lattice, the smoothness constraint \eqref{smooth constr fields 3} ensures that $\~S[A^\mu]$ encodes the familiar continuum physics. Like in \cite{Radicevic:2D}, one can define the lattice spacing and continuum coordinates,
\bel{\label{def a}
  a \equiv \frac LN,
  \quad
  x\^c \equiv a x,
  \quad
  \d\tau\^c \equiv a \d\tau,
}
and then define continuum fields and couplings via the rescaling
\bel{
  A\_c^\mu(x\^c)
    \equiv
  a^{-\Delta\^c_A} A^\mu(x),
  \quad
  g\_c
    \equiv
  a^{-\Delta\^c_g} g.
}
The exponents $\Delta\^c_A$ and $\Delta\^c_g$ will be called \emph{engineering dimensions} of continuum fields.

These cosmetic changes lead to an action of the form
\bel{
  \~S[A^\mu\_c]
    \equiv
  a^{2(\Delta\^c_A - \Delta\^c_g) - d + 1}
  \frac1{4g\_c^2}
  \int_{\Ebb} \d^{d + 1} x\^c \,
    F\_c^{\mu\nu}(x\^c) F\_c^{\mu\nu}(x\^c).
}
The standard choice in $d = 2$, $\Delta\^c_A = 2\Delta\^c_g = 1$, gives the familiar continuum action that has no explicit $a$-dependence.

The above choice is ``standard'' because $\Delta\^c_A = 1$ should be the engineering dimension of a connection on a principal bundle. In other words, in a cQFT one expects that continuum gauge fields can always be combined with continuum derivatives $\del\^c_\mu \equiv (1/a) \del_\mu$ to give covariant derivatives $D\^c_\mu = \del\^c_\mu + A^\mu\_c$ that have a definite engineering dimension.

The reader who has started absorbing the philosophy of this paper will recognize that this is not an acceptable way to assign dimensions. Indeed, the essential point of this series is to let the microscopic theory define the cQFT instead of trying to fit it to a preconceived continuum notion.

Instead of thinking about connections and bundles, recall that the analysis of the standard noncompact theory in Subsection \ref{subsec standard nc Maxwell} has shown that $g^2 \sim 1/N$ is the parameter regime in which the low-energy states are naturally tame. This in turn makes it natural to define the coupling
\bel{
  g\_c \equiv a^{-1/2} g
}
that satisfies $g_c^2 L = O(N^0)$. Thus the standard choice of engineering dimensions should be understood to follow from \eqref{scaling of g} and from asking that the action be independent of explicit factors of $a$ in continuum notation. These two requirements \emph{imply} that the continuum gauge field $A\_c^\mu(x)$ has the engineering dimension that makes it interpretable as a connection on some bundle. It is conceivable that, in more exotic cQFTs, gauge fields can still exist without satisfying $\Delta\^c_A = 1$.

An important fact now is that the fields $A^\mu(x)$ (or their trivial rescalings $A\_c^\mu(x)$) themselves do \emph{not} have well defined \emph{scaling dimensions}. In other words, they are continuum fields, but they are not \emph{scaling fields}. This means, for example, that the equal-time correlation function $\avg{A^i(\b x)  A^j(\b y)}$ does not evaluate to $\delta^{ij}/|\b x - \b y|^{2\Delta\^c_A}$. Indeed, this correlator is not even rotationally invariant. Nevertheless, the engineering dimensions $\Delta\^c_A$ remain well defined. This should serve as a reminder that engineering and scaling dimensions do not need to be the same, even in simple examples such as this one.

It is also instructive to consider the field strength $F^{12}(x) \equiv B(x)$. By \eqref{OPE B}, $\frac{\i\sqrt {4\pi}} g B(x)$ is the field with a canonically normalized two-point function. Thus this is a scaling field with scaling dimension $\Delta_{\i\sqrt {4\pi}B/g} = 3/2$. The associated continuum field $\frac{\i\sqrt {4\pi}}{g\_c} B\_c(x\^c)$ has engineering dimension $\Delta_{\i\sqrt {4\pi}B/g}\^c = \Delta\^c_B - \Delta\^c_g = 3/2$. The agreement between dimensions means that the field has no anomalous dimension.

\newpage

\section{The Higgs model and BF theory} \label{sec Higgs}

\subsection{The Higgs model and its phase structure} \label{subsec Higgs phases}

The analysis of Maxwell theory in Sections \ref{sec Maxwell} and \ref{sec Maxwell path int} heavily featured background charges. They were treated as physical, albeit time-independent, degrees of freedom --- on par with photons in terms of meaningfulness. Some authors refer to this setup as Maxwell theory coupled to matter, relegating the term ``pure Maxwell theory'' to the $\varrho_v = 0$ sector alone. This is fair, but a Maxwell-matter system must satisfy two conditions in order to precisely correspond to the pure Maxwell theory with background charges that was studied so far:
\begin{enumerate}
  \item The matter must be classical, i.e.\ the Hamiltonian should only act on the matter Hilbert space via operators built out of gauge and matter clock operators.
  \item The entire system must be subject to a gauge constraint that relates the matter charge density with the divergence of the electric fields.
\end{enumerate}

Relaxing these requirements gives further interesting generalizations of the pure Maxwell theory. Without the first requirement, the matter becomes dynamical and the Maxwell-matter system becomes a full-fledged interacting theory. Without the second requirement, it once again becomes possible to talk about different superselection sectors labeled by background charges $\varrho_v$ --- and this time they exist \emph{alongside} bona-fide matter degrees of freedom.

The purpose of this Section is to study the continuum theories that may arise from various limits of a particularly natural theory of $\Z_K$ gauge fields and matter in which the above requirements do not hold. For brevity, this theory will be referred to as the \emph{Higgs model}, though other names (e.g.\ St\"uckelberg model or affine Higgs model) might be more appropriate. Its microscopic Hamiltonian is
\gathl{\label{def H Higgs}
  H
    =
  \frac{r^2}2
  \sum_{\ell \in \Mbb}
    \left(2 - Z_\ell^{-q} \textstyle \prod_{v \in \del \ell} Z_v -
          Z_\ell^{q} \textstyle \prod_{v \in \del \ell} Z_v\+ \right)
  + \frac{1}{2r^2(\d A)^2}
  \sum_{v \in \Mbb}
    \left(2 - X_v - X_v\+ \right)
  \\
  + \frac{g^2}{2(\d A)^2}
  \sum_{\ell \in \Mbb}
    \left(2 - X_\ell - X_\ell\+ \right)
  + \frac1{2g^2}
  \sum_{f \in \Mbb}
    \left(2 - W_f - W_f\+ \right).
}
Gauge theory clock/shift operators live on links and are denoted by $Z_\ell$ and $X_\ell$, with $W_f = \prod_{\ell \in \del f} Z_\ell$ as before. Matter clock/shift operators live on sites and are denoted by $Z_v$ and $X_v$. The theory has three parameters: the gauge coupling $g$, the matter coupling $r$ (sometimes called the Higgs field radius), and the matter charge $q$. When $q = 0 \, \trm{mod}\, K$, the gauge and matter sectors are decoupled. It is possible to define this model for any $q \in \R$, but here it will be assumed that the charge is an integer satisfying $1 \leq q \leq K$.

Just like the pure Maxwell theory, the Higgs model \eqref{def H Higgs} has local symmetries. They are generated by generalized Gauss operators
\bel{\label{def gen G}
  \G_v \equiv G_v X_v^{-q} = X_v^{-q} \prod_{\ell \in \del_{-1}v} X_\ell.
}
These local symmetries exist for any integer $q$, but when $q$ divides $K$ it immediately follows that the $\Z_q$ generators
\bel{
  \G_v^{K/q} = G_v^{K/q}
}
are also local symmetries.\footnote{If $K/q$ is not an integer, one must be careful about branch cuts when raising $\G_v$ to this power. In this paper it will always be assumed that $K/q \in \Z$.} This indicates that, in some sense, there is a pure $\Z_q$ gauge theory residing within the Higgs model at charge $q > 1$. This gauge theory will be seen to emerge as the low-energy description of the system in a particular corner of parameter space.

The superselection sectors corresponding to symmetries $\G_v$ are labeled by their eigenvalues $\e^{\i \varrho_v \d A}$, where $\varrho_v \in \{0, 1, \ldots, K - 1\}$ are the background charges. There are $K^{N^2}/q$ different sectors. (Note that here it is already assumed that $q$ is an integer divisor of $K$.) When $q = K$, the number of sectors is $K^{N^2 - 1}$, just as in pure Maxwell theory.

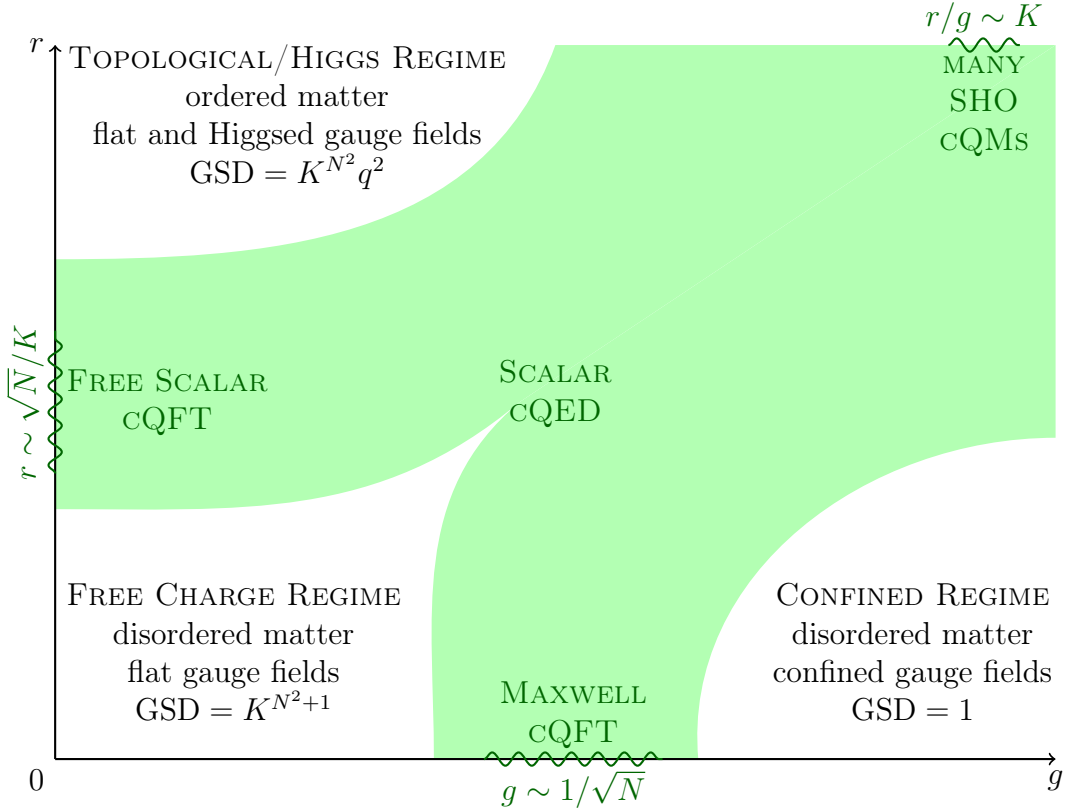
\begin{figure}
\begin{center}
\begin{tikzpicture}[scale = 0.95]
  \contourlength{1pt}

  \fill [green!30]
    (0, 3.5)
    to[out = 0, in = 220] (6.5, 5)
    to (14, 10)
    to (7, 10)
    to[out = 250, in = 0] (0, 7)
    to cycle;

  \fill [green!30]
    (5.3, 0)
    to[out = 90, in = 220] (6.5, 5)
    to (14, 10)
    to (14, 4.5)
    to[out = 180, in = 95] (9, 0)
    to cycle;

  \draw (2.5, 1.5) node[align = center]
    {\textsc{Free Charge Regime}\\
     disordered matter\\
     flat gauge fields\\
     $\trm{GSD} = K^{N^2 + 1}$
    };

  \draw (12, 1.5) node[align = center]
    {\textsc{Confined Regime} \\
     disordered matter\\
     confined gauge fields\\
     $\trm{GSD} = 1$
    };

  \draw (3.25, 9) node[align = center]
    {\textsc{Topological/Higgs Regime}\\
     ordered matter\\
     flat and Higgsed gauge fields\\
     $\trm{GSD} = K^{N^2} q^{2}$
    };

%

  \draw[->, thick] (0, 0) -- (14, 0);
  \draw (14, 0) node[below] {$g$};
  \draw[->, thick] (0, 0) -- (0, 10);
  \draw (0, 10) node[left] {$r$};
  \draw (0, 0) node[below left] {0};

  \draw[green!40!black, thick,
        style = {decorate, decoration=snake}]
    (6, 0) -- node[midway, below] {$g \sim 1/\sqrt N$}
    node[midway, above, align = center] {\textsc{Maxwell}\\ \textsc{cQFT}} ++(2.5, 0);

  \draw[green!40!black, thick,
        style = {decorate, decoration=snake}]
    (0, 4) -- node[midway, above, rotate = 90] {$r \sim \sqrt{N}/K$}
    node[midway, right, align = center] {\textsc{Free Scalar}\\ \textsc{cQFT}} ++(0, 2);

  \draw[green!40!black]
    node[above, align = center] at (7, 4.5) {\textsc{Scalar}\\ \textsc{cQED}};

  \draw[green!40!black, thick,
        style = {decorate, decoration=snake}]
    (12.5, 10) -- node[midway, above] {$r/g \sim K$}
    node[midway, below, align = center] {\textsc{many}\\ \textsc{SHO}\\ \textsc{cQMs}} ++(1, 0);

\end{tikzpicture}
\end{center}
\caption{\small A not-to-scale sketch of the parameter space of the Higgs model at fixed $q$. The unshaded areas represent three extremal regimes in which the Hamiltonian \eqref{def H Higgs} is approximately classical, i.e.\ consists of commuting operators. Here the excited states have large gaps set by the very large or very small couplings $r$ and $g$. (See the main text for descriptions of these regimes, and keep in mind that the ground state degeneracies quoted here include \emph{all} $K^{N^2}/q$ background charge sectors.) The shaded areas represent crossovers between these regimes. This is where cQFTs may emerge at low energies. The widths of these crossovers depend on $K$ and $N$. A phase transition is found whenever a crossover width goes to zero when $N$ is taken to be very large.\\
\hspace*{1ex} For $q = 1$, only the crossovers bordering the free charge regime become phase transitions as $N \gg 1$; the upper right corner remains a smooth crossover, and the topological/Higgs and confined regimes are the same phase \cite{Fradkin:1978dv, Banks:1979fi}.  For $q > 1$, a line of phase transitions extends all the way into the upper right corner.\\
\hspace*{1ex} When $K \gg 1$, the pure Maxwell cQFT studied in Sections \ref{sec Maxwell} and \ref{sec Maxwell path int} emerges as the low-energy description of the gauge fields along the crossover on the $r \rar 0$ line. The matter is decoupled from the gauge fields on this line, and remains disordered throughout the crossover. If the $g$-axis were drawn on a linear scale, the free charge regime and the vanishingly thin crossover during which the cQFT arises at $N \gg 1$ would both be concentrated in the lower left corner of the plot.\\
\hspace*{1ex} Similarly, at $K \gg 1$ and $r \sim \sqrt N/K$, the free scalar cQFT will emerge along the $g \rar 0$ line. In the interior, a plausible conjecture is that a scalar cQED (a cQFT with tame scalars and gauge fields) emerges at $r \sim g \sim 1/\sqrt N$. This gapped continuum theory will be studied in Subsection \ref{subsec noncomp scalar QED}.\\
\hspace*{1ex} Finally, the upper right corner ($r, g \rar \infty$) does not have a pithy description. The physics there is governed by the ratio $\gamma \equiv g/r$. At $\gamma \sim \d A$, many decoupled copies of the harmonic oscillator cQM \cite{Radicevic:1D} emerge at low energies.
}
\label{fig phases}
\end{figure}

The phase structure of this model is a rich and venerable subject \cite{Fradkin:1978dv, Banks:1979fi, Kitaev:1997wr}. Fig.\ \ref{fig phases} shows the different regimes accessed by the theory as $r, g \in \R^+$ are dialed. The main goal of this Section is to describe and  study the putative cQFT that arises in the middle of this parameter space. Before taking on this task, however, it is instructive to first explore the edges of the parameter space.

\begin{itemize}
  \item $r \rar 0$: \quad When $r$ is much smaller than any other parameter in the theory, the operators coupling matter and gauge fields in \eqref{def H Higgs} can be ignored. The energy eigenstates are thus also eigenstates of matter shift operators $X_v$. The matter part of the ground state is the unique, ``disordered'' state with $X_v = \1$ (or $p_v = 0$) on each site.

      \hspace*{1ex} The gauge theory Hamiltonian is the pure Maxwell one. At $g \rar \infty$, the ground state for gauge fields is confined, with $X_\ell = \1$ on each link. At $g \rar 0$, the gauge sector ground states are the $K^{N^2 + 1}$ states that obey the flatness constraints $W_f = \1$ on each face. Imposing $\varrho_v = 0$ reduces this degeneracy to $Kq$. At $q = K$, this is the topological ground state degeneracy of a pure $\Z_K$ gauge theory (``toric code'') on a torus \cite{Kitaev:1997wr}. In between these extremes, at $g \sim 1/\sqrt N$, the Maxwell cQFT emerges.

      \hspace*{1ex} The above limits on $g$ are understood to be taken after $r \rar 0$. For example, this means that $r \ll g$, no matter how small $g$ is in the topological phase. However, since the constraint $X_v = \1$ commutes with both $W_f = \1$ and $X_\ell = \1$, the order of limits does not actually matter.
  \item $r \rar \infty$, $g \rar 0$: \quad When $r$ is much bigger than other parameters, the matter shift operators can be ignored. The energy eigenstates are then necessarily eigenstates of matter clock operators $Z_v$.

      \hspace*{1ex} When $g \rar 0$, the gauge field shift operators can also be ignored. The Hamiltonian is again a sum of commuting operators. This indicates that the limits $g \rar 0$ and $r \rar \infty$ commute. The gauge fields must again be flat, $W_f = \1$, in all ground states. They must also satisfy the \emph{Higgs relation}
      \bel{
        \prod_{v \in \ell} Z_v = Z_\ell^q
      }
      on each link. In terms of eigenvalues $\e^{\i A_\ell}$ and $\e^{\i \phi_v}$ of clock operators, this relation is
      \bel{\label{Higgs relation}
        (\delta \phi)_\ell - q A_\ell = 0 \ \trm{mod}\ 2\pi.
      }
      A gauge field configuration obeying this relation --- i.e.\ largely being determined by a function $\phi_v$ on sites --- is said to be \emph{Higgsed}.

      \hspace*{1ex} The word ``largely'' in the previous sentence appears because, at $q > 1$, \eqref{Higgs relation} does not fully determine $A_\ell$ in terms of $\phi_v$. For any $A_\ell$ that satisfies the Higgs relation, $A_\ell + \frac{2\pi}q \bar n_\ell$ satisfies it as well, for any $0 \leq \bar n_\ell < q$. Thus it makes sense to define
      \bel{
        A_\ell
          \equiv
        \bar A_\ell + a_\ell,
        \quad
        \bar A_\ell
          \equiv
        \frac{2\pi}q \bar n_\ell,
        \quad
        a_\ell \in
          \left\{ -\frac \pi q, -\frac \pi q + \d A, \ldots, \frac \pi q - \d A \right\}.
      }
      The precise statement is then that Higgsing fixes the value of $a_\ell$ while leaving the $\Z_q$ gauge field $\bar A_\ell$ arbitrary. When $q = 1$, Higgsing fixes the value of the entire original gauge field $A_\ell$ in terms of a scalar field $\phi_v$. Note that this paragraph crucially relies on the assumption that $K/q \in \Z$.

      \hspace*{1ex} Not all matter configurations $\phi_v$ can satisfy the Higgs relation. The ones that do are of the form, for $0 \leq \bar n_v < K/q$,
      \bel{\label{Higgs relation scalars}
        \phi_v
          =
        \bar\phi_v + \vartheta,
        \quad
        \bar\phi_v
          \equiv
        q \bar n_v \, \d A,
        \quad
        \vartheta \in
          \left\{ - \frac q2 \d A, \left(-\frac q2 + 1\right) \d A, \ldots, \left( \frac q2 - 1 \right) \d A  \right\}.
      }

      To find the ground state degeneracy, count the states that satisfy the Higgs relation and the flatness constraint. Higgsing expresses $a_\ell$ in terms of $\bar\phi_v$ and ensures that these fields obey $(\delta a)_f = 0$. This leaves $q^{2N^2}$ gauge degrees of freedom $\bar A_\ell$ subject to $N^2 - 1$ independent flatness conditions, for a total of $q^{N^2 + 1}$ allowed gauge configurations. Meanwhile, there are $q (K/q)^{N^2}$ scalar configurations satisfying \eqref{Higgs relation scalars}. There are thus $K^{N^2} q^2$ ground states. Restricting to $\varrho_v = 0$ leaves $q^3$ ground states: $q^2$ from electric fluxes along the torus in the $\Z_q$ gauge theory, and $q$ from the allowed values of $\vartheta$.
  \item $r, g \rar \infty$: \quad In this case the remaining terms in the Hamiltonian are
      \bel{\label{large gr H}
        H
          \approx
        \frac{r^2}2
        \sum_{\ell \in \Mbb}
          \left(2 - Z_\ell^{-q}\textstyle \prod_{v \in \del \ell} Z_v -
          Z_\ell^q \textstyle \prod_{v \in \del \ell} Z_v\+ \right)
        + \frac{g^2}{2(\d A)^2}
        \sum_{\ell \in \Mbb}
          \left(2 - X_\ell - X_\ell\+ \right).
      }
      These terms do not commute with each other, so it matters whether $r$ or $g$ is taken to infinity first. In other words, the physics depends on the ratio $\gamma \equiv g/r$.

      \hspace*{1ex} Even though the Hamiltonian is not a sum of commuting operators, it is still tractable. In particular, all of its eigenstates are still eigenstates of the matter clock operators, and the gauge fields on different links are not coupled to each other. Thus for each choice of $\phi_v$ one gets $2N^2$ independent quantum mechanics (QM) theories. On each link there is a version of a simple harmonic oscillator governed by
      \bel{
        H(q, \delta \phi)
          \approx
        \frac{1}2
        \left(2 - \e^{\i \delta \phi} Z^{-q}  -
          \e^{-\i \delta \phi} Z^q \right)
        + \frac{\gamma^2}{2(\d A)^2} \left(2 - X - X\+ \right).
      }

      \hspace*{1ex} When $q = O(K^0)$, it is reasonable to assume that there exists a range of parameters $\gamma$ for which the low-energy regime of $H(q, \delta \phi)$ is described by a continuum QM (cQM) of the simple harmonic oscillator. In the first part of this series it was argued that, at $\delta \phi = 0$ and $q = 1$, this happens for $\gamma \sim \d A$ \cite{Radicevic:1D}. In this case the low-energy states are tame, but there are two taming backgrounds: one corresponding to $\phi\^{cl} = 0$ and $p\^{cl} = 0$, and another corresponding to $\phi\^{cl} = \pi$ and $p\^{cl} = K/2$. For general $q = O(K^0)$ one can expect $2q$ taming backgrounds. Meanwhile, setting $\delta\phi \neq 0$ corresponds to turning on a nontrivial $\theta$-term in this QM. If $\delta \phi \in q \, \d A \, \Z$, this merely shifts the taming backgrounds and does not affect the ground state degeneracy. However, if $\delta \phi$ is not an integer multiple of $q \, \d A$, one should expect a unique ground state at an energy higher than the energies of $\delta \phi \in q \, \d A \, \Z$ ground states. Thus, as in the $g \rar 0$ case, the ground states of the Hamiltonian \eqref{large gr H} are characterized by configurations $\phi_v$ satisfying \eqref{Higgs relation scalars}.

      \hspace*{1ex} Putting all this together, the ground state degeneracy in this corner of parameter space is
      \bel{
        q \left(K/q\right)^{N^2} (2q)^{2N^2} = (4K)^{N^2} q^{N^2 + 1}.
      }
      Upon restricting to the $\varrho_v = 0$ sector, the degeneracy can be recorded as
      \bel{
        \trm{GSD} = 2^{2N^2} q^{N^2 + 2}.
      }
      Note that the power of two counts the number of independent \emph{spin structures} in the target space of the cQM theory on each link.

      \hspace*{1ex} This result is simple but profound. In this parametric limit, the nominally $d = 2$ theory \eqref{def H} has a continuum limit described by many decoupled $d = 0$ theories.

  \item $g \rar 0$: \quad This case was mostly covered above, but there is one regime which the $r \rar 0$ and $r \rar \infty$ analyses did not touch upon --- the emergence of the free scalar cQFT. When the gauge coupling is small, the gauge fields are flat and background charges cost no energy. In this regime it is convenient to avoid imposing $\varrho_v = 0$ and to simply freeze the gauge fields into clock eigenstates $A_\ell$ satisfying $(\delta A)_f = 0$. Then the matter theory is simply a clock model in a particular classical vector potential $A_\ell$,
      \bel{
        H[A] \approx
        \frac{r^2}2
        \sum_{\ell \in \Mbb}
          \left(2
            - \e^{-\i q A_\ell} \textstyle \prod_{v \in \del \ell} Z_v
            - \e^{\i q A_\ell} \textstyle \prod_{v \in \del \ell}Z_v\+
          \right)
        + \frac{1}{2r^2(\d A)^2}
        \sum_{v \in \Mbb}
          \left(2 - X_v - X_v\+ \right).
      }

      \hspace*{1ex} It is reasonable to expect this theory to have a low-energy cQFT description at some value of $r$ between the ordered and disordered regimes. Instead of generalizing the analysis of \cite{Radicevic:2D} to $d = 2$ in great detail, it is expedient to make use of duality tools developed in Subsection \ref{subsec KW}. Consider the twist of the ordinary KW duality \eqref{KW basic} given by
      \bel{
        \zeta^{-q}_\ell \prod_{v \in \del \ell} Z_v = X^\vee_\ell, \qquad X_v = \zeta_v^\vee W_v^\vee.
      }
      The dual is a pure Maxwell theory on the lattice $\Mbb^\vee$, in the presence of a classical two-form gauge field $\zeta_v^\vee$. Though not necessary in what follows, note that consistency requires the singlet constraints
      \bel{
        \omega_c^{-q} \equiv \prod_{\ell \in c} \zeta^{-q}_\ell = T^\vee_c, \qquad \prod_{v \in \Mbb} X_v = \prod_{v \in \Mbb} \zeta^\vee_v \equiv \omega^\vee.
      }
      The flatness of the background field $\zeta_\ell$ implies that the dual gauge theory satisfies the ordinary gauge constraint $\varrho_v = 0$, or $G_f^\vee \equiv \prod_{\ell \in \del f} X_\ell^\vee = \1$.

      \hspace*{1ex} The coupling of the dual Maxwell theory is $r^\vee = \d A/r$. As long as the background fields $\zeta_v^\vee$ vary slowly enough, the analysis of Section \ref{sec Maxwell} maintains that there will be an emergent cQFT at
      \bel{
        r^\vee \sim \frac1{\sqrt N}.
      }
      This means that the Hamiltonian $H[A]$ has an emergent cQFT at
      \bel{\label{scaling of r}
        r \sim \frac{\sqrt N} K.
      }
      The assumption that $\zeta_v^\vee$ varies slowly is consistent with the assumption that only slowly varying taming backgrounds $(B\^{cl})^\vee_v$ are important at low energies. The remaining details of the duality can be filled in by using Table \ref{table KW}. The takeaway lesson here is that the $g \rar 0$ edge of parameter space hosts a cQFT at Higgs radii that satisfy \eqref{scaling of r}.
\end{itemize}

\subsection{Noncompact scalar cQED} \label{subsec noncomp scalar QED}

It is now time to gingerly proceed into the interior of parameter space. The most direct way to start would be to perform perturbation theory around the known ground states at small or large $r$ and $g$. These kinds of calculations, for instance, give information on the curvature of the crossover regions depicted on Fig.\ \ref{fig phases} \cite{Fradkin:1978dv}. Instead of attempting perturbation theory from a gapped phase, here it will be \emph{assumed} that there exists a region somewhere in the interior of parameter space in which all low-energy states are tame w.r.t.\ some taming background(s). This approach is fully analogous to the way the Maxwell cQFT was tackled in Section \ref{sec Maxwell}.

Assume for the moment that all taming backgrounds are trivial. Taming the Hamiltonian \eqref{def H Higgs} now gives a familiar theory: scalar QED with a St\"uckelberg coupling,
\bel{\label{def HT A E pi phi}
  H\_T
    \approx
  \frac{g^2}{2}
  \sum_{\b x, \, i} (E^i_{\b x})^2
    +
  \frac1{4g^2}
  \sum_{\b x, \, i, \, j} (B_{\b x}^{ij})^2
    +
  \frac{r^2}2
  \sum_{\b x, \, i} \left[(\del_i \varphi)_{\b x} - q A_{\b x}^i \right]^2
    +
  \frac1{2r^2}
  \sum_{\b x} \pi_{\b x}^2.
}
Note that under the conventions of \eqref{def lambda}, gauge transformations generated by tamed operators $(\G_{\b x})\_T$ are
\bel{
  \varphi_{\b x} \mapsto \varphi_{\b x} + q \lambda_{\b x}, \quad
  A^i_{\b x} \mapsto A^i_{\b x} + (\del_i \lambda)_{\b x},
}
with the usual caveats about how these simple maps receive corrections in states near the edges of the tame subspace.

The couplings are assumed to satisfy $g^2 \sim 1/N$ and $r^2 \sim N/K^2$. With this choice, by \eqref{scaling of g} and \eqref{scaling of r}, at $q = 0$ this just becomes a decoupled pair of known tame theories.
The main question now is whether the low-energy spectrum of \eqref{def HT A E pi phi} remains tame when $q \neq 0$.

This question can be answered because the tame Hamiltonian is quadratic, and so it can in fact be (approximately) diagonalized. The answer is that the tameness assumption indeed remains self-consistent, but the corresponding cQFT no longer has a linear spectrum with gaps $\E\_{photon} \sim 1/N$. Instead, the dispersions at low energies will be of the form
\bel{
  \sqrt{\omega_{\b k}^2 + (qgr)^2}
  \sim \sqrt{\frac 1{N^2} + \frac{q^2}{K^2}}.
}
When $N \sim K$, a theory with this dispersion can be understood as a free cQFT with a finite mass gap. This is, of course, the linchpin of the Higgs mechanism.

The derivation of this result is a straightforward repetition of the steps taken in Subsections \ref{subsec basic nc Maxwell} and \ref{subsec standard nc Maxwell}. To start, express $H\_T$ in momentum space using $A^\square_{\b k}$ and $A^\times_{\b k}$ from \eqref{Ak via A tilde},
\bel{\label{def HT A tilde E pi phi}
  H\_T
    \approx
  \sum_{\b k \in \Pbb}
  \bigg[
    \frac{g^2}{2}
    (E^i_{\b k})\+ E^i_{\b k}
      +
    \frac{\omega^2_{\b k} + (q g r)^2}{2g^2}
    (A^\square_{\b k})\+ A^\square_{\b k}
      +
    \frac{r^2}2
    \big( \omega_{\b k} \varphi_{\b k}\+ - q (A^\times_{\b k})\+ \big)
    \big( \omega_{\b k} \varphi_{\b k} - q A^\times_{\b k} \big)
      +
    \frac1{2r^2}
    \pi_{\b k}\+ \pi_{\b k}
  \bigg].
}

Before proceeding, note that the operator $(\nabla E)_{\b k}$, the Fourier transform of $(\nabla E)_{\b x}$ from \eqref{def nabla E}, was denoted by $\rho_{\b x}$ in the previous Sections. The operator $\rho_{\b x}$ also represented the taming of the the background charge density $\varrho_{\b x}$. In this Section, due to the presence of matter in the Gauss operators $\G_{\b x}$, $(\nabla E)_{\b x}$ is no longer identifiable with the taming of $\varrho_{\b x}$. This taming will still be denoted by $\rho_{\b x}$, and its Fourier transform will be $\rho_{\b k}$. The relation between electric fields and $\rho_{\b k}$, i.e.\ the tame Gauss law, is
\bel{
  \rho_{\b k} = -q \pi_{\b k} + (\nabla E)_{\b k} = - q \pi_{\b k} - \greek k^i_{- \b k} E^i_{\b k},
}
where $\b k \neq 0$, and $\greek k^i_{\b k}$ is defined in \eqref{def kappa}. This generalizes the previous definition \eqref{def rho k}. As appropriate for a gauge-invariant Hamiltonian, $\rho_{\b k}$ commutes with $H\_T$ to leading order in the taming parameters.

The electric fields in $H\_T$ can be written as
\bel{
  (E^i_{\b k})\+ E_{\b k}^i
    =
  E^\square_{\b k} E^\square_{\b k}
    +
  \frac1{\omega_{\b k}^2} (\rho_{\b k} + q \pi_{\b k} )\+ (\rho_{\b k} + q \pi_{\b k}).
}
The field $E^\square_{\b k}$ is canonically conjugate to $A^\square_{\b k}$. This sector of the pure Maxwell theory corresponded to a free, massless, basic noncompact scalar field. In the case at hand, these operators form a free, massive, basic noncompact scalar theory with Hamiltonian
\bel{\label{def HT (1)}
  H\_T^{(1)}
    =
  \sum_{\b k \in \Pbb \backslash \{0\} }
  \left[
    \frac{g^2}2 (E^\square_{\b k})\+ E^\square_{\b k}
    + \frac{\omega_{\b k}^2 + (q g r)^2}{2g^2} (A^\square_{\b k})\+ A^\square_{\b k}
  \right].
}
Just as in the massless theory, though, this is a collection of decoupled oscillators, one per momentum $\b k$, and as such it is trivial to smoothe and recast as a cQFT. The ground state is the same as in the massless case. On the other hand, the dispersion is
\bel{\label{def omega tilde}
  \~\omega_{\b k} = \sqrt{\omega_{\b k}^2 + (q g r)^2}.
}
The various OPEs, e.g.\ \eqref{OPE A tilde}, become qualitatively different from the massless ones whenever
\bel{
  q g r \sim \frac 1 K \gtrsim \frac1 N.
}
Note, in particular, that if $N \ll K$ the theory described by $H^{(1)}\_T$ becomes the usual massless and free basic noncompact scalar. However, if $N \sim K$, the dispersion can be written as $\~\omega_{\b k} = \sqrt{\omega^2_{\b k} + m^2}$ for $m \sim 1/N$, and the OPE is
\bel{\label{OPE A tilde massive}
  A^\square_{\b x} \times A^\square_{\b y}
   =
  \frac{g^2}{2N^2}
  \sum_{\b k \notin \Pbb\_S}
    \frac1{\~\omega_{\b k}}
    \e^{\frac{2\pi\i}N \b k (\b x - \b y)}
   \approx
  \frac{g^2}{4\pi} \frac{\e^{-m|\b x - \b y|}}{|\b x - \b y|}.
}

What about the remaining degrees of freedom in \eqref{def HT A tilde E pi phi}?  The other component of the gauge field, $A^\times_{\b k}$, and the scalar field $\varphi_{\b k}$ naturally combine into the operator
\bel{
  \~\pi_{\b k} \equiv \omega_{\b k} \varphi_{\b k} - q A^\times_{\b k}.
}
The operator canonically conjugate to $\~\pi_{\b k}$ takes the form
\bel{\label{def varphi tilde}
  \~\varphi_{\b k} \equiv \frac{\omega_{\b k} \pi_{\b k} + q (\nabla E)_{\b k}/\omega_{\b k}}{\omega_{\b k}^2 + q^2}.
}
Both $\~\pi_{\b k}$ and $\~\varphi_{\b k}$ commute with the tame Gauss operator $\rho_{\b k} = - q\pi_{\b k} + (\nabla E)_{\b k}$: they encode the gauge-invariant degree of freedom that is the result of the ``gauge field eating the scalar field.''

The goal now is to express the Hamiltonian in terms of the operators $\~\pi_{\b k}$, $\~\varphi_{\b k}$, and $\rho_{\b k}$, just how eq.\ \eqref{def HT A tilde} did it for pure Maxwell theory. To this end, note that \eqref{def varphi tilde} implies
\bel{
  \pi_{\b k} = \omega_{\b k} \~\varphi_{\b k} - \frac{q}{\omega_{\b k}^2 + q^2} \rho_{\b k}.
}
Inserting this into the remaining nonzero-momentum terms in the Hamiltonian,
\bel{
  H\_T^{(2)}
   =
  \sum_{\b k \in \Pbb \backslash \{0\}}
  \left[
    \frac{g^2}{2 \omega_{\b k}^2}
    \left(\rho_{\b k} + q\pi_{\b k}\right)\+
    \left(\rho_{\b k} + q\pi_{\b k}\right)
    + \frac1{2r^2} \pi\+_{\b k} \pi_{\b k}
    + \frac{r^2}2 \~\pi_{\b k}\+ \~\pi_{\b k}
  \right],
}
gives the somewhat unfamiliar expression
\bel{
  H\_T^{(2)}
   =
  \sum_{\b k \in \Pbb \backslash \{0\}}
  \left[
    \frac{r^2}2 \~\pi_{\b k}\+ \~\pi_{\b k}
    + \frac{\omega_{\b k}^2 + (qgr)^2}{2r^2}
      \~\varphi_{\b k}\+ \~\varphi_{\b k}
    + \frac{(gr)^2 \omega_{\b k}^2 + q^2}{2r^2(\omega_{\b k}^2 + q^2)^2}
      \rho_{\b k}\+ \rho_{\b k}
    + \frac{q \omega_{\b k} [(gr)^2 - 1] }{r^2(\omega_{\b k}^2 + q^2)}
      \rho_{\b k}\+ \~\varphi_{\b k}
  \right].
}
This can be significantly simplified by using $gr \sim 1/K \ll 1$. If $q \neq 0$, the result is
\bel{
  H\_T^{(2)}
   \approx
  \sum_{\b k \in \Pbb \backslash \{0\}}
  \left[
    \frac{r^2}2 \~\pi_{\b k}\+ \~\pi_{\b k}
    + \frac{\~\omega_{\b k}^2}{2r^2}
        \left(
          \~\varphi_{\b k}\+
          - \tfrac {q\omega_{\b k}}{\~\omega_{\b k}^2(\omega^2_{\b k} + q^2)} \rho_{\b k}\+
        \right)
        \left(
          \~\varphi_{\b k}
          - \tfrac {q \omega_{\b k}}{\~\omega^2_{\b k}(\omega^2_{\b k} + q^2)} \rho_{\b k}
        \right)
    + \frac{(qg)^2 \rho_{\b k}\+ \rho_{\b k}}{2\~\omega_{\b k}^2(\omega_{\b k}^2 + q^2)^2}
  \right].
}
At each $\b k \neq 0$ there is thus a harmonic oscillator of frequency $\~\omega_{\b k}$. It is centered around a $\rho_{\b k}$-dependent point in its target space, but this does not affect its energy levels. This is, therefore, the second degree of freedom in this theory with dispersion $\~\omega_{\b k}$. The fields $\~\varphi_{\b k}$ and $A^\square_{\b k}$ can thus be interpreted as a two-component (i.e.\ vector-like) massive field, as is familiar from the usual Higgs mechanism story. Note, however, that these two fields in principle have different coupling constants --- the coefficient in front of the kinetic term $(E^\square_{\b k})\+ E^\square_{\b k}$ is $g^2$, while the coefficient in front of $\~\pi_{\b k}\+ \~\pi_{\b k}$ is $r^2$.

A further simplification comes about at low momenta, when $\omega_{\b k} \sim 1/N$. The low-momentum terms in the Hamiltonian are
\bel{
  H\_T^{(2)}
   \supset
  \sum_{|\b k| \ll N}
  \left[
    \frac{r^2}2 \~\pi_{\b k}\+ \~\pi_{\b k}
    + \frac{\~\omega_{\b k}^2}{2r^2}
        \left(
          \~\varphi_{\b k}\+
          - \frac {\omega_{\b k}}{q \~\omega_{\b k}^2} \rho_{\b k}\+
        \right)
        \left(
          \~\varphi_{\b k}
          - \frac {\omega_{\b k}}{q \~\omega^2_{\b k}} \rho_{\b k}
        \right)
    + \frac{g^2}{2q^2 \~\omega_{\b k}^2}
      \rho_{\b k}\+ \rho_{\b k}
  \right].
}
The $\rho\+_{\b k} \rho_{\b k}$ term describes the energy cost of inserting a nontrivial background charge. As in \eqref{def D(x - y)}, Fourier-transforming $1/\~\omega^2_{\b k}$ gives a function $\~D(\b x - \b y)$ that controls the energy cost of changing the background charge at points $\b x$ and $\b y$. As in the massive OPE \eqref{OPE A tilde massive}, the appearance of the dispersion $\~\omega_{\b k}$ with $qgr = m \sim 1/N$ sets a characteristic length $m^{-1}$ at which the behavior of $\~D(\b x - \b y)$ qualitatively changes. When $|\b x - \b y| \ll m^{-1}$, the energy of a pair of background charges increases with the distance as $\frac{g^2}{2\pi} \log |\b x - \b y|$, as in the massless case. When $|\b x - \b y| \gg m^{-1}$, the energy of such a pair is dominated by the self-energy of background charges that is set by $\~D(0) \sim \log N$.

This means that in this scalar QED, unlike in Maxwell theory, keeping nonzero background charges at all distances greater than $m^{-1}$ costs an energy proportional to $\log N$. This is one way of quantifying how charges are screened in a Higgsed theory. Note, however, that it is meaningless to ask whether the gauge fields are confined, i.e.~whether it is a long line of electric flux that causes this high energy cost. This screening effect is a joint feature of gauge and matter fields. For example, in a sector with a nonzero $\rho_{\b x}$ that describes a pair of equal and opposite background charges, the ground state is some superposition of two kinds of states. One kind are states that have an electric flux line stretching between these background charges. The other kind are states with electric flux lines stretching from background charges to matter fields at some other points. Both kinds are present in the ground state for a generic $m \sim 1/N$. This is why, as is well known, a crisp definition of gauge field confinement is not generically available in a gauge theory coupled to matter.

In practice, most discussions of the Higgs mechanism in the theory \eqref{def HT A E pi phi} are narrowly focused on the sector $\rho_{\b k} = 0$. In this gauge-invariant sector the second massive mode has the simple free field Hamiltonian
\bel{\label{def HT (2)}
  H\_T^{(2)}
   \approx
  \sum_{\b k \in \Pbb \backslash \{0\}}
  \left[
    \frac{r^2}2 \~\pi_{\b k}\+ \~\pi_{\b k}
    + \frac{\~\omega_{\b k}^2}{2r^2}
      \~\varphi_{\b k}\+ \~\varphi_{\b k}
  \right].
}

Only zero-momentum modes remain to discuss. Both electric fields and matter momentum operators have them, and their contribution to $H\_T$ is the same as in pure gauge and matter theories,
\bel{
  H^{(3)}\_T = \frac{g^2}2 (E_0^i)\+ E_0^i + \frac1{2r^2} \pi_0\+ \pi_0.
}
Note the difference in how $r^2$ appears in the $\~\pi_{\b k}\+ \~\pi_{\b k}$ and $\pi_0\+ \pi_0$ terms in $H^{(2)}\_T$ and $H^{(3)}\_T$.

The three-part Hamiltonian
\bel{
  H\_T =  H^{(1)}\_T + H^{(2)}\_T + H^{(3)}\_T
}
derived above describes what may be called a \emph{noncompact scalar QED}. Like in pure Maxwell theory, it is also possible to distinguish between basic and standard versions of this theory. This is done by asking whether the lowest-lying sectors with nonzero $\rho_{\b x}$ have energies commensurate with the massive photon energies, which are $\~\omega_{\b k} \sim 1/N$. If so, the theory is standard; if the background charges cost much more energy than photons, the theory is basic. The choice of scaling $g^2 \sim 1/N$ in this Section is tantamount to choosing to work with a standard noncompact theory.

The noncompact scalar QED is simple to smoothe, as it only involves two free fields. The resulting cQFT of smooth fields $\~\varphi(\b x)$ and $A^\square(\b x)$ will be called a \emph{noncompact scalar cQED}. The kinds of OPEs it has have already been exhibited in eq.\ \eqref{OPE A tilde massive}. It is also possible to express the two scalar fields of this theory in a more balanced way, by treating $\~\pi(\b x)$ and $E^\square(\b x)$ as the divergence and curl of an electric field $\~E^i(\b x)$, but this will not be explored here.

Instead, consider the form this theory takes after the fields and couplings are rescaled by powers of a fiducial lattice spacing \eqref{def a}. Doing this will also provide an elegant summary of the various fields, couplings, and scaling relations that appear in this Subsection.

Start from the relations \eqref{scaling of g} and \eqref{scaling of r} that describe parts of the parameter space where a tame low-energy theory is expected. The first one motivates the definition of the continuum gauge coupling
\bel{
  g\_c \equiv a^{-1/2} g,
}
just like in pure Maxwell theory. The second one, together with the requirement $qgr \sim 1/N$ from \eqref{def omega tilde}, motivates introducing the continuum Higgs radius and the continuum target space length element,
\bel{
  r\_c \equiv a^{-1/2} r, \quad \d A\_c \equiv a^{-1} \d A.
}
Both of these quantities are understood to be $O(N^0)$ after being multiplied by the appropriate power of the spatial length $L$. Note that this is the first explicit statement that QED gauge fields have clock eigenvalues of the form
\bel{
  \e^{\i a A\_c^i(\b x)}
}
for some $O(N^0)$ variables $A\_c^i(\b x)$. This fact is often assumed as the starting point when analyzing the continuum limit of lattice gauge theory actions. Here this scaling is understood to follow from demanding that low-energy QED be tame --- or, equivalently, that both pure Maxwell theory and its dual scalar have tame low-energy spectra.

The couplings $g\_c$ and $r\_c$ can also be melded into a dimensionful mass,
\bel{
  m\_c = a^{-1} m \equiv a^{-1} q g r = q g\_c r\_c,
}
and a dimensionless coupling
\bel{
  \gamma \equiv \frac gr.
}
The latter does not appear in the spectrum of the theory, at least when $\rho_{\b x} = 0$ and $\gamma \sim 1$. It controls the asymmetry between the two massive scalar fields, and extreme values of $\gamma$ may lead to a breakdown of tameness for one of the two fields, in which case the resulting theory is no longer viewable as having massive vector-like degrees of freedom. Note that $\gamma$ plays a vital part in the $r, g \rar \infty$ regime described in Subsection \ref{subsec Higgs phases}.

The presence of the exponential falloffs $\e^{-m|\b x - \b y|}$ in the OPEs of $\~\varphi(\b x)$ and $A^\square(\b x)$ means that neither is a scaling operator. They do, however, have well defined engineering dimensions. They arise from demanding that the spatially smoothed Hamiltonians \eqref{def HT (1)} and \eqref{def HT (2)} have natural continuum expressions with engineering dimensions one. This is equivalent to demanding that smoothed actions be dimensionless, which is how engineering dimensions of fields were defined in Subsection \ref{subsec temp smoothing}. Just like in that Subsection, this leads to
\bel{
  \Delta_{A^\square}\^c = \Delta_{\~\varphi}\^c = 1.
}

This finishes the construction of the salient elements of the noncompact scalar cQED. Now take a step back to assess the situation. While the microscopic Hamiltonian \eqref{def H Higgs} is not exactly diagonalizable, assuming the tameness of its low-energy spectrum at $K \sim N$ and $g \sim r \sim 1/\sqrt N$ gives rise to this tractable continuum theory with a nonzero mass gap. The assumption of tameness is consistent with self-duality of the lattice model and with the special case $q = 0$. It is thus plausible to conjecture that the noncompact scalar cQED emerges at low energies in this one patch of the Higgs model parameter space. As $g$ and $r$ are varied across this patch, it is further plausible to imagine that the low-energy physics remains tame --- just with respect to different taming parameters.

There exist many limiting cases that can be studied by perturbing around this cQFT, e.g.\ by taking $\gamma \gg 1$ or by including corrections to subleading orders in taming parameters. A particularly interesting limit to study may be the low-energy description of a Higgs model in which $K \nsim N$. In this regime one can expect a theory that contains both tame and nontame states at low energies, e.g.\ a scalar cQFT coupled to a topological gauge theory.

Another interesting parametric regime not studied here is obtained at large $q$. It is clear that the present analysis cannot hold e.g.\ if $q = K/2$. However, even a much smaller value of $q$, say $q \sim n\_T$, is enough to ruin the tameness assumption made when writing  $\rho_{\b x} = 0$.

\subsection{Compact scalar cQED} \label{subsec comp scalar QED}

The noncompact theory \eqref{def HT A E pi phi} was obtained by taming the original Hamiltonian \eqref{def H Higgs} w.r.t.\ the trivial background
\bel{
  A\^{cl}_\ell = \varphi_v\^{cl} = 0, \quad E\^{cl}_\ell = p_v\^{cl} = 0.
}
\emph{Compact scalar QED} is obtained by including taming backgrounds that minimize the potential terms in the microscopic Hamiltonian, which are
\bel{
  \frac{r^2}2
  \sum_{\ell \in \Mbb}
    \left(2 - Z_\ell^{-q} \textstyle \prod_{v \in \del \ell} Z_v -
          Z_\ell^{q} \textstyle \prod_{v \in \del \ell} Z_v\+ \right)
  + \frac1{2g^2}
  \sum_{f \in \Mbb}
    \left(2 - W_f - W_f\+ \right).
}
As in the clock model and the pure Maxwell theory, it is a \emph{conjecture} that these are all the backgrounds that are relevant at low energies. In particular, as seems standard in field theory, nontrivial smoothing backgrounds $E\^{cl}_\ell$ and $p\^{cl}_v$ will not be considered, even though numerical diagonalization of Hamiltonians in $d = 0$ indicates that they appear at low energies \cite{Radicevic:1D}.

The classical backgrounds that minimize the potential satisfy the equations of motion
\bel{\label{taming bkgd eq QED}
  \epsilon^{ij} \del_j \sin B_{\b x}\^{cl}
  =
  q (g r)^2 \sin\left(
    \del_i \varphi\^{cl}_{\b x} - q (A\^{cl})^i_{\b x}
  \right),
}
where $\epsilon^{12} = - \epsilon^{21} = 1$ and $\epsilon^{11} = \epsilon^{22} = 0$. In general, these two equations may have highly nontrivial vortex solutions. However, the regime in which taming is valid is $(gr)^2 \sim 1/K^2$.  The l.h.s.\ of the above equation, when nonzero, must be $O(\d A) = O(1/K)$, and hence the only way the equation can be satisfied is if it splits into two separate equations
\bel{
  \del_i \sin B_{\b x}\^{cl} = 0,
  \quad
  \del_i \sin\left(
    \del_i \varphi\^{cl}_{\b x} - q (A\^{cl})^i_{\b x}
  \right) = 0.
}
After assuming spatial smoothness, the first of these becomes the same equation found in pure Maxwell theory, \eqref{taming bkgd eq 2d}. Its solutions are the ``symmetric gauge'' flux backgrounds \eqref{flux backgrounds}. The second relation is solved by the same winding solutions found in the $\Z_K$ clock model, namely
\bel{
  \varphi\^{cl}_{\b x} = \varphi\_{const} + \frac{2\pi}N w^i x^i, \quad (A\^{cl}_{\b x})^i = 0.
}
Other solutions to these equations are obtained by gauge-transforming $\varphi_{\b x}\^{cl}$ and $(A\^{cl})^i_{\b x}$ in the obvious way.

This means that compact scalar cQED is obtained by adding the already-familiar winding and magnetic flux backgrounds to the noncompact scalar cQED. There are no other taming backgrounds that need to be included. However, as soon as the couplings are increased so that $gr \gtrsim 1/\sqrt K$, nontrivial vortex configurations that solve the full equation \eqref{taming bkgd eq QED} must be included. These objects control the physics in the shaded upper right corner on Fig.\ \ref{fig phases}.

\subsection{BF theory} \label{subsec BF}

The function of this Subsection is to examine in some detail the topological theory found in the $r \rar \infty$, $g \rar 0$ limit of the Higgs model. This corresponds to the topological/Higgs regime in the upper left corner of parameter space on Fig.\ \ref{fig phases}. As described in Subsection \ref{subsec Higgs phases}, in this regime the set of ground states is parameterized by flat $\Z_q$ gauge fields $\bar A_\ell$, $\Z_K$ background charges $\varrho_v$, and an overall shift $\vartheta$ of the scalar field.

The fact of interest here is that a $\Z_q$ topological gauge theory --- a theory with a finite $q$ that may even be as small as $q = 2$ --- emerges from a lattice system with a $\Z_K$ target space. This target space is so large that it can, depending on context, be approximated by the manifolds U(1) or $\R$. Indeed, the largeness of $K$ is what allows the lattice theory \eqref{def H Higgs} to be given the cQFT description detailed in this Section. On the other hand, no matter what $K$ is, the Higgs relation comes into power when $r$ is sufficiently large, and when that happens a $\Z_q$ topological gauge theory emerges at low energies. Thus invoking the large-$r$ Higgs mechanism at $K \gg 1$ provides an appealing way to endow the $\Z_q$ topological gauge theory with an effective continuum description even if $q$ is finite. This point of view has been most clearly articulated in \cite{Banks:2010zn}. The corresponding continuum description of the topological $\Z_q$ gauge theory is called a BF theory \cite{Horowitz:1989ng}.

There is just one sticking point here: the continuum description of the Higgs model as the scalar cQED with a St\"uckelberg coupling \eqref{def HT A E pi phi} --- precisely the theory used in \cite{Banks:2010zn} --- is only valid when $r \sim \sqrt N/K \sim 1/\sqrt N \ll 1$. Meanwhile, the Higgsing was shown to happen only at \emph{large} values of $r$. Indeed, the discussion of taming backgrounds in Subsection \ref{subsec comp scalar QED} has shown that when $(gr)^2 \gtrsim 1/K$ it becomes necessary to include nontrivial vortex configurations at low energies. It is thus impossible to keep $g \sim 1/\sqrt N$ and send $r$  to infinity while still reliably using the scalar cQED description.

Now, the fact that there are no phase transitions between the cQED regime and the topological/Higgs regime means that one just \emph{might} get away with pretending that the continuum description holds up even at $r \gg 1$. Indeed, this paper has not in any sense estimated the boundaries of the topological/Higgs regime. It is possible that the salient features of its emergent $\Z_q$ gauge theory remain intact all the way down to $r \sim 1/\sqrt N$, even as other features change in more dramatic ways. The plausibility of this scenario is buttressed by the consistency arguments presented in \cite{Banks:2010zn}.

Plausible as it may be, this derivation of BF theory as a continuum description of $\Z_q$ topological gauge theory is difficult to understand at the present level of precision. What taming backgrounds should one be using?  What taming corrections can be safely neglected when $r$ is taken to be large? While it might be possible to consistently answer these questions, this paper will instead offer a completely different way of connecting BF and $\Z_q$ theories.

BF theory has a Hamiltonian equal to zero. It  only appears nontrivial in the Lagrangian formalism. Even though it looks like a continuum theory, it is \emph{not} actually a cQFT as defined in this series \cite{Radicevic:2D}. Understanding its action is still instructive, though. This paper will show how the action of a $\Z_K$ lattice gauge theory in the topological limit takes the form of a BF theory for any $K$. This can be viewed as an elementary derivation of a lattice action for the BF theory. Importantly, the configurations summed in this path integral in no way feature small, Gaussian fluctuations. In this sense there is no connection between the BF path integral and the actions of the sort \eqref{def S tilde Maxwell} that involve tame gauge field configurations.

Consider a $\Z_K$ gauge theory with Hamiltonian
\bel{\label{def H topo}
  H = \frac1{2g^2} \sum_{f \in \Mbb} \left(2 - W_f - W_f\+ \right).
}
This Hamiltonian describes the topological, $g \rar 0$ limit of the pure Maxwell theory \eqref{def H}. Its ground states in the $\varrho_v = 0$ sector are labeled by flat $\Z_K$ gauge fields.

The partition function of this theory, at low temperatures that satisfy $\beta \gg g^2$, is equal to the ground state degeneracy of the theory,
\bel{
  \trm{GSD} = K^{N^2 + 1}.
}
If restricted to the gauge-invariant sector, the degeneracy is
\bel{
  \trm{GSD}\big |_{\varrho_v = 0} = K^2.
}

These numbers can be obtained in a much more complicated way, by using the usual transfer matrix method to write down a path integral in which the flatness of gauge fields is manually imposed. The actions obtained this way are those of a BF theory.

Let $\Ebb = \Mbb \times \Sbb$ be a spacetime lattice as before. The transfer matrix procedure generates the sum
\bel{
  \Zf\_{sBF}
    \equiv
  \frac1{K^{3N^2 N_0}}
  \sum_{\{ m^\mu_x, A^i_x \}}
    \e^{
      -\i \sum_{x \in \Ebb}
      \left[ m^0_x B_x + m^i_x \del_0 A_x^i \d\tau \right]
    }.
}
Here $\{A_x^i\}_{x \in \Ebb}$ are all possible gauge field configurations, with each angular variable $A_x^i$ taking $K$ values, in increments of $\d A$, for each $(x, i)$. The magnetic field is $B_x \equiv \del_1 A_x^2 - \del_2 A_x^1$. The Lagrange multipliers $0 \leq m^\mu_x < K$ serve to impose constraints $B_x = 0$ and $\del_0 A_x^i = 0$ at each spacetime point. (The latter constraint simply means that acting by $H$ preserves $A_x^i$.) This sum can be evaluated by counting the number of gauge configurations and constraints, and the answer is
\bel{
  \Zf\_{sBF} = K^{N^2 + 1}.
}

Similarly, requiring that $\varrho_v = 0$ in all states that are inserted while constructing the path integral gives
\bel{
  \Zf\_{bBF}
  \equiv
  \frac1{K^{4N^2 N_0}}
  \sum_{\{ n_x, m^\mu_x, A^i_x \}}
    \e^{
      -\i \sum_x \left[
         m^0_x B_x + m^i_x (\del_0 A_x^i \d\tau - \del_i n_x \d A)
      \right]
    }.
}
Here there is yet another integer-valued field $0 \leq n_x < K$. It can be understood as the gauge parameter that enters the definition \eqref{def G n} of the global Gauss operator $G[n]$. The value of the sum is
\bel{
  \Zf\_{bBF} = K^2.
}

These two sums both represent a type of BF theory. The first one includes all background charges, and can be called the \emph{standard BF theory}. The second one considers only states without background charges, and can be called the \emph{basic BF theory}.

The action of the basic BF theory can be brought to a more familiar form by rescaling the Lagrange multipliers so that they look like angular variables,
\bel{\label{def A0 BF}
  b_x^i \equiv  m^i_x \d A,
  \quad
  b_x^0 \equiv m^0_x \frac{\d A}{\d \tau},
  \quad
  A_x^0 \equiv n_x \frac{\d A}{\d \tau}.
}
The resulting action is
\bel{\label{def S BF}
  S[b^\mu, A^\mu]
   \equiv
  \frac{\i K}{2\pi}
  \sum_{x \in \Ebb}
    \left[b_x^0 B_x + b_x^i (\del_0 A_x^i - \del_i A_x^0) \right] \d\tau.
}
The Lagrangian can be compactly written by letting $b^i_x \mapsto \epsilon^{ij} b^j_x$ and using the totally antisymmetric symbol $\epsilon^{\mu\nu\lambda}$, so
\bel{
  \L =
  \frac{\i K}{4\pi} \epsilon^{\mu\nu\lambda}\, b^\mu_x F^{\nu\lambda}_x
  \equiv \frac{\i K}{2\pi} b \wedge F.
}
This is the standardly quoted form of the BF action that corresponds to a $\Z_K$ gauge theory in $d = 2$. This derivation holds for any $K \geq 1$. In particular, one can replace $K \mapsto q$ everywhere and say that this action corresponds to a finite-$q$ topological $\Z_q$ theory that has emerged from a $\Z_K$ Higgs model  whose $r \rar \infty, g \rar 0$ Hamiltonian had the term like \eqref{def H topo}.

The crucial fact about the action \eqref{def S BF} is that its variables are neither tame nor spatially smooth. Indeed, the lack of spatial smoothness is \emph{precisely} the indicator that this action is topological, as it does not depend on the underlying lattice structure in any significant way. The lack of tameness also means that these fields cannot be Fourier-transformed in a simple way. It may, however, be possible to manually remove nonsmooth or nontame configurations from the sum and to argue that the correct answer is hidden somewhere in the universal ($N$- and $N_0$-independent) part of this adulterated sum. This kind of assumption is equivalent to assuming that the cQED regime persists at $r \gg 1$. This leap of faith will not be made here, and BF theory will always be understood to refer to the noncontinuum expression \eqref{def S BF}.

\newpage

\section{Quantum electrodynamics} \label{sec QED}

\subsection{Dirac fermions in $(2 + 1)$D} \label{subsec Dirac fermions}

The goal of this Section is to study the continuum behavior of ordinary QED, i.e.\ of gauge fields coupled to fermions. However, pure fermion theories in $d = 2$ are not without interest. This Subsection is dedicated to studying them and the Dirac cQFT they give rise to.

Consider a $2N \times 2N$ toric lattice $\Mbb$ with sites labeled by $\b v = (v^1, v^2)$.  On each site there is a complex, spinless fermion $\psi_{\b v}$. A natural lattice theory is built out of nearest neighbor hopping terms,
\bel{\label{def H ferm naive}
  H
   =
  \i \sum_{i = 1}^2 \sum_{\b v \in \Mbb} \left(
    \psi_{\b v + \b e_i}\+ \psi_{\b v}  - \psi_{\b v}\+ \psi_{\b v + \b e_i}
  \right).
}
Going to Fourier space via
\bel{
  \psi_{\b v} \equiv \frac1{2N} \sum_{ k^i = - N}^{N - 1} \psi_{\b k} \, \e^{\frac{2\pi\i}{2N} \b k \b v}
}
gives
\bel{
  H
   =
  2 \sum_{k^i = -N}^{N - 1} \sum_{i = 1}^2
    \sin \frac{\pi k^i}{N}  n_{\b k},
  \quad
  n_{\b k} \equiv \psi_{\b k}\+ \psi_{\b k}.
}
Like its $d = 1$ analogue, this theory is free and has an obvious precontinuum basis generated by the $\psi_{\b k}$'s. Unlike the $d = 1$ case, though, the dispersion $2 \big(\! \sin\frac{\pi k^1}N + \sin\frac{\pi k^2}N \big)$ has an extensive number of nodes. In other words, this theory has a square-shaped Fermi surface parameterized by $k^2 = - k^1 \, \trm{mod}\, 2N$ and $k^2 = k^1 + N \, \trm{mod}\, 2N$. There are thus $O(N)$ species of light, effectively $d = 1$ excitations in this theory. Interesting as it may be, this result shows that the na\"ive Hamiltonian \eqref{def H ferm naive} does not give rise to a $d = 2$ Dirac fermion cQFT.

However, this Hamiltonian is just one member of an entire family of interesting quadratic Hamiltonians. This family is parameterized by what are sometimes called \emph{Kawamoto-Smit signs} \cite{Kawamoto:1981hw, Burden:1986by}. A general Hamiltonian of this type is
\bel{
  H(\b s_i)
   \equiv
  \i \sum_{i = 1}^2 \sum_{\b v \in \Mbb}
    (-1)^{\b s_i \b v}
    \left(
      \psi_{\b v + \b e_i}\+ \psi_{\b v}  - \psi_{\b v}\+ \psi_{\b v + \b e_i}
    \right),
}
where the two two-component vectors $\b s_i$ have entries 0 or 1. There are thus $2^{d^2} = 16$ different choices for $\b s_i$. In principle, each leads to a different solvable theory, and hence a different cQFT at low energies. Kawamoto and Smit have shown that a dispersion with a finite number of nodes is found when
\bel{
  \b s_1 = \b 0, \quad \b s_2 = (1, 0) \equiv \b e_1.
}

This choice leads to what at first is a strange-looking theory,
\bel{\label{def H ferm}
  H
   =
  \i \sum_{\b v \in \Mbb} \left(
    \psi_{\b v + \b e_1}\+ \psi_{\b v}  + (-1)^{v^1} \psi_{\b v + \b e_2}\+ \psi_{\b v} - \trm{H.c.}
  \right).
}
After Fourier-transforming this becomes
\bel{
  H
   =
  2 \sum_{k^i = -N}^{N - 1} \left(
    \psi\+_{\b k} \psi_{\b k} \sin \frac{\pi k^1}{N} +
    \psi\+_{\b k} \psi_{\b k + N \b e_1}\sin \frac{\pi k^2}{N}
  \right).
}
The precontinuum basis for this theory is \emph{not} generated by the na\"ive Fourier transforms $\psi_{\b k}$ and $\psi\+_{\b k}$. Instead, for each $\b k$ in the momentum space
\bel{
  \Pbb \equiv \{ \b k \}_{- N/2 \leq k^i < N/2}
}
one can diagonalize the Hamiltonian density and get two pairs of precontinuum ladder operators. One pair, denoted $\Psi_{\b k}$, contains linear combinations of $\psi_{\b k}$ and $\psi_{\b k + N \b e_1}$. The other pair, denoted $\Psi_{\b k + N \b e_2}$, contains linear combinations of $\psi_{\b k + N \b e_2}$ and $\psi_{\b k + N \b e_1 + N \b e_1}$.

In theory \eqref{def H ferm}, the two-component objects $\Psi_{\b k}$ and $\Psi_{\b k + N \b e_2}$ can be mapped to two uncoupled Dirac fermions. The fact that there are two of them is an example of doubling in the staggered fermion formalism \cite{Susskind:1976jm}. Note, however, that other choices of Kawamoto-Smit signs may couple all four modes $\psi_{\b k}$ related by shifting the momentum by $N$. Alternatively, as in theory \eqref{def H ferm naive}, all four of these modes may be decoupled from each other.

Focus on just the $\Psi_{\b k}$ degrees of freedom for now. The $\b k = 0 \, \trm{mod}\, N$ modes all have zero energy and can be treated separately, so assume that $\b k \in \Pbb\backslash \{0\}$. The components of $\Psi_{\b k}$ are
\algns{\label{def Psi k}
  \Psi_{\b k}^+
   &=
  \psi_{\b k} \cos \frac{\theta_{\b k}}2
   +
  \psi_{\b k + N \b e_1} \sin \frac{\theta_{\b k}}2  ,
  \\
  \Psi_{\b k}^-
   &=
  - \psi_{\b k}\, \sgn \theta_{\b k} \sin \frac{\theta_{\b k}}2
   +
  \psi_{\b k + N \b e_1} \sgn \theta_{\b k} \cos \frac{\theta_{\b k}}2  .
}
The angle $\theta_{\b k} \in [-\pi, \pi)$ is defined via
\bel{\label{def theta}
  \cos \theta_{\b k} \equiv \frac{2}{\omega_{\b k}} \sin \frac{\pi k^1}N,
   \quad
  \sgn \theta_{\b k} \equiv \sgn k^2,
}
where the dispersion is
\bel{
  \omega_{\b k} \equiv 2 \sqrt{\sin^2 \frac{\pi k^1}N + \sin^2 \frac{\pi k^2}N }.
}
From this convention it further follows that $\sin \theta_{\b k} = \frac{2}{\omega_{\b k} } \sin \frac{\pi k^2}N$.
\newpage

In this notation, the part of the Hamiltonian \eqref{def H ferm} that depends only on the $\Psi_{\b k}$ degrees of freedom becomes
\bel{\label{def H ferm Dirac}
  H\_{Dirac}
   =
  \sum_{\b k \in \Pbb}
    \omega_{\b k}
    \left[
      (\Psi_{\b k}^+)\+ \Psi_{\b k}^+ - (\Psi_{\b k}^-)\+ \Psi_{\b k}^-
    \right].
}
As expected, this is a free theory that can be trivially smoothed. The result is called the Dirac cQFT. Even though the starting Hamiltonian \eqref{def H ferm} seemed to violate rotation invariance through the Kawamoto-Smit signs, the precontinuum basis turns out to treat $k^1$ and $k^2$ on an equal footing. There is no mystery here: the momentum space $\Pbb$ is only a quarter of the Fourier dual of the original lattice $\Mbb$. Said differently, the position space fields $\Psi^\alpha_{\b x}$ live on a lattice with double the lattice spacing compared to $\Mbb$. This makes the oscillations of the form $(-1)^{\b s_i \b v}$ on the original space invisible to the Dirac fermions.

The two-component objects \eqref{def Psi k} should not themselves be referred to as Dirac fermions. Their Hamiltonian in position space, even if only written in terms of smooth fields, does not show the conventional structure with Dirac matrices and a single derivative. To get this structure, define the two-component objects
\bel{
  \psi^\alpha_{\b k} \equiv \bcol{\psi_{\b k}}{\psi_{\b k + N \b e_1}}.
}
\emph{These} are Dirac fermions (or spinors). Their Hamiltonian density at low momenta is approximately
\bel{
  K^{\alpha \beta}_{\b k} (\psi^\alpha_{\b k})\+ \psi^\beta_{\b k},
  \quad \trm{for} \quad
  K_{\b k} \equiv \frac{2\pi}N  \bmat{k^1}{k^2}{k^2}{-k^1}.
}
Now it is possible to define $\gamma$-matrices
\bel{
  \gamma^0 \equiv \bmat0{-1}{1}0, \quad \gamma^1 \equiv -\bmat0110, \quad \gamma^2 \equiv \gamma^0 \gamma^1 = \bmat100{-1},
}
and to let
\bel{
  \overline \psi_{\b k} \equiv \psi_{\b k}\+ \gamma^0.
}
This is one choice for which the Hamiltonian density at low momenta can be written as
\bel{
  \frac{2\pi}N \overline \psi_{\b k} \gamma^i k^i \psi_{\b k}
   \equiv
  \frac{2\pi}N \overline \psi_{\b k} \, \c k \, \psi_{\b k}.
}
Note that the precontinuum generators $\Psi_{\b k}$ on their own know nothing about this spinorial structure. All the information about it is contained in the linear transformation  \eqref{def Psi k} that takes $\psi_{\b k}$ to $\Psi_{\b k}$. The topological properties of this map are one of the central objects of study in condensed matter theory. This subject will be taken up in Section \ref{sec CS}.

A few simple observations are in order. The precontinuum particle number operators in this theory are
\bel{
  n^\alpha_{\b k} \equiv (\Psi_{\b k}^\alpha)\+ \Psi_{\b k}^\alpha.
}
All ground states have a Dirac sea structure with $\avg{n^-_{\b k}} = 1$ and $\avg{n^+_{\b k}} = 0$ for all $\b k \in \Pbb \backslash \{0\}$. The $\b k = 0$ modes are degenerate and decoupled from each other, but one can still define $\Psi_{\b k = 0}^\alpha \equiv \psi_{\b k = 0}^\alpha$. The four ground states differ by the occupation numbers of $n^\alpha_0$. The ground states are not na\"ive analogues of the $d = 1$ situation \cite{Radicevic:2019jfe}, where the occupation numbers of $n^\alpha_k$ depended on $\sgn k$. It is also easy to check that
\bel{
  \left[
    n_{\b k}^\alpha,
    (\psi_{\b k}^\beta)\+ \psi^\beta_{\b k}
  \right]
  \neq 0,
}
with no summation implied. This means that there is no way to express the ground state of $H\_{Dirac}$ as an eigenstate of the densities of individual Dirac fermions.

Dirac fermions in position space $\Mbb^\star = \{\b x\}_{1 \leq x^i \leq N}$ can be defined as
\bel{
  \psi_{\b x}^\alpha
   \equiv
  \frac1{N}
  \sum_{\b k \in \Pbb}
    \psi_{\b k}^\alpha \, \e^{\frac{2\pi\i}N \b k \b x}.
}
The corresponding smooth fields $\psi^\alpha(\b x)$ are obtained by projecting to the smooth momentum subspace
\bel{
  \Pbb\_S \equiv \{\b k\}_{-k\_S \leq k^i < k\_S}.
}
An OPE that is rather simple to evaluate is
\algns{
  (\psi^+_{\b x})\+ \times \psi^+_{\b y}
   &=
  \frac1{N^2}
  \sum_{\b k \notin \Pbb\_S}
    \sin^2 \frac{\theta_{\b k}}2 \,
    \e^{\frac{2\pi\i}N (\b y - \b x) \b k} \\
   &=
  \frac1{2N^2}
  \sum_{\b k \notin \Pbb\_S}
    \left(1 - \frac{2}{\omega_{\b k}} \sin \frac{\pi k^1}N \right)
    \e^{\frac{2\pi\i}N (\b y - \b x) \b k}
   \approx
  \frac12 \delta_{\b x,\, \b y}
  - \frac1{4\pi\i}
  \frac{(\gamma^i)^{++} (x^i - y^i)}{|\b x - \b y|^3}.
}
Calculating the other components is also straightforward. Concretely, this shows that the general OPE at nonzero distances is
\bel{
  (\psi^\alpha_{\b x})\+ \times \psi^\beta_{\b y}
   \approx
  -\frac1{4\pi\i}
  \frac{\boldsymbol \gamma^{\alpha\beta} \cdot (\b x - \b y)}{|\b x - \b y|^3}
   \sim
  \frac \i {|\b x - \b y|^2}.
}
The appearance of $\gamma$-matrices in the OPE means that, as with gauge fields in the standard noncompact Maxwell theory, the Dirac fermions are not proper scaling operators. Nevertheless, they can be given an engineering dimension of $\Delta_\psi\^c = 1$.

Actual scaling operators can be built out of fermion bilinears. The simplest example is $\overline\psi_{\b x}\psi_{\b x}$. Their OPEs will not be calculated here.

There is also a natural mass term that can be added to the theory. In momentum space it is given by
\bel{
  \i \overline \psi_{\b k} \psi_{\b k}
   =
  - \i \left(
    (\psi_{\b k}^+)\+ \psi_{\b k}^- -
    (\psi_{\b k}^-)\+ \psi_{\b k}^+
  \right).
}
The massive Dirac Hamiltonian is therefore
\bel{\label{def H ferm massive}
  H\_{Dirac}(m)
   \equiv
  H\_{Dirac}
   +
  \i m \sum_{\b k \in \Pbb}
    \overline \psi_{\b k} \psi_{\b k},
}
and its dispersion is
\bel{
  \~\omega_{\b k} = \sqrt{m^2 + \omega_{\b k}^2}.
}

It is interesting to consider the limit $|m| \rar \infty$, where the mass is so large that
\bel{\label{def H ferm inf mass}
  H\_{Dirac}(m) \approx \i m \sum_{\b k \in \Pbb}
    \overline \psi_{\b k} \psi_{\b k}.
}
In this limit the physics is completely controlled by the sign of the mass. The precontinuum generators are
\bel{\label{def massive fermions}
  \Psi_{\b k}^\pm
   =
  \frac1{\sqrt 2}
  \left( \pm \psi_{\b k}^+ - \i \, \sgn(m) \, \psi_{\b k}^- \right).
}
Of course, this precontinuum basis does not generate a continuum basis, as the dispersion is flat ($\omega_{\b k} = \pm m$) and hence there is no natural ordering of momenta in $\Pbb$. Still, this parametric regime will be important when the analysis turns to Chern-Simons theory in Section \ref{sec CS}.

There are three more operators that can be added to the Hamiltonian while keeping different momenta uncoupled. In momentum space they are given by
\bel{
  \overline \psi_{\b k} \gamma^\mu \psi_{\b k}.
}
They can be interpreted as chemical potential terms. At low momenta, adding the $\mu = 1, 2$ terms simply changes the dispersion and the precontinuum operators \eqref{def Psi k} by shifting the corresponding $k^i$ by a constant. Adding the $\mu = 0$ term shifts both momenta $k^i$ by the same amount, but in different directions. These operators therefore all introduce a Fermi surface.

To end this Subsection, note that there is no conceptual difficulty with defining a single Dirac spinor on a lattice, despite the widely advertised doubling problems. One way to think about this is to recall an old observation by Banks and Windey \cite{Banks:1982ut}: imposing a continuity requirement on lattice fermion fields eliminates doublers. More technically, the starting theory \eqref{def H ferm} does contain doublers, but as long all interactions are judiciously chosen and local (on the scale of the ``string length'' $N/k\_S$), it will be possible to write perfectly reasonable lattice theories without doubling issues.

\newpage

\subsection{Do gauge theories gauge symmetries?} \label{subsec geom anom}

As hinted by their very name, a common perspective on gauge theories is that they arise when a symmetry of some starting theory is \emph{gauged}. Gauging is a procedure in which a starting theory is coupled to gauge fields in such a way that the new theory has an extensive number of symmetries generated by certain operators $\G_v$ analogous to \eqref{def gen G}. These are built from gauge-theoretic Gauss operators $G_v$ and from local charge densities of the symmetry of the starting theory that is to be gauged. This new theory is then subjected to gauge constraints like $\G_v = \1$, and as a result the degrees of freedom of the original theory end up restricted to the zero-charge (singlet) sector of the gauged symmetry.

The Higgs model \eqref{def H Higgs} can be understood through this lens as a clock model whose $\Z_K$ shift symmetry was gauged by introducing gauge fields on links. The singlet constraint that gauging imposes comes from multiplying all the operators $\G_v$ and takes the form $\prod_{v \in \Mbb} X_v = \e^{\frac{2\pi\i}K \sum_{v \in \Mbb} \varrho_v} \1$. In other theories the local charge densities may live on links or other parts of the lattice, and in this case the gauge fields must have a correspondingly higher rank. For example, the pure Maxwell theory \eqref{def H} has a one-form symmetry with local charges $X_\ell$ that can be gauged by introducing $\Z_K$ gauge fields on plaquettes.

This familiar story shortchanges pure gauge theories like \eqref{def H}. Not all gauge theories need to arise from gauging a symmetry in some already given system. Conversely, not all symmetries can be gauged. If operators $\G_v$ on different sites (or edges, etc) fail to commute, the symmetry has an \emph{'t Hooft anomaly} and cannot be gauged \cite{Wen:2013oza}.

There also exist more subtle issues that can be encountered when gauging. The foremost example is Witten's SU(2) anomaly \cite{Witten:1982fp}. A gauge theory with this anomaly has a vanishing partition function, indicating that there are actually no states that satisfy all gauge constraints on $\Mbb$, even if there is nothing preventing the gauge constraints in any subset of $\Mbb$ from being satisfied. Anomalies of this type are called \emph{global}.

Global anomalies have traditionally been understood as somewhat subtle nonperturbative phenomena \cite{Witten:1982fp, Wang:2018qoy}. However, a class of global anomalies can be found very explicitly in any system where the symmetry-to-be-gauged has more superselection sectors than there are degrees of freedom on each site. Whether there exist states that satisfy all gauge constraints in that case delicately depends on the charge that enters the gauge constraint and on the size of the system. When no states satisfy all gauge constraints, the associated global anomaly was called a \emph{geometric anomaly} in \cite{Radicevic:2018zsg}.

QED, the subject of this Section, will be \emph{defined} as a theory of Dirac fermions in which the fermion number symmetry has been gauged. This symmetry has no 't Hooft anomaly that prevents this gauging, but it \emph{does} have a geometric anomaly. This Subsection will show how this global anomaly limits the possible choices that go into the definition of lattice QED.

The fermion number symmetry of the Dirac fermion is generated by the operator
\bel{
  N\^F
   \equiv
  - \sum_{\b x \in \Mbb^\star}
    \overline \psi_{\b x} \gamma^0 \psi_{\b x}
   =
  \sum_{\b k \in \Pbb} \left[
    (\psi_{\b k}^+)\+ \psi_{\b k}^+ + (\psi_{\b k}^-)\+ \psi_{\b k}^-
  \right].
}
It is convenient to work with symmetry generators that are products of local operators at each $\b x \in \Mbb^\star$, for instance
\bel{
  \e^{\i \eps N\^F}
   =
  \prod_{\b x \in \Mbb^\star}
    \e^{-\i \eps \overline \psi_{\b x} \gamma^0 \psi_{\b x}}
}
for some $\eps \in \R$.  Further, it is more instructive to generalize this slightly and consider the fermion number symmetry generated by
\bel{
  Q_\eps\^F(q_1, q_0)
   \equiv
  \prod_{\b x \in \Mbb^\star}
    \e^{\i \eps \left[
      (q_0 - q_1) \overline \psi_{\b x} \gamma^0 \psi_{\b x} + q_0 \1
    \right]}
   =
  \e^{\i \eps (q_1 - q_0) N\^F + \i \eps q_0 N^2 \1}.
}
Informally, one can say that $Q_\eps\^F(q_1, q_0)$ assigns a charge $q_0$ to the state with no fermions and increases the charge by $q_1 - q_0$ for each fermionic excitation. Most applications set $q_0 = 0$, pick some fiducial infinitesimal $\eps$, and say that $q_1$ is \emph{the} charge of fermions under the symmetry $Q_\eps\^F(q_1, 0)$. Nevertheless, it will soon become apparent that keeping a nonzero $q_0$ is convenient for pedagogical purposes.

The symmetry generator $Q_\eps\^F(q_1, q_0)$ has at most $2N^2 + 1$ superselection sectors. This is how many different eigenvalues the operator $N\^F$ has. The number of superselection sectors actually distinguished by $Q_\eps\^F(q_1, q_0)$ depends on the choice of $\eps$, $q_0$, and $q_1$. For example, when $\eps (q_1 - q_0) \in 2\pi\Z + \pi$, there will only be two distinct superselection sectors labeled by $N\^F \, \trm{mod}\, 2$. This choice of $Q\^F_\eps(q_1, q_0)$ generates the \emph{fermion number parity}, which is a $\Z_2$ symmetry because $[Q_\eps\^F(q_1, q_0)]^2 \propto \1$. At the other extreme, if $\eps(q_1 - q_0) \rar 0$, or if $\eps(q_1 - q_0)$ is an irrational multiple of $\pi$, $Q_\eps\^F(q_1, q_0)$ will be able to distinguish between all $2N^2 + 1$ sectors. In this case it will be possible to define arbitrary powers $[Q_\eps\^F(q_1, q_0)]^m$, but when $m > 2N^2 + 1$ any such operator will be a linear combination of powers $[Q_\eps\^F(q_1, q_0)]^{m'}$ for $m' \leq 2N^2 + 1$.

It is thus important to be careful when assigning a symmetry group to the fermion number symmetry. It is tempting to say that $Q_\eps\^F(q_1, q_0)$ for $\eps(q_1 - q_0) \rar 0$ generates a U(1) symmetry, but this is imprecise. In the presence of a lattice, the faithfully acting fermion number symmetry is \emph{at most} $\Z_{2N^2 + 1}$. There are no U(1) number symmetries for lattice fermions.

In the following it will be assumed that
\bel{
  \eps = \frac{2\pi}{2N^2} \quad \trm{and} \quad q_0, q_1 \in \Z.
}
Such a $Q_\eps\^F(q_1, q_0)$ generates a $\Z_{2N^2}$ symmetry whose singlet states are the two ``insulator'' states, one with no fermions ($N\^F = 0$) and one with no holes ($N\^F = 2N^2$).

The number of superselection sectors distinguished by this $Q_\eps\^F(q_1, q_0)$ is generally much larger than the number of degrees of freedom on each site. (The word ``generally'' is used because there exist special cases like $q_1 - q_0 = 2N^2$; for the purposes of the following argument, assume that $q_1$ and $q_0$ have no ``nice'' relation to $N^2$.) This means that some superselection sectors will be much bigger than others, and in particular some sectors might turn out to be \emph{empty}. This is a manifestation of the geometric anomaly.

As a concrete illustration, consider the simple choice
\bel{\label{ex QF}
  Q_\eps\^F(3, 1) = -\, \e^{\frac{2\pi \i}{N^2} N\^F}.
}
Now imagine that gauging this symmetry results in imposing the singlet constraint
\bel{
  Q\^F_\eps(3, 1) = \1.
}
This is equivalent to demanding
\bel{
  N\^F
   =
  \frac{2m + 1}2 N^2
}
for some $m \in \Z$. This is only possible if $N$ is even.  If $N$ is odd, no state satisfies the singlet constraint, and so gauging the fermion number symmetry \eqref{ex QF} leads to an empty state space. This symmetry has a geometric anomaly!

This example also clarifies why the anomaly is called ``geometric:'' it is detected by the size of the lattice. Specifically, the geometric anomaly arises whenever $q_0$ is odd and $q_1 - q_0$ is such that
\bel{
  N\^F = \frac{N^2}{q_1 - q_0} (2m + 1) \1
}
has no solutions for $m \in \Z$.

The above example relied on having $q_0 \neq 0$. It is possible to construct examples of a geometric anomaly when $q_0 = 0$, but then the anomaly is only seen when demanding that gauging impose a constraint like
\bel{
  Q\^F_\eps(q_1, 0) = \e^{\frac{\pi \i}{N^2} q_1 N\^F} = \e^{\frac{\pi \i}{N^2} \varrho} \1, \quad \varrho \in \Z.
}
There are many choices of $q_1$ and $\varrho$ for which this constraint cannot be fulfilled. One example is $\varrho = 1$, $q_1 = 2$. In this case there are no solutions to the global gauge constraint regardless of what $N$ is.

A borderline case happens when $q_0 = 0$, $q_1 = 1$, with the vanilla gauge constraint $Q\^F = \1$. This may seem like the natural choice when defining QED. However, the only states that obey the constraint are the insulators with $N\^F = 0$ and $N\^F = 2N^2$. No few-fermion excitations are allowed in this theory. This situation was called \emph{quasi-anomalous} in \cite{Radicevic:2018zsg}.

These purely kinematical considerations already suggest that one must be careful when defining QED by gauging the fermion number symmetry. Hamiltonian lattice gauge theory literature typically sidesteps this entire discussion by focusing on a superselection sector with a nontrivial background charge density chosen such that the global gauge constraint has solutions. For a general $q_0$ and $q_1$, the background charges can be chosen to ensure
\bel{
  Q\^F_\eps(q_1, q_0) = (-1)^{q_1 - q_0} \1.
}
This is satisfied whenever
\bel{\label{QED half filling}
  N\^F = N^2 \1.
}
This can be understood as a ``half-filling'' constraint. In a sense, this is the natural constraint to impose because the overall ground state of the free Dirac fermion has precisely $N^2$ fermions in its Dirac sea (after the $\b k = 0$ modes are required to be half-filled too).

This is the route this paper will take to construct versions of QED that do not suffer from the geometric anomaly. The concrete generalized Gauss operator that generates local symmetries in QED will be
\bel{\label{def gen G QED}
  \G_{\b x}
   =
  \e^{\i q\, \overline \psi_{\b x} \gamma^0 \psi_{\b x} \, \d A}
  G_{\b x},
}
where $G_{\b x}$ is the Gauss operator in a $\Z_K$ gauge theory with
\bel{
  K = 2N^2.
}
To avoid geometric anomalies, focus on the sector defined by the constraints
\bel{\label{QED gauge constraint}
  \G_{\b x} = \e^{-\i q \, \d A} \1.
}
Taking a product over all $\b x \in \Mbb^\star$ gives the global constraint
\bel{
  Q\^F_\eps(q, 0) \equiv \e^{-\i q N\^F \d A} = \e^{-\i q N^2 \d A} \1 = (-1)^q \1.
}
When the fermion charge is $q = 1$, this implies that the matter is always at half-filling, as in \eqref{QED half filling}. When $q > 1$, however, there may exist solutions at fractional fillings, with
\bel{\label{QED fractional filling}
  N\^F = \frac {2m + q} {q} N^2 \1, \quad -\frac q2 \leq m \leq \frac q2, \quad m \in \Z.
}
Such solutions exist if $2m N^2/q$ is an integer. The absence of solutions for some choices of $N$ and $q$ can be viewed as a mild manifestation of the geometric anomaly. The important thing is that, with the definition \eqref{def gen G QED} and the Gauss law \eqref{QED gauge constraint}, the above constraint is at least guaranteed to always have the half-filled states ($m = 0$) as solutions.

The story so far has focused on making sure that the QED Hamiltonian, once introduced at long last in Subsection \ref{subsec conv QED}, does not suffer from global anomalies in the background charge sector of interest. However, the Gauss operators $\G_{\b x}$ in \eqref{def gen G QED} also have interesting \emph{local} properties that arise from the seemingly trivial fact that the fermion theory has much fewer than $K = 2N^2$ degrees of freedom per site. To see how this factoid impacts things, consider higher powers of the generalized Gauss operators,
\bel{
  \G^m_{\b x} = \e^{- \i m q \, j^0_{\b x}\, \d A} G^m_{\b x},
  \quad
  j^0_{\b x}
   \equiv
  - \overline \psi_{\b x} \gamma^0 \psi_{\b x}
   =
  (\psi^+_{\b x})\+ \psi^+_{\b x} + (\psi^-_{\b x})\+ \psi^-_{\b x}.
}
Take one spinor component, say the one with density $j^+_{\b x} \equiv (\psi^+_{\b x})\+ \psi^+_{\b x}$. For any $m \in \Z$ it is possible to write
\bel{
  \e^{-\i m q \, j_{\b x}^+\, \d A}
   =
  \alpha_m \1 + \beta_m \e^{-\i q\, j_{\b x}^+\, \d A},
}
where
\bel{
  \alpha_m
   \equiv
  \frac{\greek w^{-mq} - \greek w^{-q}}{1 - \greek w^{-q}},
  \quad
  \beta_m
   \equiv
  \frac{1 - \greek w^{-mq}}{1 - \greek w^{-q}},
  \quad
  \greek w \equiv \e^{\i\, \d A}.
}
In other words, any power of the exponentiated fermion number density $\e^{-\i q j^+_{\b x} \d A}$ can be written as a linear combination of the identity and the original exponential itself. Putting the two spinor components together gives the relation
\bel{
  \G_{\b x}^m =
  \left[
    \alpha_m^2 \1 +
    \alpha_m \beta_m \left(\e^{-\i q\, j_{\b x}^+\, \d A} + \e^{-\i q\, j_{\b x}^-\, \d A} \right) +
    \beta_m^2 \e^{-\i q\, (j_{\b x}^+ + j_{\b x}^-)\, \d A}
  \right]
  G^m_{\b x}.
}

Once a particular gauge constraint like \eqref{QED gauge constraint} is imposed, these local relations then imply further constraints on the pure gauge sector of the theory. This is particularly clear after expanding the exponentials of $j^\alpha_{\b x}$ to first order in $\d A$, assuming that $q \ll N$. This is always allowed since the operators $j^\pm_{\b x}$ have eigenvalues $0$ and $1$. This gives
\bel{
  \G_{\b x}^m
   \approx
  \left[
    \alpha_m \1 +
    \beta_m\, \e^{-\i q\, j_{\b x}^0\, \d A}
  \right]
  G^m_{\b x}
   =
  \alpha_m G^m_{\b x} +
  \beta_m\, \G_{\b x} G^{m - 1}_{\b x}.
}
In a world with $\G_{\b x} = \e^{-\i q \, \d A} \1$, this implies a family of constraints on Gauss operators,
\bel{\label{QED consistency conditions}
  \alpha_m G_{\b x}^m + \beta_m\, \e^{-\i q\, \d A} G_{\b x}^{m - 1}
   \approx
  \e^{-\i q m \, \d A} \1.
}

There is a trivial solution to these constraints, $G_{\b x} = \e^{-\i q \, \d A}\1 = \G_{\b x}$. This operator equation implies $j^0_{\b x} = 0$, and so it is only satisfied in the insulator state with no fermions.\footnote{Note that this is inconsistent with $N\^F = N^2$. The local and global constraints must be satisfied separately. Taking the product of \eqref{QED consistency conditions} over all $\b x$ does not yield the global constraint because $(\d A)^2$ corrections become important here.} But this is just a boring special case of a much more exciting class of solutions: \emph{tame states}. Indeed, any state for which $G_{\b x} \approx \1 + \i \, \d A\, (\nabla E)_{\b x}$ solves the above equation for any $m \ll N$.

It is tempting to rephrase this finding as follows: self-consistency of the natural QED gauge constraint \eqref{QED gauge constraint} forces the gauge fields to be tame. This is not quite correct, of course. There are nontame solutions to the conditions \eqref{QED consistency conditions}. For example, any operator of the form $G_{\b x} \approx \1 + \O_{\b x} \d A$ where $\O_{\b x}$ has $O(K^0)$ eigenvalues also solves this constraint.  Nevertheless, this is a nontrivial signal that it is going to be consistent to \emph{assume} that the low-energy states of QED involve tame gauge fields. It is remarkable that this follows from purely kinematic considerations. The same argument would work in any QED-like theory in which the eigenvalues of the local matter density $j^0_{\b x}$ are all much smaller than the size $K$ of the gauge field target space.

The constraint \eqref{QED consistency conditions} is also approximately satisfied by $G_{\b x} = \1$. Thus any state in which the gauge fields are decoupled from the fermions and obey their usual Gauss law will also be consistent. This solution implies that $j_{\b x}^0 = \1$, meaning that the matter sector is forced to have exactly one fermion per site. (Thus this solution to \eqref{QED consistency conditions} also satisfies the global constraint \eqref{QED half filling}.) This does not make the matter sector trivial, however. At each site there exists a two-dimensional subspace of states with exactly one fermion. This means that another set of simple-to-describe states that solve \eqref{QED consistency conditions} are those in which the gauge fields are decoupled from matter and not necessarily tame, while the Dirac fermions are reduced to ``electrically neutral'' spinless fermions. These are often called \emph{spinons} in the condensed matter literature.

If the gauge fields \emph{are} tame, however, the gauge constraint \eqref{QED gauge constraint} can be rewritten as
\bel{
  (\nabla E)_{\b x} \approx q (j_{\b x}^0 - \1).
}
This is precisely the familiar Gauss law that associates sources of electric fields to points at which the fermion number is not unity, i.e.\ where the fermions are \emph{not} precisely at half-filling. This is, of course, a staple of lattice gauge theory formulations of QED. Nevertheless, the analysis of consistency conditions leading to \eqref{QED consistency conditions} and justifying the tameness assumption seems to be absent from the literature.

To recap, imposing a natural QED gauge constraint \eqref{QED gauge constraint} forces both gauge fields and fermions to obey separate consistency conditions. The fermions must obey the global constraint \eqref{QED half filling} at $q = 1$ (or \eqref{QED fractional filling} at $q > 1$), and the gauge fields must obey the local constraints \eqref{QED consistency conditions}. Changing the background charge density $\varrho_{\b x}$ or introducing a nonzero charge $q_0$ can drastically change the situation and lead to an empty Hilbert space in the gauged theory. To draw the familiar ``basic vs standard'' distinction between theories, focus only on gauge constraints of the form
\bel{
  \G_{\b x} = \e^{\i \varrho_{\b x} \d A},
   \quad
  \varrho_{\b x} \equiv -q + \delta\varrho_{\b x},
}
where $\sum_{\b x \in \Mbb^\star} \delta\varrho_{\b x} = 0$ and $\delta\varrho_{\b x} = O(N^0)$. Tame gauge-theoretic states and fermions at half-filling satisfy the consistency conditions that arise from any such $\varrho_{\b x}$.

\subsection{Conventional QED} \label{subsec conv QED}

It is now time to ask what kind of Hamiltonian has local symmetries generated by operators \eqref{def gen G QED}. The discussion so far merely motivated using Dirac fermions and $\Z_K$ gauge fields with $K = 2N^2$, but no dynamical details were given. This Subsection will rectify this.

When gauging the $\Z_K$ shift symmetry to get the Higgs model \eqref{def H Higgs}, it is natural to define the Higgs Hamiltonian by simply substituting nearest-neighbor clock operators with their gauge-invariant versions,
\bel{
  \prod_{v \in \del\ell} Z_v
   \mapsto
  Z_\ell^{-q} \prod_{v \in \del\ell} Z_v.
}
Now it may feel similarly natural to replace
\bel{
  \psi_{\b v}\+ \psi_{\b v + \b e_i}
   \mapsto
  (Z_{\b v}^i)^{q} \psi_{\b v}\+ \psi_{\b v + \b e_i}
}
in the original fermion Hamiltonian \eqref{def H ferm} on the lattice $\Mbb$. The temptation to do so should be resisted at all costs, however. The resulting theory would generically couple the doublers and make a pig's breakfast out of the entire construction of Dirac fermions in Subsection \ref{subsec Dirac fermions}. Instead, to get a theory that is guaranteed to have only a single Dirac fermion in the matter sector, place the gauge fields on links of the lattice $\Mbb^\star$. This amounts to replacing
\bel{
  (\psi_{\b x}^\alpha)\+ \psi^\beta_{\b x + \b e_i}
   \mapsto
  (Z_{\b x}^i)^{q} (\psi^\alpha_{\b x})\+ \psi^\beta_{\b x + \b e_i}
}
in the position space version of the Dirac Hamiltonian \eqref{def H ferm Dirac}.

There is a price to pay here, though. In position space, the Hamiltonian \eqref{def H ferm Dirac} is \emph{not} just composed of simple nearest-neighbor fermion hopping terms. It has the form
\bel{\label{def H ferm Dirac pos space}
  H\_{Dirac}
   =
  2 \sum_{\b k \in \Pbb}
    \overline \psi_{\b k} \gamma^i \psi_{\b k} \sin \frac{\pi k^i}N
   \equiv
  \sum_{\b x,\, \b y \in \Mbb^\star}
    d_i(\b x - \b y) \overline \psi_{\b x} \gamma^i \psi_{\b y},
}
for
\bel{
  d_i(\b r)
   \equiv
  \frac2{N^2}
  \sum_{\b k \in \Pbb}
    \sin\frac{\pi k^i}N \, \e^{-\frac{2\pi\i}N \b k \, \b r}
   =
  \frac{\i}{N^2}
  \sum_{\b k \in \Pbb} \left[
    \e^{-\frac{2\pi\i}N \b k (\b r + \frac12 \b e_i)} -
    \e^{-\frac{2\pi\i}N \b k (\b r - \frac12 \b e_i)}
  \right].
}
The shifts by half a lattice spacing prevent $d_i(\b r)$ from becoming a simple difference of Kronecker deltas. Instead, it is
\bel{
  d_1(\b r) \approx \frac{2\i \delta_{r^2, 0}}{\pi} \frac{(-1)^{r^1} r^1}{(r^1)^2 - 1/4}
}
and \emph{mutatis mutandis} for $d_2(\b r)$. Here it is assumed that $r^i \ll N$. The part of $H\_{Dirac}$ that acts on spatially smooth fields $\psi(\b x)$ is just the usual single-derivative density $-\i\overline\psi(\b x) \c \del \psi(\b x)$. This agrees with the momentum space analysis of Subsection \ref{subsec Dirac fermions}.

The Dirac Hamiltonian is thus not nearest-neighbor in position space, even if it is local in the original space $\Mbb$. This of course does not stop this theory from having a perfectly local cQFT description. The nonlocality is more obviously important when gauging, because now one needs a rule for how to replace bilinears like $(\psi^\alpha_{\b x})\+  \psi^\beta_{\b y}$ with gauge-invariant operators when $\b x$ and $\b y$ are not separated by a single link.

In general, there would be no natural way to do this. There are many Wilson lines (products of $Z^q_\ell$) that connect two given points $\b x$ and $\b y$. In the case at hand, however, there \emph{is} a natural way forward. The extenuating circumstance is that $d_i(\b r)$ always contains a $\delta$-function in one direction. Thus the only bilinears that appear in $H\_{Dirac}$ have the form $(\psi^\alpha_{\b x})\+ \psi^\beta_{\b x + r \b e_i}$, and the natural choice for a Wilson line connecting the two fermions is the shortest one, parallel to the $i$'th direction. The resulting Hamiltonian is
\bel{\label{def H QED}
  H =
  \sum_{\b x, \, \b y \in \Mbb^\star}
    d_i (\b x - \b y) \overline \psi_{\b x} \gamma^i \psi_{\b y}
    \, W^q_{\b x,\, \b y}
  +
  H\_{Maxwell},
}
where $H\_{Maxwell}$ is the gauge theory Hamiltonian with coupling $g$ given by \eqref{def H Maxwell}. The  elementary Wilson line connecting $\b x$ and $\b y = \b x + r \b e_i$ is defined as
\bel{
  W_{\b x, \, \b y}
   \equiv
  \left\{
    \begin{array}{ll}
      \textstyle \prod_{r' = 0}^{r - 1} Z_{\b x + r' \b e_i}, & r > 0, \medskip \\
      \textstyle \prod_{r' = 1}^{|r|} Z_{\b x - r' \b e_i}\+, & r < 0.
    \end{array}
  \right.
}

Analyzing the Hamiltonian \eqref{def H QED} appears hopeless, both because of the nontrivial function $d_i(\b x - \b y)$ and because the fermions are coupled to many clock operators in the gauge theory. Indeed, it may seem that the most one can say about this $H$ is that it has local symmetries generated by $\G_{\b x}$ from \eqref{def gen G QED}. But Subsection \ref{subsec geom anom} has shown that a Gauss law of the form \eqref{QED gauge constraint} imposes powerful constraints on the gauge degrees of freedom. In this Subsection it will be assumed that the gauge field states are tame, which is one way of satisfying the consistency conditions \eqref{QED consistency conditions}. This considerably simplifies the analysis, even though the tame theory is still not solvable.

The fermion-dependent piece of \eqref{def H QED} is the sum of two terms labeled by the direction $i$,
\bel{
  H^i
   \equiv
  \sum_{\b x \in \Mbb^\star}
  \sum_{r = 1}^{N/2}
    \left[
      d_i(-r \b e_i)
      \overline \psi_{\b x} \gamma^i \psi_{\b x + r \b e_i}
      W^q_{\b x,\, \b x + r \b e_i}
       +
      d_i(r \b e_i)
      \overline \psi_{\b x + r \b e_i} \gamma^i \psi_{\b x}
      W^q_{\b x + r \b e_i,\, \b x}
    \right].
}
Now let $Z_{\b x}^i \approx \1 + \i A_{\b x}^i$ for each clock operator in the Wilson lines. This gives
\bel{\label{def H i}
  H^i\_T
   \approx
  H^i\_{Dirac}
   -
  \i q \sum_{\b x \in \Mbb^\star}
  \sum_{r = 1}^{N/2}
    d_i(r \b e_i)
    \left[\,
      \overline \psi_{\b x} \gamma^i \psi_{\b x + r \b e_i}
       +
      \overline \psi_{\b x + r \b e_i} \gamma^i \psi_{\b x}
    \, \right]
    \textstyle \sum_{r' = 0}^{r - 1} A_{\b x + r' \b e_i}^i.
}

A few remarks about this approximation are in order:
\begin{itemize}
  \item A more general taming Ansatz is
      \bel{
        Z_{\b x}^i \approx \e^{\i (A\^{cl})^i_{\b x}} \left( \1 + \i A_{\b x}^i \right).
      }
      The consistency conditions \eqref{QED consistency conditions} in no way limit the allowed taming backgrounds $(A\^{cl})^i_{\b x}$. In fact, tameness of low-energy states w.r.t.\ any background is a conjecture. It can be further conjectured that the only taming backgrounds that need to be analyzed at low energies minimize the potential terms in $H$. Unlike in the Higgs model, such equations of motion do not just involve c-numbers, so this na\"ive approach cannot work without further postulating how to deal with the fermions. A reasonable way to do this would be to diagonalize the Dirac Hamiltonian in the presence of the background field $(A\^{cl})_{\b x}^i$ and use its ground state energy as the fermionic contribution to the potential term for the background fields. 
  \item Expanding the Wilson line $W_{\b x, \, \b y}$ into $\1 + \i \sum_{\b z} A^i_{\b z}$ is justified only if there are few terms in the sum over $\b z$. In other words, if the Wilson line is very long, taming corrections may need to be included. The above expression for $H^i$ assumes that $d_i(\b r)$ decays fast enough so that these corrections are not important for any $\b r$. This is reasonable: $d_i(\b r)$ behaves as $1/|\b r|$ at large distances and therefore counteracts the large number $|\b r|$ of factors $Z_{\b z}^i$ in $W_{\b x, \, \b y}$. However, this argument has not been made rigorous here.
\end{itemize}

The main feature of the approximation \eqref{def H i} is that it only couples fermion bilinears to one gauge field operator at a time. This does not make the theory exactly solvable, but (at least in principle) it allows perturbative access to it. Still, this is a big ask. Each gauge field between $\b x$ and $\b x + r \b e_i$ enters the Hamiltonian with a different coupling, and this makes perturbation theory impractical. While the pure Dirac fermion at least had a simple expression in momentum space, the QED Hamiltonian is complicated even there.  There does not exist a simple way to find even an approximate precontinuum basis for this theory.

It is possible to make progress towards a slightly less ambitious goal. Consider the space of spatially smooth states acted upon by operators $\psi(\b x)$ and $A^i(\b x)$. These smooth fields are defined the same way as in the pure fermion or gauge theories, by taking their Fourier transforms and projecting out ladder operators at momenta outside a subset $\Pbb\_S$ of momentum space. The corresponding ``string lengths'' $\ell\_S$ can even be different for fermions and gauge fields. The tamed QED Hamiltonian then becomes a theory of smooth fields perturbed by high-momentum terms,
\bel{\label{QED split}
  H\_T =\,  \nord {H\_T} + \Delta H\_T.
}
The new goal now is to understand $\nord{H\_T}$ while ignoring $\Delta H\_T$ for as long as possible.

For the interaction Hamiltonian in \eqref{def H i}, this kind of restriction to low momenta is achieved by sending
\bel{
  A^i_{\b x + r' \b e_i} \mapsto A^i(\b x + r' \b e_i),
  \qquad
  \overline \psi_{\b x} \gamma^i \psi_{\b x + r \b e_i}
   \mapsto
  \overline \psi(\b x) \gamma^i \psi(\b x + r \b e_i).
}
In momentum space, this corresponds to removing all momentum modes with at least one momentum component exceeding $k\_S^A$ or $k\_S^\psi$ in absolute value. These cutoffs, defined separately for gauge fields and fermions, are assumed to satisfy $1 \ll k\_S^A \sim k\_S^\psi \ll N$. Such a removal of high-momentum modes is precisely what was identified with normal ordering in the smoothing framework developed in \cite{Radicevic:2019jfe, Radicevic:2D}. This is why the low-momentum Hamiltonian is denoted by $\nord{H\_T}$ in \eqref{QED split}.

It may be instructive to compare this projection to smoothing. The two are not \emph{quite} the same, because smoothing would replace the fermion bilinear $\overline \psi_{\b x} \gamma^i \psi_{\b x + r \b e_i}$ with
\bel{
  \overline \psi \gamma^i \psi(\b x, \b x + r \b e_i)
   \equiv
  \overline \psi(\b x) \gamma^i \psi(\b x + r \b e_i)
   +
  (\gamma^0 \gamma^i)^{\alpha\beta}\,  (\psi^{\alpha}_{\b x})\+ \times \psi^\beta_{\b x + r \b e_i}.
}
In a cQFT, the OPE term contains all particle number operators $n_{\b k}^\alpha$ at $\b k \notin \Pbb\_S$. These can in turn be replaced by their expectation values in the superselection sector that contains the ground state. In the case at hand, however, these number operators are not symmetries. The approximation scheme for QED that is being developed now does not purport to say anything about what happens at high momenta. The goal is simply to write down how low-momentum modes interact with each other. This information is contained in $\nord {H\_T}$.

This restriction to low momenta now makes it possible to simplify the sums over $r$ and $r'$ appearing in \eqref{def H i}. Consider the sum
\bel{
  \sum_{r = 1}^{N/2} \sum_{r' = 0}^{r - 1}
    d_i(r \b e_i)
    \left[\,
      \overline \psi(\b x) \gamma^i \psi(\b x + r \b e_i)
       +
      \overline \psi(\b x + r \b e_i) \gamma^i \psi(\b x)
    \, \right]
    A^i(\b x + r' \b e_i).
}
When $r \ll \ell\_S \sim N/k\_S^\psi$, to first order in $1/\ell\_S$ the summand can be written as
\bel{
  2 d_i(r \b e_i)\, \overline \psi(\b x) \gamma^i \psi(\b x)\, A^i (\b x).
}
The sums over $r$ and $r'$ thus become the simple prefactor
\bel{
  \sum_{r \ll \ell\_S} \sum_{r' = 0}^{r - 1} 2 d_i (r \b e_i)
   =
  \sum_{r \ll \ell\_S} 2 r d_i (r \b e_i)
   \approx
  \i.
}
Directly performing this sum is delicate. A consistent way to evaluate it is by demanding agreement between the momentum and position space versions of $\nord{H\_{Dirac}}$ in \eqref{def H ferm Dirac pos space}.

What happens when $r \gtrsim \ell\_S$? There is no simple smoothness relation that relates $\psi(\b x)$ and $\psi(\b x + \ell\_S \b e_i)$, for example. Fields at points a ``string length'' apart should be treated as independent. Thus the sum over $r$ and $r'$ should be divided into sums over multiples of $\ell\_S$, with a generic term of the form $\overline \psi(\b x) \gamma^i \psi(\b x + n \ell\_S \b e_i) A^i(\b x + m \ell\_S \b e_i)$ for some integers $n$, $m$. The prefactors of such terms in the Hamiltonian decay as the inverse of the distance $n$ between fermion insertions. The reason is simple: the sum over $r'$ can boost a term with fixed $n$ and $m$ by at most $\ell\_S$, and summing over the $r$'s spread around $n \ell\_S$ will give a function that behaves as $1/(n\ell\_S)$. Together these effects give $1/n$. Thus to a first approximation the $r \gtrsim \ell\_S$ terms can be neglected. To be truer to the original theory, the nearest-neighbor ones (with $n = 1$) can be kept as leading corrections.

This argument is not fully rigorous. It does not estimate what happens when $r \sim \ell\_S/2$, say. This is a gap that will not be filled in this paper. From now on, it will be assumed that only the terms with $r \ll \ell\_S$ need to be kept in $\nord{H\_T}$.

Putting these results together and adding the two normal-ordered Hamiltonians $\nord{H\_T^i}$ for $i = 1, 2$ now gives the familiar expression
\algns{\label{def H QED conv}
  \nord{H\_T}
   \ &\approx \
  \nord{H\_{Dirac}}
   +
  \nord{H\_{Maxwell}}
   +\,
  q
  \sum_{\b x \in \Mbb^\star}
    \overline \psi(\b x) \gamma^i A^i(\b x) \psi(\b x) \\
   &\approx\
  \nord{H\_{Maxwell}}
   +
  \sum_{\b x \in \Mbb^\star}
    \overline\psi(\b x) \gamma^i \left[-\i \del_i + q A^i(\b x)\right] \psi(\b x) \\
  &\equiv\
  \sum_{\b x \in \Mbb^\star} \left[
    \frac{g^2}2 E^i(\b x)^2+ \frac1{2g^2} B(\b x)^2
  \right]
   -\,
  \i \sum_{\b x \in \Mbb^\star}
    \overline \psi(\b x) \c D \psi(\b x).
}
The low-momentum theory described by this Hamiltonian will be called \emph{conventional QED}.

Several approximations were crucial in getting to this result. Here is an overview of the unjustified ones, ranked in descending order of plausibility:
\begin{itemize}
  \item The gauge fields were assumed to be tame. The consistency conditions \eqref{QED consistency conditions} lent credence to this, but there are other solutions to these conditions and they may also appear at low energies.
  \item The large-$r$ terms in \eqref{def H i} were neglected because their strength decreases as a power law with distance.
  \item The high-momentum degrees of freedom were completely ignored by focusing on $\nord{H\_T}$. This approximation makes it impossible to probe what happens at distances below $\ell\_S$. Worse, there exist terms that couple high- and low-momentum modes, and it is not obvious that they can be ignored. One motivation for doing so comes from RG: integrating out high-momentum modes is mainly expected to renormalize $g$, so $H\_T$ and $\nord{H\_T}$ show the same physics.
\end{itemize}

\subsection{Observations on QED dynamics}

The interaction term in the restricted Hamiltonian \eqref{def H QED conv} can be expressed in momentum space as
\bel{
  -\i \sum_{\b x \in \Mbb^\star}
    \overline \psi(\b x) \c D \psi(\b x)
   \approx
  \sum_{\b k, \, \b l \in \Pbb\_S} \left[
    \frac{2\pi k^i}N \delta_{\b k, \, \b l}
    + q A^i_{\b k - \b l}
  \right]
  \overline \psi_{\b k} \gamma^i \psi_{\b l}.
}
Depending on the choice of $k\_S^\psi$ and $k\_S^A$ it may happen that the bilinear $\overline \psi_{\b k} \gamma^i \psi_{\b l}$ has no gauge field $A^i_{\b k - \b l}$ to couple to. This is avoided if
\bel{
  k\_S^A \geq 2k\_S^\psi.
}
For concreteness, assume that this inequality is saturated, and let $k\_S \equiv k\_S^\psi$. Then \eqref{def H QED conv} defines interactions in a theory of $(2k\_S)^2$ Dirac fermion modes and $(4k\_S)^2$ gauge field modes. The hard question is whether this theory has anything to do with the low-energy physics of \eqref{def H QED}. Answering this is equivalent to justifying (or falsifying) the assumptions listed above. But regardless of the status of this question, the Hamiltonian \eqref{def H QED conv} can be studied on its own. It is perfectly well defined and it exhibits nontrivial physics. Indeed, it is this conventional QED Hamiltonian (though with an even number of two-component fermions) that is studied by most literature on QED in $(2+1)$D \cite{Pisarski:1984dj, Appelquist:1986fd, Appelquist:1988sr, Gusynin:1998kz, Hands:2004bh, Strouthos:2008kc, Giombi:2015haa, DiPietro:2015taa, Chester:2016wrc, DiPietro:2017kcd}.

It is instructive to examine some special cases of the theory \eqref{def H QED conv}. The simplest one is $q = 0$. The fermions and gauge fields are decoupled in this case. The ground state of the fermions is precisely the Dirac sea described in Subsection \ref{subsec Dirac fermions}. The $\b k = 0$ mode can be either occupied or unoccupied; there is no nontrivial half-filling constraint like \eqref{QED half filling} in this special case. Meanwhile, the gauge theory is just the pure Maxwell theory at low momenta. As per the results of Section \ref{sec Maxwell}, this theory emerges from the $q = 0$ version of lattice QED \eqref{def H QED} when $g^2 \sim 1/N$. Therefore this is the coupling regime at which the $q = 0$ version of \eqref{def H QED conv} can be reliably said to correspond to the low energy states of the starting lattice theory.

A more interesting --- but still relatively simple --- situation arises when $q \neq 0$ and $g \rar 0$. Here it is important to distinguish between studying this limit in the original lattice theory \eqref{def H QED} and in the conventional QED \eqref{def H QED conv}.

Consider first taking this limit in the starting lattice theory. All low-energy states must have flat gauge fields in this case, and when acting on them the Hamiltonian is approximately
\bel{\label{H QED g0 first}
  H
   \approx
  \sum_{\b x, \, \b y \in \Mbb^\star}
    d_i (\b x - \b y) \overline \psi_{\b x} \gamma^i \psi_{\b y}
    \, W^q_{\b x,\, \b y},
}
namely just a theory of Dirac fermions in a flat background field. If this field is small and slowly varying, this is an easily solvable perturbation of the Dirac fermion cQFT.

Even if the background field is not small or slowly varying, the fermion Hamiltonian \eqref{H QED g0 first} is still quadratic and therefore straightforward to diagonalize, at least in principle. Thus in the $g \rar 0$ limit the lattice theory is always solvable.

This is fortunate, as the gauge fields are decidedly \emph{not} tame in this limit. It is impossible to write $X_{\b x}^i \approx \1 + \i E_{\b x}^i \, \d A$, as any application of the operator $X_{\b x}^i$ will necessarily take the state out of the flat subspace. This does not mean, however, that the Gauss operators $G_{\b x}$ must stop being of form $\1 + \i \O_{\b x} \d A$, as needed to most simply satisfy \eqref{QED consistency conditions}. It just means that the operator $\O_{\b x}$ should not be interpreted as the divergence of a tame electric field.

Taking $g \rar 0$ in conventional QED \eqref{def H QED conv} is a different kettle of fish altogether. In fact, this very notation is sloppy. It insinuates that $g$ is taken to be much smaller than all other relevant scales like $1/N$ or $1/K$, but this is incompatible with the fact that the derivation of \eqref{def H QED conv} assumed that all states at a given $g$ were tame. Strictly speaking, then, the ``small $g$'' limit in the context of conventional QED must be taken with the understanding that $g$ still scales in a certain way with $N$ and/or other taming parameters.

Even with these caveats, it may be reasonable to postulate that the low-energy states of conventional QED at $g \rar 0$ have energies that are obtained by diagonalizing the fermion Hamiltonian
\bel{\label{H QED g0 second}
  \frac1{2g^2}
  \sum_{\b x \in \Mbb^\star}
    B(\b x)^2
   -\,
  \i \sum_{\b x \in \Mbb^\star}
    \overline \psi(\b x) \c D \psi(\b x).
}
Such an approximation will not be examined (or used) in this paper. Instead, here are some warnings about using it even if it is justified.

The effective fermion theories  \eqref{H QED g0 first} and \eqref{H QED g0 second} are vastly different. The former is a bona fide theory of fermions on the $N \times N$ lattice $\Mbb^\star$. Its Hilbert space is $4^{N^2}$-dimensional. It is parameterized by gauge fields $A_{\b x}^i$ that are flat but otherwise arbitrary integer multiples of $\d A$ on each link. The latter theory, on the other hand, is a theory of smeared fermion fields. Taking into account the smoothness constraints, these fermions effectively live on a $2k\_S \times 2k\_S$ lattice. Their Hilbert space is $4^{(2k\_S)^2}$-dimensional. This theory is parameterized by tame and spatially smooth flat gauge field configurations. These configurations may be labeled by $A_{\b x}^i$, just as in the theory \eqref{H QED g0 first}, but these labels must satisfy many requirements. In addition to being smooth, so that $A_{\b x + \b e_j}^i \approx A_{\b x}^i$, distinct gauge fields are labeled by $A_{\b x}^i$'s that are integer multiples of $\frac{2\pi}{2E\_S} \gg \d A$ while still obeying $|A_{\b x}^i| < A\_T \ll 1$.

One might try to access the na\"ive $g \rar 0$ limit of conventional QED by ``rescaling the fields,'' i.e.\ letting $A^i_g \equiv A^i/g$. This eliminates the explicit $g$ prefactors from $\nord{H\_{Maxwell}}$ and inserts a $g$ into the covariant derivative, which is now $D_i = \del_i + \i g q A^i_g$. Then $g \rar 0$ appears equivalent to simply setting $q = 0$. Appearances deceive, however. The $g$-dependence does not go away just because it is invisible in a choice of notation. The new gauge field eigenvalues $A_g^i$ are multiples $\d A/g$, which becomes large at $g \rar 0$.

However, these new variables $A_g^i$ are useful when $g$ is not \emph{too} small. In particular, they can be used when $g^2 = O(1/N)$, like in \eqref{scaling of g}. In this case the rescaled gauge field eigenvalues are multiples of $\d A/g \sim 1/N^{3/2}$. Since this is still tiny, with an appropriate choice of taming parameters the fields $A_g^i$ may still be regarded as tame, the electric field $E_g^i \equiv g E^i$ can be identified with the derivative w.r.t.~$A_g^i$, and $\nord{H\_{Maxwell}}$ can be viewed as a set of harmonic potentials, one for each momentum $\b k \in \Pbb\_S$. It is now reasonable to view the theory $\nord{H\_T}$ as a tame Maxwell theory coupled to a Dirac fermion via the term $g \overline \psi \gamma^i \psi A_g^i$.

This is also the parametric regime with the greatest prominence in the literature. It is only in this regime that the gauge coupling can be assigned an engineering dimension of $1/2$. This assumption has been the staple of essentially all studies of $(2+1)$D QED, as they invariably boil down to studying what happens when momenta get much smaller than the energy scale set by $g^2$. It is difficult to imagine that any of these works can be safely extended beyond the $g^2 \sim 1/N$ sliver of parameter space.

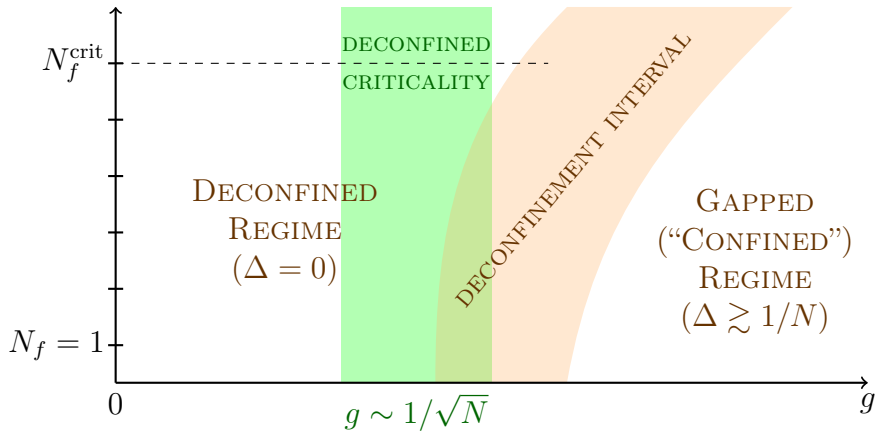
\begin{figure}
\begin{center}
\begin{tikzpicture}[scale = 1]
  \contourlength{1pt}

  \fill [green!30]
    (3, 0)
    to (5, 0)
    to (5, 5)
    to (3, 5)
    to cycle;
  \draw[green!40!black] (4, 0) node[below] {$g \sim 1/\sqrt N$};

  \fill [orange!60, opacity = 0.3]
    (4.25, 0)
    to (6, 0)
    to[out = 80, in = 225] (9, 5)
    to (6, 5)
    to[out = 225, in = 90] cycle;
  \draw[orange!40!black] (5.9, 3) node[below, align = center, rotate = 50] {\small \textsc{deconfinement interval}};

  \draw[green!40!black] (4, 4.25) node[align = center] {\footnotesize \textsc{deconfined} \\ \footnotesize \textsc{criticality}};

  \draw[orange!40!black] (8.5, 0.5) node[above, align = center]
    {\textsc{Gapped}\\
     \textsc{(``Confined'')}\\
     \textsc{Regime}\\
     ($\Delta \gtrsim 1/N$)
    };
  \draw[orange!40!black] (2.25, 2) node[align = center]
    {\textsc{Deconfined}\\
     \textsc{Regime}\\
     ($\Delta = 0$)
    };

  \draw[->, thick] (0, 0) -- (10, 0);
  \draw (10, 0) node[below] {$g$};
  \draw[->, thick] (0, 0) -- (0, 5);
  \draw (0, 0) node[below] {0};

  \foreach \x in {0.5, 1.25,..., 4.5}
    \draw[thick] (-0.1, \x) -- (0.1, \x);

  \draw[dashed] (0, 4.25) -- (5.75, 4.25);
  \draw[left] (0, 4.25) node {$N_f\^{crit}$};
  \draw[left] (0, 0.5) node {$N_f = 1$};

\end{tikzpicture}
\end{center}
\caption{\small A tentative phase diagram for conventional QED \eqref{def H QED conv} at different numbers $N_f$ of fermion flavors. The green region is where the gauge coupling has engineering dimension 1/2. The orange region denotes the crossover between the gapped, strongly interacting soup of fermions and photons (``confined regime'') and the ensemble of superselection sectors featuring generically gapless excitations (``deconfined regime''). The extreme, $g \rar 0$ side of the diagram is a ``free charge''  regime with free fermions moving in the background of flat, tame gauge fields. The $g \sim 1/\sqrt N$ part of the deconfined regime may be very different from a free fermion theory and is often called a ``deconfined critical phase'' \cite{Senthil:2004, Hermele:2004}. At $N_f = 0$ and $g \sim 1/\sqrt N$, the pure Maxwell theory is approximately gapless, as per Subsections \ref{subsec standard nc Maxwell} and \ref{subsec temp smoothing}. The critical fermion number $N_f\^{crit}$ is defined so that for $N_f \geq N_f\^{crit}$ the theory is deconfined for every dimension-$1/2$ gauge coupling $g$.
}
\label{fig deconf}
\end{figure}

The present analysis still leads to a few new remarks about physics within this sliver:
\begin{itemize}
  \item Recall the nontrivial steps that were used to extract the conventional QED Hamiltonian \eqref{def H QED conv} from a Dirac theory with a gauged fermion number. Since this Hamiltonian only involves degrees of freedom at low momenta, it should \emph{not} be considered to define a full-fledged cQFT. Indeed, this is why the conventional QED is not referred to as a ``cQED,'' as was done in Section \ref{sec Higgs} for scalar QED after smoothing. Due to its agnosticism about the dynamics of degrees of freedom at high momenta, one should not try to use conventional QED to calculate OPE coefficients or other data that reflects short-distance ``singularities.'' Conversely, to calculate OPEs in a well defined way, one should go back to the lattice theory \eqref{def H QED} and find an actual precontinuum basis for it.
  \item Despite these warnings, one may still ask about the spectrum of the Hamiltonian \eqref{def H QED conv}. There is a dearth of research on QED with an odd number of two-component fermions. Such theories do not have a notion of chiral symmetry whose breaking can be used to describe universal features of the ground state. Nevertheless, by blithely extrapolating the existing studies to the case of one fermion flavor, it is reasonable to assume that the low-energy spectrum of \eqref{def H QED conv} is not approximately linear but instead features a dispersion relation like \eqref{def omega tilde}, $\~\omega_{\b k} = \sqrt{\left(\frac{2\pi}{N} \right)^2 |\b k|^2 + \Delta^2(g)}$. If the gap is greater than the $g = 0$ level spacing, i.e.\ $\Delta(g) \gtrsim 1/N$, the theory can be said to be in a massive/gapped phase. More generally, the gapped condition may be written as $\Delta \equiv \lim_{\b k \rar 0} \~\omega_{\b k} \gtrsim 1/N$. This definition makes sense even in the absence of a chiral symmetry.

      It would now be natural to calculate the dispersion $\~\omega_{\b k}$ instead of just guessing it. Unfortunately, the length of this paper is already past all reasonable bounds, so this investigation will be reported elsewhere.
  \item It is also possible to use this language to discuss the nature of the ``confinement transition'' in conventional QED. This is the crossover that happens as $g$ is varied. As discussed above, at $g^2 \ll 1/N$ it is natural to assume that the theory breaks up into superselection sectors, with each sector describing fermions moving in a different flat background field. These sectors will generically have excitations with a linear dispersion (i.e.\ without a gap); this is certainly what happens in the sector labeled by $A_{\b x}^i = 0$. If the theory has gap $\Delta \gtrsim 1/N$ at $g^2 \sim 1/N$, there must exist a crossover between the free charge regime and the gapped regime. (When the crossover width vanishes at large $N$, it can be more properly called a transition.) Signatures of this crossover should be visible for couplings $g \sim 1/N^\alpha$ for all  $\alpha$ in some interval $I$ between $1/2$ and $-\infty$.

      Based on available evidence, it is further reasonable to assume that this ``deconfinement interval'' moves in $g$-space as the number of fermion flavors is increased. This way, the ``critical'' number of flavors is defined as that $N_f\^{crit}$ for which all couplings in the corresponding $I$ satisfy $g \gg 1/\sqrt N$. (See Fig.\ \ref{fig deconf}.) This paper will not try to estimate the infamous quantity $N_f\^{crit}$. This discussion merely illustrates how to frame the confinement transition in a way that does not require assuming any kind of symmetry breaking or even invoking RG notions.
\end{itemize}

\newpage

\section{Chern-Simons theory} \label{sec CS}

\subsection{Two roads to Chern-Simons}

There is no need to stress the importance of Chern-Simons (CS) theory in this day and age. Its Abelian version is given by the continuum Lagrangian
\bel{\label{def L CS}
  \L = \frac{\i \kappa}{4\pi} \epsilon^{\mu \nu \rho} A^\mu(x) \del_\nu A^\rho(x), \quad \kappa \in \Z.
}
The quadratic nature of the Lagrangian makes many calculations possible. Despite this, however, CS theory has never been defined in the finitary way advocated in this series. In other words, there is no available formulation of CS theory as a precise low-energy limit of a large but finite-dimensional $d = 2$ quantum system with an explicit Hamiltonian.

This claim may sound extraordinary. The following remarks should make it more palatable while simultaneously providing a brief overview of the relevant literature.

\begin{itemize}
  \item A well known approach to CS theory uses the Lagrangian \eqref{def L CS} to formally induce a lower-dimensional quantum theory that can often be given a precise Hamiltonian definition \cite{Witten:1988hf, Elitzur:1989nr}. When CS is placed on a closed three-manifold $\Ebb = \Sbb \times \Mbb$, this procedure results in a QM ($d = 0$) theory whose Hilbert space dimension depends on $\kappa$ and the topology and punctures of the spatial manifold $\Mbb$. When $\Mbb$ has a boundary, the resulting theory is a $d = 1$ CFT. These connections are remarkable, but they say little about an intrinsic $d = 2$ definition of CS that is independent of the global properties of the lattice it lives on.
  \item An enormous body of work studies rigorous constructions of CS and other topological QFTs using the language of cobordisms, group cohomology, modular tensor categories, etc (see \cite{Atiyah:1988, Reshetikhin:1991tc, Turaev:1992hq, Freed:1994ad, Baez:1995xq, Hansen:1997, Kapustin:2010ta, Walker:2011, Kong:2014} for a broad but incomplete selection of references). Some of these approaches may be understood in a finitary setting (as so-called ``state sums''), but they do not shed light on the Hamiltonian origin of CS theory.
  \item Some authors prefer to study na\"ive lattice actions of a CS form, with various tweaks that ensure that the resulting path integral is parity-odd, that it has the desired symmetries, or that its equations of motion give familiar commutation relations and gauge constraints \cite{Eliezer:1991qh, Alekseev:1994pa, Bietenholz:2002mt, Sun:2015hla}. Such approaches do not have regulated (finite-dimensional) gauge field variables, and just like state sums they do not obviously arise from any precisely defined Hamiltonian system. Using them is just as unsatisfactory as using the continuum BF theory to access the topological phase of the $\Z_K$ Maxwell theory \cite{Banks:2010zn}.
  \newpage
  \item A longstanding result in the study of conventional $(2 + 1)$D QED is that integrating out massive Dirac fermions gives rise to an effective CS action for the remaining gauge field \cite{Niemi:1983rq, Redlich:1983dv, Redlich:1983kn}. Indeed, this is the same phenomenon that underlies the integer quantum Hall effect, except there the gap comes from a classical magnetic flux background, not from an intrinsic fermion mass \cite{Thouless:1982zz}. Versions of this phenomenon have also been found in lattice gauge theory \cite{Coste:1989wf, Golterman:1992ub, Aoki:1993rg}. As seen in Subsection \ref{subsec Dirac fermions}, there is no obstacle to formulating a finitary definition of the massive Dirac fermion cQFT. With the exception of some recent work on CS-matter dualities \cite{Chen:2017lkr, Chen:2018vmz}, however, massive fermions do not seem to have been used to \emph{define} a finitary CS theory.
  \item It was only recently realized that gauge theories can be endowed with flux attachment directly on the lattice, by working with Hamiltonians whose local symmetries combine Gauss operators $G_{v}$ and Wilson loops $W_f$ in a consistent way \cite{Chen:2017fvr}. Such theories were shown to have CS-like actions, but the connection to continuum CS was never fleshed out.
\end{itemize}

At present, the last two bullet points appear to be the most promising approaches to a satisfactory definition of a CS theory, as understood in this paper. The rest of this Section will flesh out where both of them lead.

Perhaps surprisingly, it will turn out that the massive fermion approach can be used to define a CS theory \emph{only in a framework of temporally smooth path integrals}. Recall that temporal smoothing is a truncation performed on a microscopically defined path integral. It involves simply discarding the path integral variables at high Matsubara frequencies in order to make the integral actually doable. This procedure must be accompanied by the introduction of various additional terms into the action in order for the temporally smooth integral to reproduce the correct microscopic answer (or at least its universal part). It will here be shown that CS is one of the terms that must be added when smoothing fermion path integrals in order to reproduce the correct correlations of current operators. If no temporal smoothing is performed, the fermion path integral is guaranteed to give the correct answer on the nose, without the introduction of any CS terms.

Meanwhile, the flux-attached gauge theory will prove to have a CS-like action in the \emph{confined} phase. This will follow from the same argument that was used to derive the BF action \eqref{def S BF} from the topological phase of the $\Z_K$ gauge theory. In fact, the idea presented here can be used to show that there exist different microscopic theories that all lead to what one might call a CS action. This indicates that it is \emph{unacceptable} to be cavalier about the regularization of CS theory: theories with very different topological properties may give the same smoothed CS actions.

\newpage

\subsection{A toy example} \label{subsec toy}

The connection between CS and temporal smoothing is clearest in $(0+1)$D, where there are no distractions posed by spatial smoothing. Consider the massive fermion QM, an utterly trivial system with
\bel{
  H = m \psi\+ \psi, \quad \psi \equiv \bmat0100, \quad m \in \R.
}
The eigenstates of the Hamiltonian will be denoted by $\qvec 0$ and $\qvec 1$. Their energies are, respectively, $0$ and $m$. The ground state is $\qvec 0$ for $m > 0$, $\qvec 1$ for $m < 0$, and the two are degenerate at $m = 0$.

The analogue of the higher-dimensional fermion current operators is simply $j \equiv \psi\+ \psi$. Its finite-temperature expectation is
\bel{\label{1D avg j}
  \avg j_\beta = \frac 1 {1 + \e^{m \beta}}.
}
For any $\beta \in \R^+$, this interpolates between $\avg j_\beta = 0$ at $m \rar \infty$ and $\avg j_\beta = 1$ at $m \rar - \infty$.

The zero-temperature expectation can be written as
\bel{\label{1D avg j inf m}
  \avg j = \frac12 \big(1 - \sgn(m) \big).
}
Here it is assumed that $\avg j$ calculates the expectation of $j$ in the equal mix of all ground states, and by convention $\sgn(0) \equiv 0$. If $m \neq 0$, this result can be obtained from \eqref{1D avg j} by taking the temperature low enough so that $\beta \gg 1/|m|$. If $m = 0$, the zero-temperature limit is not meaningful. In this case, since $m$ is the only energy scale in $H$, $\avg j_\beta$ must be $\beta$-independent, and the zero-temperature result is obtained for any $\beta$.


Given some finite (possibly very large) number $\beta$, it is possible to express $\avg j_\beta$ as a Berezin integral,
\bel{\label{1D avg j path int}
  \avg j_\beta
   \approx
  \frac{1}\Zf
  \int \d\eta_{\d\tau} \d\bar\eta_{\d\tau} \cdots \d\bar\eta_\beta \,
  \bar\eta_{\d\tau} \eta_{\d\tau} \,
  \e^{
    \sum_{\tau = \d\tau}^{\beta}
      \d\tau \left[
        \bar\eta_\tau (\del \eta)_{\tau - \d\tau} - m\,\bar\eta_\tau \eta_\tau
      \right]
    },
}
where $\eta_0 \equiv - \eta_\beta$, and $\d \tau \equiv \beta/N_0$ for some large $N_0$. This expression for $\avg j_\beta$ is valid to leading order in $m\d\tau$. The derivation can be found by slightly  adapting the methods of \cite{Radicevic:1D}, which were focused on the time-ordered case of correlations like $\avg{\psi\+ \psi_\tau}_\beta$ for $\tau > 0$.

There are many interesting facts about this path integral, but the one that is important now is that the Grassmann fields at nearby times, like $\eta_\tau$ and $\eta_{\tau + \d\tau}$, are not ``close'' to each other. The situation is analogous to the one encountered in Subsection \ref{subsec temp smoothing} with bosonic fields in the path integral. Of course, Grassmann variables shouldn't be thought of as fields taking different values in some set. They are better thought as fixed matrices acting on an auxiliary vector space, and so the precise statement here is that $\eta_\tau$ and $\eta_{\tau + \d\tau}$ are linearly independent from each other.

This observation is particularly important when defining the Fourier transforms
\bel{\label{def eta n}
  \eta_\tau
   \equiv
  \frac1{\sqrt{N_0}}
  \sum_{n \in \Fbb}
    \eta_n\, \e^{\i \omega_n \tau},
  \quad
  \bar \eta_\tau
   \equiv
  \frac1{\sqrt{N_0}}
  \sum_{n \in \Fbb}
    \bar \eta_n\, \e^{- \i \omega_n \tau},
  \quad
  \Fbb
    \equiv
  \left\{
    -\frac12N_0, \ldots, \frac12N_0 - 1
  \right\},
}
with the usual fermionic Matsubara frequencies $\omega_n \equiv \frac{2\pi}{\beta}(n + \frac12)$. In frequency space, the action appearing in \eqref{1D avg j path int} is
\bel{
  S
   =
  \sum_{n \in \Fbb}
    \bar\eta_n \eta_n \left[
      \e^{- \i \omega_n \d\tau} - 1 + m \, \d \tau
    \right].
}
The crucial point here is that expanding the exponential is \emph{not} justified, even though $\d\tau$ is small. The reason is that at generic values of $n$, the frequency is $\omega_n \sim N_0/\beta = 1/\d\tau$. The high-frequency contributions are necessary for \eqref{1D avg j path int} to give the correct result.

To make progress it is thus necessary to define the temporally smoothed action
\algns{\label{1D action}
  \~S
   &\equiv
  \sum_{n \in \Fbb\_S}
    \bar\eta_n \eta_n \left[
      \e^{- \i \omega_n \d\tau} - 1 + m \, \d \tau
    \right]
   \approx
  \sum_{n \in \Fbb\_S}
    \bar\eta_n \eta_n
    \left(
      m - \i \omega_n
    \right)
    \d\tau \\
   &\equiv
  \sum_{\tau \in \Sbb}
    \d\tau  \,
    \bar\eta(\tau)
    \left(
      m - \del
    \right)
    \eta(\tau).
}
The smooth Grassmann fields $\eta(\tau)$ and $\bar\eta(\tau)$ involve only Matsubara frequencies from the restricted space $\Fbb\_S \equiv \{-n\_S, -n\_S + 1, \ldots, n\_S - 1\}$. Their smoothness relation can be recorded as e.g.
\bel{
  \eta(\tau + \d\tau) = \eta(\tau) + \d\tau \, \del\eta(\tau) + O\left(n\_S^2/N_0^2 \right).
}

The partition function $\~\Zf$ obtained from this action differs from the exact partition function by a factor proportional to $\e^{\beta m / 2}$. (This may be called the universal part of the proportionality factor; the remainder depends on $n\_S$ and $N_0$.) This means that the temporally smoothed Lagrangian  $\~\L \equiv \bar\eta(\tau) (m - \del) \eta(\tau)$ must be supplemented by the counterterm $\~\L\_{ct} \equiv - \frac12 m$ in order to ensure that universal parts of $\~\Zf$ and $\Zf$ match.

The U(1) phase rotation invariance of $\~S$ generates a Noether current given by
\bel{
  J(\tau) \equiv \bar\eta(\tau) \eta(\tau).
}
(See \cite{Radicevic:2D} for the Noether theorem in the context of temporally smooth lattice actions.) This Grassmann object is a smeared version of the canonical current at time $\tau$, which is given by
\bel{
  j_\tau \equiv \e^{-\tau H} j \, \e^{\tau H} = \psi_\tau\+ \psi_\tau.
}
CS theory lives in the difference between the expectations of these currents.

By time-translation invariance, $\avg{J(\tau)}_\beta \equiv \avg J_\beta$ is independent of $\tau$. It is given by
\bel{
  \avg{J}_\beta
   =
  \frac1{N_0}
  \sum_{n, m \in \Fbb\_S}
    \avg{\bar\eta_n \eta_m}_\beta
   \approx
  \frac1\beta
  \sum_{n \in \Fbb\_S}
    \frac1{\i \omega_n - m}.
}
At leading order in $\d\omega/|m| \equiv 2\pi/|m|\beta$, which is for now assumed to be small, this sum can be rewritten as an integral, giving
\algns{
  \avg{J}_\beta
   &\approx
  \sum_{n = 0}^{n\_S - 1}
    \frac{\d\omega}{2\pi}
    \left[\frac1{\i\omega_n - m} - \frac1{\i\omega_n + m} \right] \\
   &\approx
  \int_{\pi/\beta}^{\omega\_S}
    \frac{\d\omega}{2\pi}
    \frac{-2m}{\omega^2 + m^2}
   =
  \frac1\pi
  \left[
    \arctan \frac\pi{\beta m} - \arctan \frac{\omega\_S}{m}
  \right].
}
Assuming that $n\_S$ is large enough so that $\omega\_S \gg |m|$, or equivalently $n\_S \gg |m| \beta$, the correlation function can finally be written as
\bel{\label{1D avg J inf m}
  \avg{J}_\beta
   \approx
  - \frac12\, \sgn(m).
}
On the other hand, if $n\_S \ll |m|\beta$, the result is vanishingly small,
\bel{
  \avg{J}_\beta
   \approx
  - \frac{2n\_S}{m\beta}.
}

If $|m|\beta \lesssim 1$, the sum over frequencies cannot be approximated by an integral in this fashion. In particular, if $|m|\beta \ll 1$, the computation reduces to the sum $\sum_{n = 0}^{n\_S - 1} \frac1{(n + 1/2)^2} \approx \frac{\pi^2}2$. In this regime the expectation is
\bel{\label{1D avg J 0m}
  \avg{J}_\beta
   \approx
  -\frac{m\beta}4,
}
and is hence again negligibly small.

It is instructive to compare the results \eqref{1D avg J inf m} and \eqref{1D avg J 0m} to the canonical result \eqref{1D avg j}. At both large and small $|m|\beta$, the two expectations are related by
\bel{\label{1D avg j avg J}
  \avg j_\beta = \avg J_\beta + \frac12.
}
In fact, it is easy to numerically verify that this relation holds for all $|m|\beta \ll n\_S$. It is also clear that the two answers cannot agree beyond this, since the path integral answer $\avg J_\beta$ changes at $|m|\beta \sim n\_S$ while the canonical answer $\avg j_\beta$ knows nothing about the cutoff $n\_S$. This means that taking the low-temperature limit $|m|\beta \rar \infty$ in the (smoothed) path integral formalism is fraught with peril: at best, it can be done correctly only by taking $n\_S \rar \infty$ before $|m|\beta \rar \infty$.

The most remarkable fact about the relation \eqref{1D avg j avg J} is the universal \emph{contact term} $1/2$. The fact that the two answers differ by an additive term is unsurprising on its own. As argued in \cite{Radicevic:1D}, an expectation value of temporally smoothed fields, e.g.\ $\avg J_\beta = \avg{\bar\eta(\tau) \eta(\tau)}_\beta$, is a nontrivial smearing of the two-point function $\avg{\bar\eta_{\tau'} \eta_{\tau''}}_\beta$ over all values $\tau - \frac{\beta}{2n\_S} \leq \tau', \tau'' \leq \tau + \frac{\beta}{2n\_S}$. The diagonal terms $\tau' = \tau''$ all contribute coherently to this smearing, but the off-diagonal terms are harder to analyze. The primary difficulty is that $(\tau', \tau'')$ and $(\tau'', \tau')$ cannot both be time-ordered, and hence the path integral correlators $\avg{\bar\eta_{\tau'}\eta_{\tau''}}_\beta$ and $\avg{\bar\eta_{\tau''}\eta_{\tau'}}_\beta$ correspond to canonical correlation functions $\avg{\psi_{\tau'}\+ \psi_{\tau''}}_\beta$ and $-\avg{\psi_{\tau'} \psi_{\tau''}\+}_\beta$. Thus $\avg J_\beta$ calculates a complicated smearing of canonical correlation functions weighted with $\sgn(\tau' - \tau'')$. It is therefore noteworthy that the entire effect of this temporal smearing is to shift $\avg j_\beta$ by $1/2$, regardless of other physical parameters (as long as $|m|\beta \ll n\_S$).

Another interesting empirical observation is that the relation \eqref{1D avg j avg J} has no multiplicative prefactor relating the two expectation values. This is in contrast to the relation between partition functions $\Zf$ and $\~\Zf$, which required a finite counterterm to be inserted into the Lagrangian $\~\L$ just to ensure the universal parts of the partition functions matched.

Now, what do all these observations have to do with CS theory? In $(0+1)$D, the CS action has a Lagrangian given by
\bel{\label{1D CS}
  \~\L\_{CS} = \i \kappa A(\tau),
}
where $A(\tau)$ is a smooth real field.  It is convenient to introduce it by generalizing the action \eqref{1D action} to
\bel{\label{1D gen fnal}
  \~S[A]
   =
  \sum_{\tau \in \Sbb} \d\tau\, \left[
    \bar\eta(\tau) \big(m - \del + \i A(\tau)\big) \eta(\tau) + \i \kappa A(\tau)
  \right].
}
As the notation suggests, $A(\tau)$ is a kind of gauge field. Though this will not be used here, $\~S[A]$ is invariant under smooth gauge transformations defined by a smooth single-valued field $\lambda(\tau)$,
\bel{
  \eta(\tau) \mapsto \e^{\i \lambda(\tau)} \eta(\tau),\quad
  \bar\eta(\tau) \mapsto \e^{-\i \lambda(\tau)} \bar\eta(\tau),  \quad
  A(\tau) \mapsto A(\tau) + \del\lambda(\tau).
}

Let $\~\Zf[A]$ be the partition function obtained by integrating out fermions with this action. With this definition the canonical expectation value is given by the formal derivative
\bel{
  \avg {j}_\beta
   =
  \frac\i{\d\tau} \left[ \fder{}{ A(\tau)} \log \~\Zf[A] \right]_{A(\tau) = 0}, \quad \trm{with} \quad
  \kappa = \frac 12.
}
The CS term thus serves to capture the contact term that the na\"ive temporally smooth correlator would not know of. (See \cite{Closset:2012vp} for a similar take in 3D.) Note that it is possible to replace the formal functional derivatives with discrete differences by choosing an appropriate target space for $A(\tau)$. There is, however, nothing in this derivation that forces $A(\tau)$ or $\lambda(\tau)$ to be angular variables, and hence there is no a priori reason for $\kappa$ to be ``quantized.''

\subsection{CS from massive fermions} \label{subsec CS mass ferm}

The previous Subsection has shown that, in $(0 + 1)$D, the expectation of the microscopic current operator in a fermion theory can be computed from a temporally smoothed path integral. The only caveat was that, due to temporal smoothing, the path integral result had to be shifted by $1/2$. This shift was tantamount to including a $\kappa = 1/2$ CS term into the generating functional for this current operator. The infinite mass result \eqref{1D avg J inf m} indicates that integrating out the fermion fields in the generating functional gives rise to an effective CS term that shifts the starting level $\kappa$ by $\Delta \kappa = -\frac12\, \sgn(m)$.

A dramatic way to interpret this result is to say that CS theory exists \emph{only} within temporally smooth path integrals. After all, the canonical formalism gives \emph{the} correct answer for $\avg j_\beta$. It was only the path integral that had to play catch-up by including an ad hoc CS term in the effective action.

Here is a plausible conservative reaction to this dramatic statement: maybe the $\kappa = 1/2$ CS Lagrangian that gives rise to the $1/2$ shift in $\avg J_\beta$ has no analogue in the canonical formalism, but surely the notion of integrating out a fermion and generating a $\Delta \kappa$ CS term can be given currency in the Hamiltonian formalism? Concretely, say that you couple the fermion QM to a tame gauge field $a$ (i.e.\ a position operator in a bosonic QM), with the Hamiltonian containing a term of the form $j a$. Would it not be possible to integrate out the fermion and find a canonical analogue of the CS term for $a$?

Unfortunately, this proposal does not work. One argument is that, for a given $a$, the term $ja$ in the Hamiltonian can simply be canceled by a shift of the mass, $m \mapsto m - a$. This effectively decouples the fermion and the boson. And if the gauge field is small, then shifting the large number $m$ by this amount would not change the expectation of $\avg j_\beta$, which only depends on the sign of $m$. Thus, from a canonical perspective, this kind of coupling cannot give rise to an interesting effective theory.

Another argument against such a proposal is that there is no coupling that can be added to the Hamiltonian in order to get $\i a(\tau) J(\tau)$ in the Lagrangian, which is what \eqref{1D gen fnal} requires. In other words, this is a coupling that may appear in a generating functional, but not in an action that follows from a Hamiltonian.

This is strong evidence that $(0+1)$D CS is germane to temporally smooth path integrals. In other words, the $\Delta \kappa$ shift is nontrivially encoded in the canonical formalism, mainly by the simple fact that the ground state changes with the change of $\sgn(m)$.

The situation appears slightly different in $(2 + 1)$D. As shown in Subsection \ref{subsec conv QED}, here there are nontrivial couplings between a gauge field and a fermion. Nevertheless, it will now be shown that the story is mostly analogous to the $(0+1)$D one: a CS action appears in the path integral to make it correctly reproduce the universal parts of current-current correlators.

The focus in the massive $(2+1)$D Dirac theory \eqref{def H ferm massive} will be on the three-component current
\bel{\label{def j}
  j^\mu_{\b x}
   =
  - \overline\psi_{\b x} \gamma^\mu \psi_{\b x}
   \equiv
  -  (\sigma^\mu)^{\alpha\beta} (\psi^\alpha_{\b x})\+ \psi_{\b x}^\beta,
}
where a convenient choice of $\sigma$-matrices is
\bel{
  \sigma^0 \equiv \bmat{-1}{}{}{-1}, \quad
  \sigma^1 \equiv \bmat{1}{}{}{-1}, \quad
  \sigma^2 \equiv \bmat{}{1}{1}{}, \quad
  \sigma^3 \equiv \bmat{}{-\i}{\i}{}.
}
One complication that did not exist in $(0+1)$D is the need for spatial smooting. The current \eqref{def j} is
\bel{
  j_{\b x}^\mu
   =
  -\frac1{N^2}
  \sum_{\b k, \b l \in \Pbb}
    \overline \psi_{\b l} \gamma^\mu \psi_{\b k} \,
    \e^{\frac{2\pi\i}N \b x(\b k - \b l)}
   =
  -\frac1{N^2}
  \sum_{\b k, \b l \in \Pbb}
    \overline \psi_{\b l} \gamma^\mu \psi_{\b k + \b l} \,
    \e^{\frac{2\pi\i}N \b x \b k},
}
and so it is natural to define the momentum-space current
\bel{
  j_{\b k}^\mu
   \equiv
  - \sum_{\b l \in \Pbb}
    \overline \psi_{\b l} \gamma^\mu \psi_{\b k + \b l}
   =
  - (\sigma^\mu)^{\alpha\beta}
  \sum_{\b l \in \Pbb}
    (\psi^\alpha_{\b l})\+ \psi^\beta_{\b k + \b l}.
}
For any $\b k \neq 0$, projecting this operator onto the smooth algebra gives the field
\bel{\label{def j smooth}
  j^\mu(\b k)
   \equiv
  - \sum_{\b l \in \Pbb\_S(\b k)}
    \overline \psi_{\b l} \gamma^\mu \psi_{\b k + \b l},
}
where $\Pbb\_S(\b k) \subset \Pbb\_S$ is the set of momenta $\b l$ in $\Pbb\_S$ whose shifts $\b l + \b k$ also belong to $\Pbb\_S$. If $\b k = 0$, the smoothing does not actually restrict the sum over $\b l$, so it makes sense to set $\Pbb\_S(0) \equiv \Pbb$.

An analogous smooth current can be defined when studying fermions in $(1+1)$D. In \cite{Radicevic:2019jfe} it was shown that the fact $\Pbb\_S(\b k) \neq \Pbb\_S$ precisely implies that commutators of $j^\mu(\b k)$ form a Kac-Moody structure. (Various versions of this observation date back to the seminal work of Lieb and Mattis \cite{Mattis:1964wp}.) In the case at hand, this $\b k$-dependence will not be crucial, but it will still be important that the momenta in $j^\mu(\b k)$ are limited by $k\_S$.

The quantity of greatest interest here is the current-current correlator. But before tackling this, consider the expectation value of the smooth current,
\bel{\label{3D avg j}
  \avg{j^\mu(\b k)}
   =
  - (\sigma^\mu)^{\alpha\beta}
  \sum_{\b l \in \Pbb\_S(\b k)}
    \avg{(\psi^\alpha_{\b l})\+ \psi^\beta_{\b k + \b l}}.
}
This is the analogue of the quantity $\avg j$ considered in Subsection \ref{subsec toy}. The correlator can be evaluated by transforming to the precontinuum basis in which the Hamiltonian is diagonal. In the massless theory, this was achieved by the map \eqref{def Psi k}. It will be useful to carefully generalize this map to the massive theory.

\newpage

The precontinuum generators in the massive theory are
\bel{
  \Psi_{\b k}^\alpha \equiv (U_{\b k})_{\alpha\beta} \, \psi_{\b k}^\beta
}
for
\bel{\label{def U k}
  U_{\b k}
   \equiv
  \bmat
    { \cos\frac{\theta_{\b k}}2 }
    { \e^{-\i\vartheta_{\b k}} \left|\sin\frac{\theta_{\b k}}2\right| }
    { - \left|\sin\frac{\theta_{\b k}}2\right| }
    { \e^{-\i\vartheta_{\b k}} \cos\frac{\theta_{\b k}}2 }.
}
Generalizing \eqref{def theta}, the angles $\theta_{\b k}$ and $\vartheta_{\b k}$ are given by
\gathl{
  \cos\theta_{\b k}
   \equiv
  \frac{2}{\~\omega_{\b k}} \sin\frac{\pi k^1}N
   \approx
  \frac {\frac{2\pi}N k^1} {\sqrt{\left(\frac{2\pi}N k^1\right)^2 + \left(\frac{2\pi}N k^2\right)^2 + m^2}},
   \\
  \e^{-\i\vartheta_{\b k}}
   \equiv
  \frac {2\sin\frac{\pi k^2}N - \i m} {\sqrt{4\sin^2 \frac{\pi k^2}N + m^2}}
   \approx
  \frac {\frac{2\pi}N k^2 - \i m} {\sqrt{\left(\frac{2\pi}N k^2\right)^2 + m^2}}.
}
The approximations hold in the smooth subspace to leading order in $k\_S/N$. At $m = 0$, these relations reduce to \eqref{def theta}. At $|m| \rar \infty$, or more precisely at $|m| \gg k\_S/N$, the matrix $U_{\b k}$ becomes
\bel{
  U_{\b k}
   \approx
  \frac1{\sqrt 2}
  \bmat 1 {-\i \, \sgn(m)} {-1} {-\i \, \sgn(m)},
}
which is in agreement with \eqref{def massive fermions}. For any nonzero  momentum, the ground state is characterized by
\bel{
  \avg{(\Psi^\alpha_{\b k})\+ \Psi^\beta_{\b k}} = \delta_{\alpha, -} \delta_{\beta, -}.
}
For simplicity, it may be assumed that the $\b k = 0$ modes were chosen so that this holds there too. These zero-modes have no effect on the rest of this story.

This means that the correlator \eqref{3D avg j} is
\algns{
  \avg{j^\mu(\b k)}
   &=
  -
  (\sigma^\mu)^{\alpha\beta}
  \sum_{\b l \in \Pbb\_S(\b k)}
  (U_{\b l})_{\alpha'\alpha}
  (U\+_{\b l + \b k})_{\beta\beta'}
    \avg{(\Psi^{\alpha'}_{\b l})\+ \Psi^{\beta'}_{\b k + \b l}}  \\
   &=
  - \delta_{\b k, 0} \sum_{\b l \in \Pbb}
  \left(U_{\b l} \sigma^\mu U_{\b l}\+ \right)_{--}
   \stackrel{|m| \rar \infty}\approx
  \delta_{\b k, 0} \delta_{\mu, 0} N^2.
}
There is no nontrivial physics in this expectation value, compared to the $(0+1)$D case, eq.\ \eqref{1D avg j inf m}, with its delicate dependence on the sign of $m$. This is not surprising, given that translation invariance requires the $\b k \neq 0$ correlator to vanish. The value $N^2$ is simply the number of fermions in the massive ground state.  Nevertheless, this is a good example of correlation function computation on which to cut one's teeth.

With this preparation, consider the correlator
\algns{\label{avg jj}
  \avg{j^\mu(\b k) j^\nu(-\b k)}
   &=
  \sum_{\substack{\b l \in \Pbb\_S(\b k)\\ \b l' \in \Pbb\_S(-\b k)}}
    \avg{\overline \psi_{\b l} \gamma^\mu \psi_{\b l + \b k}
         \overline \psi_{\b l'} \gamma^\nu \psi_{\b l' - \b k}} \\
   &=
  \sum_{\substack{\b l \in \Pbb\_S(\b k)\\ \b l' \in \Pbb\_S(-\b k)}}
    \left(U_{\b l} \sigma^\mu U_{\b l + \b k}\+ \right)_{\alpha\beta}
    \left(U_{\b l'} \sigma^\nu U_{\b l' - \b k}\+ \right)_{\alpha'\beta'}
    \avg{ (\Psi^\alpha_{\b l})\+ \Psi^\beta_{\b l + \b k}
          (\Psi^{\alpha'}_{\b l'})\+ \Psi^{\beta'}_{\b l' - \b k} }.
}
Some simple fermion gymnastics shows that
\bel{
  \avg{ (\Psi^\alpha_{\b l})\+ \Psi^\beta_{\b l + \b k}
          (\Psi^{\alpha'}_{\b l'})\+ \Psi^{\beta'}_{\b l' - \b k} }
   =
  \delta_{\alpha, -} \delta_{\beta, +} \delta_{\alpha', +} \delta_{\beta', -}
  \delta_{\b l + \b k, \b l'}
   +
  \delta_{\alpha, -} \delta_{\beta, -} \delta_{\alpha', -} \delta_{\beta', -}
  \delta_{\b k, 0}.
}
The interesting physics will lie at $\b k \neq 0$. Assuming this, the correlator becomes the simple expression
\bel{
  \avg{j^\mu(\b k) j^\nu(-\b k)}
   =
  \sum_{\b l \in \Pbb\_S(\b k)}
    \left(U_{\b l} \sigma^\mu U_{\b l + \b k}\+ \right)_{-+}
    \left(U_{\b l + \b k} \sigma^\nu U_{\b l}\+ \right)_{+-}.
}
By rearranging the matrix multiplication this can be brought to a more familiar form,
\bel{
  \avg{j^\mu(\b k) j^\nu(-\b k)}
   =
  \sum_{\b l \in \Pbb\_S(\b k)}
    \trm{tr} \left[G^-_{\b l} \sigma^\mu G^+_{\b l + \b k} \sigma^\nu \right],
}
where the ``propagators'' are defined as
\bel{
  (G^\pm_{\b k})_{\alpha \beta}
   \equiv
  (U_{\b k}\+)_{\alpha \pm} (U_{\b k})_{\pm \beta}.
}
This is reminiscent of the one-loop Feynman diagram for the above correlation function. However, the two calculations should not be conflated: this one is done in the canonical formalism, the sum runs over spatial momenta only, and the answer is exact without any need to include further loop corrections.

The propagators turn out to have a very simple form for any $m$. They are given by
\bel{
  G^\pm_{\b k}
   =
  \frac1{2 \~\omega_{\b k}} \left[
    -\~\omega_{\b k}\sigma^0
     \pm
    \left( \frac{2\pi}N k^1 \sigma^1 + \frac{2\pi}N k^2 \sigma^2 + m \sigma^3 \right)
  \right].
}
This expression has some far-reaching implications. Observe that the only purely imaginary $\sigma$-matrix is $\sigma^3$. Thus the imaginary part of $\avg{j^\mu(\b k) j^\nu(-\b k)}$ must be assembled by picking a $\sigma^3$ term from one propagator and a $\sigma^\lambda$ term from the other propagator, with $\lambda \in \{0, 1, 2\}$ being a spacetime index. For the trace to be nonzero, the indices $\mu$ and $\nu$ must then be different from $\lambda$ and from each other.

Concretely, this means that
\algns{\label{j0j1}
  \trm{Im} \avg{j^0(\b k) j^1(-\b k)}
   &=
  \sum_{\b l \in \Pbb\_S(\b k)}
   \frac {\i m}{4\~\omega_{\b l} \~\omega_{\b l + \b k}}
   \frac{2\pi}N
   \left(
     (l^2 + k^2) \trm{tr}\left[\sigma^3 \sigma^0 \sigma^2 \sigma^1 \right]
     +
     l^2\, \trm{tr}\left[\sigma^2 \sigma^0 \sigma^3 \sigma^1 \right]
   \right)\\
   &=
  -\frac{2\pi k^2}N
  \sum_{\b l \in \Pbb\_S(\b k)}
    \frac {m}{2\~\omega_{\b l} \~\omega_{\b l + \b k}}
   \stackrel{|m| \rar \infty}\approx
  -\sgn(m) \frac{2\pi k^2}N
  \frac{\trm{Vol} \left\{\Pbb\_S(\b k)\right\}}{2|m|},
}
and similarly
\algnl{\label{j2j0}
  \trm{Im} \avg{j^2(\b k) j^0(-\b k)}
   &=
  -\frac{2\pi k^1}N
  \sum_{\b l \in \Pbb\_S(\b k)}
    \frac {m}{2\~\omega_{\b l} \~\omega_{\b l + \b k}}
   \stackrel{|m| \rar \infty}\approx
  -\sgn(m)
  \frac{2\pi k^1}N
  \frac{\trm{Vol} \left\{\Pbb\_S(\b k)\right\}}{2|m|},
   \\ \label{j1j2}
  \trm{Im} \avg{j^1(\b k) j^2(-\b k)}
   &=
  -
  \sum_{\b l \in \Pbb\_S(\b k)}
    \frac {m(\~\omega_{\b l} + \~\omega_{\b l + \b k})}{2\~\omega_{\b l} \~\omega_{\b l + \b k}}
   \stackrel{|m| \rar \infty}\approx
  -\sgn(m)
  \, 2|m|
  \frac{\trm{Vol} \left\{\Pbb\_S(\b k)\right\}}{2|m|}.
}
The other choices of $\mu \neq \nu$ are related to the above ones by a sign change.

These results deserve some comment. To start, note that each of them has been represented as a product of three terms. The first is the familiar mass-dependent overall sign. The second is a component of the momentum three-vector, which in the massive limit ($|m| \gg k\_S/N$) is
\bel{
  \~k^\mu \equiv \left(2|m|, \frac{2\pi}N k^1, \frac{2\pi}N k^2 \right),
}
as appropriate for two massive particles moving at a total momentum $\b k$. And the third term is a large number when $k^i \ll k\_S$, a ``regulated UV divergence'' approximately equal to the kinetic energy-like quantity $\E\_S \equiv \frac1{2|m|}(2k\_S)^2$. Thus the general result for the imaginary part of the current-current correlator in the very massive theory is
\bel{\label{canonical contact term}
  \frac1{\E\_S} \trm{Im}\avg{j^\mu(\b k) j^\nu(-\b k)}
   =
  - \sgn(m)\, \epsilon^{\mu\nu\lambda} \~k^\lambda.
}
The r.h.s.\ is the ``parity-odd'' structure of the current two-point function that is familiar from the early continuum studies of $(2 + 1)$D fermions \cite{Niemi:1983rq, Redlich:1983dv, Redlich:1983kn}. The normalization factor on the l.h.s.\ is less familiar, but it can be eliminated by a rescaling of the momentum-space operators $j^\mu(\b k)$. Such rescalings, as shown in Subsection \eqref{subsec temp smoothing}, are needed to turn smooth canonical operators into familiar continuum fields.

It takes little effort to generalize \eqref{avg jj} to a correlator at different Euclidean times,
\bel{\label{avg jj tau}
  \avg{j^\mu_\tau(\b k) j^\nu_{\tau'}(-\b k)}_\beta
   =
  \frac{\e^{-\beta \E_0}}{\Zf}
  \sum_{\b l \in \Pbb\_S(\b k)}
    \trm{tr} \left[G^-_{\b l} \sigma^\mu G^+_{\b l + \b k} \sigma^\nu \right]
    \e^{- \left(\~\omega_{\b l} + \~\omega_{\b l + \b k}\right)(\tau' - \tau)} ,
}
where $\E_0$ is the ground state energy. This expression will be used later.

\newpage

Now consider the analogous calculation in the temporally smoothed path integral formalism. The smoothed action for $\b k \in \Pbb\_S$ modes is
\algns{
  \~S
   &=
  \sum_{\substack{\b k \in \Pbb\_S \\ \tau \in \Sbb}}
    \d\tau \,
    \bar\eta_k^\alpha(\tau)
    \left[
      \sigma^0 \del_0 + \frac{2\pi}N k^i \sigma^i + m\sigma^3
    \right]^{\alpha\beta}
    \eta_k^\beta(\tau) \\
   &=
  \sum_{\substack{\b k \in \Pbb\_S \\ n \in \Fbb\_S}}
    \d\tau\,
    \bar\eta_{k, n}^\alpha
    \left[
      \i \omega_n \sigma^0  + \frac{2\pi}N k^i \sigma^i + m\sigma^3
    \right]^{\alpha\beta}
    \eta_{k, n}^\beta.
}
The Noether current is
\bel{
  J^\mu(\b x, \tau)
   \equiv
  - \bar\eta^\alpha(\b x, \tau) (\sigma^\mu)^{\alpha\beta} \eta^\beta(\b x, \tau).
}
The smooth momentum space current analogous to \eqref{def j smooth} is given by
\bel{
  J^\mu(\b x, \tau)
   \equiv
  \frac1{N^2N_0}
  \sum_{\b k, n}
    J^\mu(\b k, n) \,
    \e^{\frac{2\pi\i}N \b k \b x + \frac{2\pi\i}\beta m \tau },
   \quad
  J^\mu(\b k, n)
   =
  - \!\!\sum_{\substack{\b l \in \Pbb\_S(\b k) \\ m \in \Fbb\_S(n)}}\!\!
    \bar\eta_{\b l, m}^\alpha
    (\sigma^\mu)^{\alpha\beta}
    \eta_{\b l + \b k, m + n}^\beta,
}
The first sum runs over all momenta with $-2k\_S \leq k^i < 2k\_S$ and frequencies with $-2n\_S \leq n  < 2n\_S$. In the sum over $\b l$ and $m$,  $\Fbb\_S(n)$ is a subset of $\Fbb\_S$ such that, for each $m \in \Fbb\_S(n)$, $m + n$ belongs to $\Fbb\_S$. This is the current that is usually used in path integral calculations.

The correlator $\avg{j^\mu(\b k) j^\nu(- \b k)}$ was evaluated at equal times in \eqref{avg jj}, so it should be compared to
\bel{\label{avg JJ}
  \avg{J^\mu(\b k) J^\nu(-\b k)},
}
where
\bel{
  J^\mu(\b k)
   \equiv
  J^\mu(\b k, \tau = 0)
   =
  \sum_{\b x \in \Mbb^\star}
    J^\mu(\b x, 0)\, \e^{-\frac{2\pi}N \b k \b x}
   =
  - \frac1{N_0}
  \sum_{\substack{\b l \in \Pbb\_S(\b k) \\ n, m \in \Fbb\_S}}
    \bar\eta_{\b l, n}^\alpha
    (\sigma^\mu)^{\alpha\beta}
    \eta_{\b l + \b k, m}^\beta.
}
As in $(0 + 1)$D, the fact that all frequencies in this sum remain in $\Fbb\_S$ means that $J^\mu(\b k)$ is smeared in the temporal direction and hence contains Grassmann variables of both possible time-orderings. This is ultimately why nontrivial contact terms may be needed to make \eqref{avg JJ} agree with \eqref{avg jj}.

Evaluating \eqref{avg JJ} using path integral techniques is straightforward --- it is essentially the usual derivation of Feynman rules for fermions. Substituting $\eta^\alpha_{\b k, n} \equiv (U_{\b k}\+)_{\alpha\beta} {\eta'}^\beta_{\!\b k, n}$, with the same $U_{\b k}$ as in \eqref{def U k}, diagonalizes the kernel in the action $\~S$ and gives
\bel{
  \~S
   =
  \sum_{\substack{\b k \in \Pbb\_S \\ n \in \Fbb\_S}}
  \sum_{\alpha \in \{\pm\}}
    \d\tau\,
    {\bar \eta'}_{\b k}{}^{\!\alpha}_{\!,\, n}
    \left[
      -\i \omega_n  + \alpha \~\omega_{\b k}
    \right]
    {\eta'}_{\!\b k, n}^\alpha.
}

Converting to the $\eta'$ variables in the path integral expression for \eqref{avg JJ} and performing the integrals eventually gives
\algnl{\notag
  \avg{J^\mu(\b k) J^\nu(-\b k)}_\beta
   &=
  \frac1{\~\Zf N_0^2}
  \sum_{\substack{\b l \in \Pbb\_S(\b k) \\ n, m \in \Fbb\_S}}
  \sum_{\substack{\b l' \in \Pbb\_S(-\b k) \\ n', m' \in \Fbb\_S}}
  \int [\d\eta\d\bar\eta] \,
    \bar\eta_{\b l, n}^\alpha
    (\sigma^\mu)^{\alpha\beta}
    \eta_{\b l + \b k, m}^\beta
    \bar\eta_{\b l', n'}^{\alpha'}
    (\sigma^\nu)^{\alpha'\beta'}
    \eta_{\b l' - \b k, m'}^{\beta'}
    \, \e^{- \~S} \\ \notag
   &=
  \frac1{\beta^2}
  \sum_{\substack{\b l \in \Pbb\_S(\b k) \\ n, m \in \Fbb\_S}}
  \sum_{\alpha, \beta \in \{\pm\}}
    \frac1{\alpha \~\omega_{\b l} - \i \omega_n}
    (U_{\b l} \sigma^\mu U\+_{\b l + \b k})_{\alpha\beta}
    \frac{-1}{\beta \~\omega_{\b l + \b k} - \i \omega_m}
    (U_{\b l + \b k} \sigma^\nu U\+_{\b l})_{\beta\alpha} \\ \label{JJ calc}
     &=
  -\frac1{\beta^2}
  \sum_{\substack{\b l \in \Pbb\_S(\b k) \\ n, m \in \Fbb\_S}}
  \sum_{\alpha, \beta \in \{\pm\}}
  \tr\left[
    \frac{G_{\b l}^\alpha}{\alpha \~\omega_{\b l} - \i \omega_n}
    \sigma^\mu
    \frac{G^\beta_{\b l + \b k}}{\beta \~\omega_{\b l + \b k} - \i \omega_m}
    \sigma^\nu
  \right] \\ \notag
   &\hspace{-2em} =
  -\frac1{\beta^2}
  \sum_{\substack{\b l \in \Pbb\_S(\b k) \\ n, m \in \Fbb\_S}}
  \tr\left[
    \frac
      {-\i \omega_n \sigma^0 + \frac{2\pi}N l^i \sigma^i + m \sigma^3}
      {\~\omega_{\b l}^2 + \omega_n^2}
    \sigma^\mu
    \frac
      {-\i \omega_m \sigma^0 + \frac{2\pi}N (l^i + k^i) \sigma^i + m \sigma^3}
      {\~\omega_{\b l + \b k}^2 + \omega_m^2}
    \sigma^\nu
  \right].
}
Note that the sum over $\alpha$'s converts the ``canonical'' propagators $G_{\b l}^\alpha/(\alpha\~\omega_{\b l} - \i \omega_n)$ into conventional spacetime ones. The final result is, up to conventions for $\sigma$-matrices, precisely the expression one would have gotten by evaluating a one-loop Feynman diagram.

Performing the sums over Matsubara frequencies reduces the above expression to one that is readily compared to the sums in eqs.\ \eqref{j0j1}--\eqref{j1j2}. Assuming that $\omega\_S \equiv \omega_{n\_S} \gg \~\omega_{\b k}$ for all $\b k$, these sums give
\bel{
  \avg{J^\mu(\b k) J^\nu(-\b k)}_\beta
   \approx
  -
  \sum_{\b l \in \Pbb\_S(\b k)}
  \tr\left[
    \frac
      {\frac{2\pi}N l^i \sigma^i + m \sigma^3}
      {2\~\omega_{\b l}}
    \sigma^\mu
    \frac
      {\frac{2\pi}N (l^i + k^i) \sigma^i + m \sigma^3}
      {2\~\omega_{\b l + \b k}}
    \sigma^\nu
  \right].
}
It now becomes apparent how the path integral result will differ from the canonical one. The lack of a $\sigma^0$ matrix in the numerators means that the imaginary part of this expression necessarily vanishes. In fact, even without asking whether the answer is real or imaginary, it is evident that a $\epsilon^{\mu\nu\lambda} \~k^\lambda$ structure cannot be contained in this answer. This means the na\"ive path integral knows \emph{nothing} about the contact term \eqref{canonical contact term}. This whole structure has to come from an ad-hoc counterterm in the generating functional.

An alternative computation in the path integral formalism focuses on the correlator
\algns{\label{avg JJ conv}
  \avg{J^\mu(\b k, n) J^\nu(-\b k, -n)}_\beta
   &= \\
   &\hspace{-11em} =
  -\frac1{(\d \tau)^2}
  \sum_{\substack{\b l \in \Pbb\_S(\b k) \\ m \in \Fbb\_S(n)}}
  \!\!\tr\left[
    \frac
      {-\i \omega_m \sigma^0 + \frac{2\pi}N l^i \sigma^i + m \sigma^3}
      {\~\omega_{\b l}^2 + \omega_m^2}
    \sigma^\mu
    \frac
      {-\i \omega_{m + n} \sigma^0 + \frac{2\pi}N (l^i + k^i) \sigma^i + m \sigma^3}
      {\~\omega_{\b l + \b k}^2 + \omega_{m + n}^2}
    \sigma^\nu
  \right].
}
Summing over all $-2n\_S \leq n < 2n\_S$ and rescaling by $N_0^2$ gives back the sum \eqref{JJ calc}.

A quick inspection shows that,  unlike \eqref{avg JJ},  \eqref{avg JJ conv} will have a term of form $\epsilon^{\mu\nu\lambda} \~k^\lambda$. When $\mu, \nu \neq 0$, this term will be real, because the factor of $\i$ by one of the $\sigma^0$ matrices will be multiplied by a factor of $\i$ inside one of the $\sigma^3$ matrices. This can be understood as a consequence of the Euclidean nature of this path integral. (Note that the canonical calculation of the correlation function \eqref{avg jj} was done on a single time slice and is independent on whether time is Euclidean or not.)

The actual calculation of the sum over $\b l$ and $m$ reduces to a familiar integral presented in most reviews of CS theory \cite{Dunne:1998qy}. There is nevertheless a small subtlety here that is not often emphasized. For any choice of $\mu \neq \nu$, the correlator \eqref{avg JJ conv} becomes proportional to
\bel{
  \sum_{\substack{\b l \in \Pbb\_S(\b k) \\ m \in \Fbb\_S(n)}}
    \frac1{\~\omega_{\b l}^2 + \omega_{m}^2}
    \frac1{\~\omega_{\b l + \b k}^2 + \omega_{m + n}^2}
}
By assembling $\omega_m$ and $\frac{2\pi}N \b l$ into a three-vector $l^\mu$, and likewise $\omega_n$ and $\frac{2\pi}N \b k$ into $k^\mu$, this sum can be rewritten as
\bel{
  \sum_{l^\mu \in \Pbb\_S(\b k) \times \Fbb\_S(n)}
    \frac1{m^2 + l^2}\frac1{m^2 + (l + k)^2}
   \approx
  \beta N^2
  \int_{\Pbb\_S \times \Fbb\_S} \frac{\d^3 l}{(2\pi)^3}
    \frac1{m^2 + l^2}\frac1{m^2 + (l + k)^2}.
}
In the last step it was assumed that $\b k$ and $n$ were small enough, so that $\Pbb\_S(\b k)$ and $\Fbb\_S(n)$ could be approximated as the entire smooth subspaces $\Pbb\_S$ and $\Fbb\_S$ when doing the integral.

Now, if the domain of integration ranged over all of $\R^3$, this integral would be straightforward to evaluate using Feynman parameters. Doing this gives the standard result quoted in the literature, namely $1/{8\pi|m|}$. Unfortunately, the domain $\Pbb\_S \times \Fbb\_S$ is actually a cube of size $2l\_S \times 2l\_S \times 2\omega\_S$, where $l\_S \equiv \frac{2\pi}N k\_S \ll 1$. Thus the standard result only holds in a formal sense. Said another way, $1/8\pi|m|$ should be understood to be just a universal part of some $l\_S$- and $\omega\_S$-dependent answer in the limit where $|m| N \gg 1$ and $|m|\beta \gg 1$.

This remark on universality, of course, does not falsify existing results; it merely hints at the difficulty of explicitly connecting the CS effective action with a finite fermion theory. One may still ask whether the structure $\epsilon^{\mu\nu\lambda} \~k^\lambda$ from \eqref{avg JJ conv} may be found in some similarly formal calculation on the canonical side. The obstacle here is that the correlator $\avg{J^\mu(\b k, n) J^\nu(-\b k, -n)}_\beta$ has no well behaved canonical counterpart. The one natural candidate would be the Fourier transform of the time-dependent correlation function \eqref{avg jj tau}. Unfortunately, this function is not periodic in Euclidean time, so its Fourier transform can only be taken in a formal sense --- much like the above integral. These formal extensions will not be studied here. The conclusion of this Subsection must be that there is no straightforward way to extract a canonically defined CS theory out of a massive fermion cQFT.

\subsection{CS from flux attachment}

Implicit in the previous Subsection was the desire to define CS theory as a cQFT. However, as argued in the Introduction and when discussing BF theory in Subsection \ref{subsec BF}, this is really not the right framework for topological field theories. Instead of discussing temporal smoothing and the slippery concept of universality, this Subsection will present an alternative way to define certain CS-like actions. The idea is a direct analogue of using $\Z_K$ gauge theories at weak couplings to define BF theory actions \eqref{def S BF}.

Consider a gauge theory with $\Z_K$ degrees of freedom on links of a lattice $\Mbb$. Instead of having local symmetries generated by Gauss operators $G_v$ from \eqref{def G}, suppose that this theory has symmetry generators
\bel{\label{def G flux attach}
  \G_v \equiv G_v W_{f(v)},
}
where $f(v)$ is some function that assigns a face to each vertex. For concreteness, take $\Mbb$ to be a square lattice and let
\bel{\label{def f}
  f(v) = \trm{NW}(v),
}
meaning that $f(v)$ is the plaquette just northwest of the vertex $v$. The operators \eqref{def G flux attach} can then be written as
\bel{
  \G_{\b x} = G_{\b x} W_{\b x - \b e_1}.
}

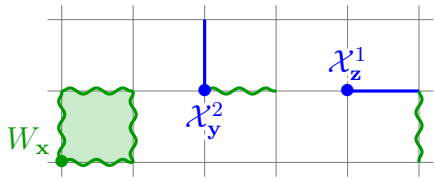
\begin{figure}[b]
\begin{center}
\begin{tikzpicture}[scale = 0.95]
  \contourlength{1.5pt}

  \draw[step = 1, gray] (-2.2, -1.2) grid (3.2, 1.2);

  \fill[fill=green!60!black, fill opacity = 0.2]  (-2, -1) rectangle +(1, 1);
  \draw[green!60!black, very thick,
    style = {decorate, decoration = {snake, amplitude = 0.4mm}}]
    (-2, -1) -- (-1, -1);
  \draw[green!60!black, very thick,
    style = {decorate, decoration = {snake, amplitude = 0.4mm}}]
    (-1, -1) -- (-1, 0);
  \draw[green!60!black, very thick,
    style = {decorate, decoration = {snake, amplitude = 0.4mm}}]
    (-2, -1) -- (-2, 0);
  \draw[green!60!black, very thick,
    style = {decorate, decoration = {snake, amplitude = 0.4mm}}]
    (-2, 0) -- (-1, 0);
  \draw (-2, -0.7) node[left, green!60!black] {\contour{white}{$W_{\b x}$}};
  \draw[green!60!black, thick] (-2, -1) node {$\bullet$};

  \draw[blue, very thick]
    (0, 0) -- (0, 1);
  \draw[green!60!black, very thick,
    style = {decorate, decoration = {snake, amplitude = 0.4mm}}]
    (0, 0) -- (1, 0);
  \draw (0, 0) node[below, blue] {\contour{white}{$\X_{\b y}^2$}};
  \draw[blue, thick] (0, 0) node {$\bullet$};

  \draw[blue, very thick]
    (2, 0) -- (3, 0);
  \draw (2, 0) node[above, blue] {\contour{white}{$\X_{\b z}^1$}};
  \draw[blue, thick] (2, 0) node {$\bullet$};
  \draw[green!60!black, very thick,
    style = {decorate, decoration = {snake, amplitude = 0.4mm}}]
    (3, 0) -- (3, -1);

\end{tikzpicture}
\end{center}
\caption{\small Generators of the gauge-invariant algebra in a flux-attached gauge theory. Straight blue lines represent shift operators $X_{\b x}^i$, and jagged green lines represent clock operators $Z_{\b x}^i$ or $(Z_{\b x}^i)\+$.}
\label{fig flux attachment}
\end{figure}

A theory whose local symmetries are generated by these new operators $\G_v$ is called a \emph{gauge theory with flux attachment} \cite{Chen:2017fvr}. Magnetic excitations in this theory are electrically charged. The gauge-invariant algebra is generated by the Wilson loops $W_{\b x}$ and the flux-attached shift operators
\bel{
  \X_{\b x}^1 \equiv X_{\b x}^1 (Z_{\b x + \b e_1 - \b e_2}^2)\+,
   \quad
  \X_{\b x}^2 \equiv X_{\b x}^2 Z_{\b x}^1.
}
See Fig.\ \ref{fig flux attachment} for an illustration. The Gauss law $\G_v = \1$ can correspondingly be interpreted as a constraint that pairs of magnetic fluxes be connected with lines of electric flux. The specific choice of $f(v)$ in \eqref{def f} means that every gauge-invariant state with equal and opposite elementary magnetic fluxes at plaquettes $f_1$ and $f_2$ must also contain an electric flux line between sites SE$(f_1)$ and SE$(f_2)$, the southeastern corners of $f_1$ and $f_2$.

The topological limit of a Maxwell theory, described by the Hamiltonian \eqref{def H topo}, is invariant under flux-attached gauge symmetries. There is thus no new physics to be found by studying this extreme limit in the presence of flux attachment.

The situation is quite different if one asks for the analogue of the confined phase of a flux-attached gauge theory. A na\"ive generalization of the Maxwell Hamiltonian at large coupling would suggest studying
\bel{\label{def H CS naive}
  H
   =
  \frac{g^2}{2 (\d A)^2}
  \sum_{\b x, i} \left[
    2 - \X_{\b x}^i - (\X_{\b x}^i)\+
  \right]
}
as the epitome of the confined phase in this situation. However, since operators $\X_{\b x}^1$ and $\X_{\b y}^2$ do not commute for all $\b x, \b y \in \Mbb$, this task is difficult. As an alternative, take the commuting projector Hamiltonian
\bel{\label{def H CS}
  H\_{CS}^\star
   =
  \frac{g^2}{2 (\d A)^2}
  \sum_{\b x \in \Mbb} \left[
    2 - \X_{\b x}^2 - (\X_{\b x}^2)\+
  \right].
}
This model will \emph{almost} give rise to a CS action at low temperatures.

The ground state subspace of $H\_{CS}$ consists of states obeying the constraint
\bel{\label{CS conf constraint}
  \X_{\b x}^2 = \1.
}
In terms of the original clock and shift operators, this is the statement that
\bel{\label{CS conjugate operators}
  X_{\b x}^2 = (Z_{\b x}^1)\+.
}
In other words, in this subspace, $Z_{\b x}^1$ and $Z_{\b x}^2$ generate a full clock algebra at each site $\b x$. They are a conjugate pair of clock and shift operators.

Now consider the partition function at temperatures low enough that only the contribution from the ground states needs to be included. Without any gauge constraints in play, there are just $N^2$ independent constraints \eqref{CS conf constraint} to impose upon a Hilbert space of dimension $K^{2N^2}$. The partition function is equal to the resulting ground state degeneracy
\bel{
  \trm{GSD} = K^{N^2}.
}
Restricting to the gauge-invariant sector $\G_{\b x} = \1$ introduces a further $N^2 - 1$ constraints (if $N$ is odd), or $N^2 - 2$ constraints (if $N$ is even). The corresponding degeneracy is
\bel{\label{CS gsd}
  \trm{GSD}\big |_{\varrho_{\b x} = 0} = K^{2 - N\, \trm{mod}\, 2}.
}
Eigenvalues of flux-attached Gauss operators are here labeled by $\e^{\i \varrho_{\b x} \d A}$ for $0\leq \varrho_{\b x} < K$. This result already indicates that $H^\star\_{CS}$ will not quite define a straightforward CS theory.

To see more precisely how this theory departs from CS, it is instructive to write out the partition function leading to the result \eqref{CS gsd} using the transfer matrix formalism. The procedure is the same as in Subsection \ref{subsec BF}, only the constraints are different. Since neither $\X^2_{\b x} = \1$ nor $\G_{\b x} = \1$ are familiar constraints, this computation will be presented in detail.

The partition function may be written as
\bel{\label{CS part fn}
  \Zf
   =
  \sum_{\{A_\tau^i\}}
  \prod_{\tau = \d\tau}^\beta
    \qmat{A^i_{\tau + \d\tau}} {\trm P\_{Gauss} \trm P\_{conf}} {A^i_\tau}.
}
The sum goes over all configurations $A_{\b x, \tau}^i$ that label eigenstates of $Z^i_{\b x}$ at different times $\tau$. At each $x \equiv (\b x, \tau)$, the field $A_x^i$ takes $K$ different values between $0$ and $2\pi$. This is all the same as in Section \ref{sec Maxwell path int}. Since the interest is in the ground-state and gauge-invariant subspace, in lieu of $\e^{-\d\tau H\_{CS}^\star}$ it is sufficient to insert two projection operators to ensure (flux-attached) gauge invariance and the ``confinement'' constraint \eqref{CS conf constraint}. These projectors are
\bel{
  \trm P\_{Gauss}
   =
  \prod_{\b x \in \Mbb} \left(
    \frac1K \sum_{n_{\b x} = 1}^K \G_{\b x}^{n_{\b x}}
  \right)
   =
  \frac1{K^{N^2}}
  \sum_{\{n\}}
    \prod_{\b x \in \Mbb} G_{\b x}^{n_{\b x}} W_{\b x - \b e_1}^{n_{\b x}}
}
and
\bel{
  \trm P\_{conf}
   =
  \prod_{\b x \in \Mbb} \left(
    \frac1K \sum_{m_{\b x} = 1}^K (\X^2_{\b x})^{m_{\b x}}
  \right)
   =
  \frac1{K^{N^2}}
  \sum_{\{m\}}
    \prod_{\b x \in \Mbb} (X^2_{\b x})^{m_{\b x}} (Z^1_{\b x})^{m_{\b x}}.
}

The actions of these projectors can be recorded as
\gathl{
  \trm P\_{conf} \qvec{A^i_\tau}
   =
  \frac1{K^{N^2}}
  \sum_{\{m_\tau\}}
    \e^{\i \sum_{\b x \in \Mbb} m_{\b x, \tau} A^1_{\b x, \tau}}
    \qvec{A^i_\tau - \delta_{i, 2} \, m_\tau  \d A}
   \\
  \qvecconj{A^i_{\tau + \d\tau}} \, \trm P\_{Gauss}
   =
  \frac1{K^{N^2}}
  \sum_{\{n_{\tau}\}}
    \qvecconj{A^i_{\tau + \d\tau} -  (\del_i n_{\tau}) \d A} \,
    \e^{-\i \sum_{\b x \in \Mbb} n_{\b x + \b e_1, \tau} B_{\b x, \tau + \d\tau}}.
}
This calculation discards terms quadratic in $m$, as they will not survive any smoothing that is performed at the end. Taking the inner product of the above two results for each $\tau$ gives the matrix element
\algns{\label{CS mat elem}
  \qmat{A^i_{\tau + \d\tau}} {\trm P\_{Gauss} \trm P\_{conf}} {A^i_\tau}
   =
  \frac1{K^{2N^2}}
  \sum_{\{n_\tau, m_\tau\}}
  & \Big[
    \e^{
      \i \sum_{\b x \in \Mbb} \left(
        m_{\b x, \tau} A^1_{\b x, \tau} - n_{\b x + \b e_1, \tau} B_{\b x, \tau + \d\tau}
      \right)
    } \times \\
  &\quad \times
  \delta_{\del_0 A^i_{\b x, \tau} \d\tau  - (\del_i n)_{\b x, \tau} \d A + \delta_{i, 2} m_{\b x, \tau} \d A,\, 0} \Big].
}
The Kronecker delta imposes two relations between fields,
\gathl{\label{CS eom}
  \del_0 A^1_x \d\tau - \del_1 n_x \d A = 0, \quad
  \del_0 A^2_x \d\tau - \del_2 n_x \d A + m_x \d A = 0.
}

The goal is in sight. Let
\bel{
  A_x^0 \equiv  n_x \der A \tau,
}
which is the same definition of $A_x^0$ as in BF theory, cf.\ \eqref{def A0 BF}. By the second equation in \eqref{CS eom}, $m_x$ is fixed to \bel{
  m_x = (\del_2 A^0_x - \del_0 A_x^2) \der \tau A.
}
Substituting this, and the definition of $A_x^0$, into \eqref{CS mat elem} and then back into \eqref{CS part fn} gives
\bel{
  \Zf
   =
  \frac1{K^{2N^2 N_0}}
  \sum_{\{A^\mu_\tau\}}
    \e^{\i \sum_{x \in \Ebb} \der\tau A \left[
      (\del_2 A_x^0 - \del_0 A_x^2) A_x^1 - A_{x + e_1 - e_0}^0 (\del_1 A_x^2 - \del_2 A_x^1)
    \right]}.
}
By the first equation in \eqref{CS eom}, there is no obstacle to adding $0 = A_x^2 (\del_0 A^1_x - \del_1 A^0_x) \der \tau A$ to the action. Thus the partition function becomes
\bel{
  \Zf
   =
  \frac1{K^{2N^2 N_0}}
  \sum_{\{A^\mu_\tau\}}
    \e^{-S\_{CS}[A^\mu] - \Delta S[A^\mu]},
}
where
\gathl{\label{def S CS}
  S\_{CS}
   =
  -\frac{\i\kappa}{4\pi}
  \sum_{x \in \Ebb} \d\tau \left[
    A^0_{x + e_1 - e_0} F^{12}_x + A_x^1 F^{20}_x + A^2_x F^{01}_x
  \right],  \\
  \Delta S
   = \frac{\i \kappa}{2\pi} \sum_{x \in \Ebb} \d\tau A^0_{x + e_1 - e_0} F^{12}_x, \quad \kappa \equiv 2K.
}
This explicitly shows how the above theory differs from a ``pure'' CS action. Beyond the obvious discrepancy in the term $\Delta S$, there is a number of points worth making:
\begin{itemize}
  \item An immediate departure from the familiar CS form is the shift of $A^0_x$ along the $0$- and $1$-directions in the first term. This shift is due to particular choices made en route to this result, e.g.\ the choice of $f(v)$ in \eqref{def f}, or the use of $\X^2_{\b x}$ as opposed to $\X^1_{\b x}$ in \eqref{def H CS}. Different choices will lead to slightly different answers. This issue is not quite trivial, as the dependence on these choices reflects the more general dependence of the theory on a branching structure on the lattice \cite{Chen:2017fvr}. However, it is also evident that if the action is (spatially and temporally) smoothed, these choices will become invisible.
  \item Just like with the BF action \eqref{def S BF}, the path integral variables are very clearly \emph{not} smooth, real-valued fields. For $K = 2$, say, each field $A^\mu_x$ takes just two different values. There is no justified approximation that would allow these fields to become smoothly varying. Of course, it is still possible to formally extend the domain of all gauge fields to the whole $\R$ (or to a circle). Once this is done, the canonical interpretation of the action in terms of a simple Hamiltonian like \eqref{def H CS} is lost.
  \item Cognoscenti will not be surprised that the CS level $\kappa$ turned out to be even in this construction. It is a well know fact that odd-level Abelian CS requires a choice of a spin structure. (See \cite{Witten:2003ya} for a clear and concise exposition.) Practically, a spin structure is a specific background $\Z_2$ gauge field \cite{Radicevic:2018okd}, and no such choice was consciously made here. It remains unknown how to obtain something that looks like a spin (odd $\kappa$) CS theory using this approach.
  \item Another --- perhaps even more fundamental --- reason why a theory like \eqref{def H CS} could not have given a proper CS action is the lack of any explicit breaking of the time reversal symmetry.
  \item The particular theory \eqref{def H CS} is dual to an uncoupled stack of parafermions that live in one spatial dimension. In this case the spatial manifold of these parafermions is a circle parallel to the $1$-direction. This situation has a certain appeal, as introducing a boundary parallel to this direction would result in a parafermionic edge mode. In particular, if $K = 2$, this would give a theory of a Majorana fermion on the edge. This is not a free chiral fermion, however; this is another sense in which one can see that \eqref{def H CS} does not lead to a proper CS theory. There are other unappealing features too. For example, a boundary parallel to the $2$-direction would result in an edge mode localized in space. This action is therefore far from what one might call ``diffeomorphism-invariant.''
  \item The torus GSD of the theory \eqref{def H CS} is $K = \kappa/2$ when $N$ is odd. This means that this theory misses exactly half of the CS ground states, as its torus GSD is supposed to be $\kappa$. Roughly speaking, the missed states are the excitations that can go in the $2$-direction.
  \item Another perspective comes from studying the topological excitations of this theory \cite{Kitaev:2006lla}. While gauge invariance is enforced, the only available topological excitation is described by a state in which the condition \eqref{CS conf constraint} is violated. These massive particles must move along NW-SE diagonals by the application of a $\X^1_{\b x}$ operator, and they may be created in fours by a $W_{\b x}$ operator. This limited mobility is another sense in which the theory \eqref{def H CS} has only half the excitations as a regular CS theory. In fact, this is quite reminiscent of fracton theories \cite{Nandkishore:2018sel}.
  \item Finally, it is rather surprising that any kind of CS theory might be found as the ``confining'' limit of some lattice gauge theory. This means that as $g^2$ is decreased in a Maxwell-like generalization of \eqref{def H CS} there may exist an interesting deconfinement transition.
\end{itemize}

This all means that the model \eqref{def H CS} was a bit too simplistic, but that the flux-attached theories have the potential to give rise to CS actions in carefully chosen models. An analysis of more complicated but related theories will be presented elsewhere.

\newpage

\section{Remarks}

This paper has shown how various Abelian continuum gauge theories in $(2 + 1)$D may be defined directly in terms of lattice variables. The smoothing used in this definition was designed to work at the level of entire operator algebras, with the proximity to the familiar continuum behavior governed by a small parameter $k\_S/N$. At no point was it necessary to take a ``formal'' continuum limit $a \rar 0$; the cQFTs defined here all ultimately live on large but finite lattices. This approach clarifies many foundational questions in both the canonical and the path integral formalism. While this paper does not pretend to be up to a mathematician's standards, the definitions of the relevant cQFTs are all rigorous, and the various tameness and smoothness bounds (e.g.\ the limit on the winding number below \eqref{quantization pi vee}) can be made totally explicit if needed.

This study can be extended along two very natural directions. The first one concerns nonabelian gauge theories. These theories do not have ready discretizations: there exists no sequence of finite groups that converges to, say, SU(2). This means that a precise notion of taming, as introduced in \cite{Radicevic:1D} and applied to gauge theories in Subsection \ref{subsec tame gauge transf}, will not be available in a nonabelian setup. This issue, together with generalizing to theories with multiple matter flavors, will be tackled in the next part of this series, which will simultaneously increase the spacetime dimension to four in an attempt to become more relevant to our world \cite{Radicevic:4D}.

The second natural way to extend this series is to systematically address the various concepts of universality that were touched upon. More than the previous two installments, this paper has shown how much contemporary field theorists rely on universality when modifying the microscopic theory in order to get something more tractable, for example:
\begin{itemize}
  \item Including configurations with nontrivial Polyakov loops (Subsection \ref{subsec standard nc Maxwell}).
  \item Temporally smoothing all path integrals (Subsection \ref{subsec temp smoothing}).
  \item Replacing the interacting lattice QED with conventional QED (Subsection \ref{subsec conv QED}).
  \item Treating topological QFTs like BF theory (Subsection \ref{subsec BF}) and CS theory (Subsection \ref{subsec CS mass ferm}) as ordinary cQFTs with smoothly varying fields.
\end{itemize}
This paper made a concerted effort to separate issues of universality from other statements that can be made about the lattice-continuum correspondence. After all, even without using universality to justify any of the modifications of the microscopic theories, each modified theory was still well defined. A future study of universality would, ideally, precisely explain how each of these modifications tracks (or deviates from) the original theory.

This paper has also focused only on foundational questions --- how to consistently define various theories, how to justify the tameness or smoothness of their algebras, and how to reproduce the most familiar continuum hallmarks. This sets the stage for a number of new analyses of many dynamical questions. For example, does the conventional QED really behave as depicted on Fig.\ \ref{fig deconf}?  Does the smoothing approach at all help improve the precision and efficiency of numerical simulations? How to quantify the importance of uniquely 3D solitonic configurations in a smooth path integral?\footnote{In $(0+1)$D, this question was addressed in some detail in the first paper of this series \cite{Radicevic:1D}. There it was shown that the path integral of a theory with  smooth but untame low energy states must include nontrivial configurations that wind along the thermal circle in order to reproduce the correct partition function. The connection between the lack of target space tameness and the necessity of including nontrivial path integral configurations presumably extends to higher dimensions, but it has not been explored here.} How to microscopically understand the bulk-edge correspondence of CS and other topological theories?  Most importantly, how to \emph{prove} the tameness assumptions?  These are just some of the many studies that can be built atop the definitions presented here.

\section*{Acknowledgments}

It is a pleasure to thank Nathan Benjamin, Lorenzo Di Pietro, Matt Headrick, and David Poland for useful conversations. This work was completed with the support from the Simons Foundation through \emph{It from Qubit: Simons Collaboration on Quantum Fields, Gravity, and Information}, and from the Department of Energy Office of High-Energy Physics grant DE-SC0009987 and QuantISED grant DE-SC0020194.

\bibliographystyle{ssg}
\bibliography{Refs}

\end{document}